\documentclass{emulateapj}
\usepackage{graphicx}
\usepackage{epstopdf}
\usepackage{epsfig}
\usepackage{natbib}
\begin{document}
\newcommand{\z}{\emph{z} $\sim$ }
\newcommand{\kms}{km s$^{-1}$}
\newcommand{\lya}{Ly$\alpha$ }
\makeatletter
\newcommand{\Rmnum}[1]{\expandafter\@slowromancap\romannumeral #1@}
\makeatother

\title{Fine-Structure Fe~II$^*$ Emission and Resonant Mg~II \\ Emission in \z1 Star-Forming Galaxies\altaffilmark{1}}
\author{\sc Katherine A. Kornei and Alice E. Shapley\altaffilmark{2}}
\affil{Department of Physics and Astronomy, University of California, Los Angeles, CA 90025, USA}
\author{\sc Crystal L. Martin}
\affil{Physics Department, University of California, Santa Barbara, CA 93106, USA}
\author{\sc Alison L. Coil\altaffilmark{3}}
\affil{Center for Astrophysics and Space Sciences, Department of Physics, University of California, San Diego, CA 92093, USA}
\author{\sc Jennifer M. Lotz}
\affil{Space Telescope Science Institute, Baltimore, MD 21218, USA}
\author{\sc Benjamin J. Weiner}
\affil{Steward Observatory, University of Arizona, Tucson, AZ 85721, USA}

\altaffiltext{1}{Based, in part, on data obtained at the W.M. Keck Observatory, which is operated as a scientific partnership among the California Institute of Technology, the University of California, and NASA, and was made possible by the generous financial support of the W.M. Keck Foundation.} 
\altaffiltext{2}{Packard Fellow.}
\altaffiltext{3}{Alfred P. Sloan Fellow.}

\begin{abstract} We present a study of the prevalence, strength, and kinematics of ultraviolet Fe~II and Mg~II emission lines in 212 star-forming galaxies at \z1 selected from the DEEP2 survey. We find Fe~II$^*$ emission in composite spectra assembled on the basis of different galaxy properties, indicating that Fe~II$^*$ emission is common at \z1. In these composites, Fe~II$^*$ emission is observed at roughly the systemic velocity. At \z1, we find that the strength of Fe~II$^*$ emission is most strongly modulated by dust attenuation, and is additionally correlated with redshift, star-formation rate, and [O~II] equivalent width, such that systems at higher redshifts with lower dust levels, lower star-formation rates, and larger [O~II] equivalent widths show stronger Fe~II$^*$ emission. We detect Mg~II emission in at least 15\% of the individual spectra and we find that objects showing stronger Mg~II emission have higher specific star-formation rates, smaller [O~II] linewidths, larger [O~II] equivalent widths, lower dust attenuations, and lower stellar masses than the sample as a whole. Mg~II emission strength exhibits the strongest correlation with specific star-formation rate, although we find evidence that dust attenuation and stellar mass also play roles in the regulation of Mg~II emission. Future integral field unit observations of the spatial extent of Fe~II$^*$ and Mg~II emission in galaxies with high specific star-formation rates, low dust attenuations, and low stellar masses will be important for probing the morphology of circumgalactic gas.\end{abstract}

\section{Introduction}

The transport of gas into and out of galaxies has been recorded at a range of redshifts \citep[e.g.,][]{heckman1990,steidel1996,franx1997,martin1999,pettini2000,pettini2001,shapley2003,martin2005,veilleux2005,rupke2005,tremonti2007,weiner2009,steidel2010,coil2011}. This cycling of baryons is an integral component of galaxy evolution as galactic winds are thought to drive the mass-metallicity relation \citep[e.g.,][]{tremonti2004,erb2006}, enrich the intergalactic medium in metals \citep[e.g.,][]{bordoloi2011,menard2011}, and regulate both star formation and black hole growth \citep[e.g.,][]{tremonti2007,gabor2011}. 

In the local universe, galactic winds are revealed through H$\alpha$ and X-ray imaging of high surface brightness gas seen in emission around the disks of starburst galaxies \citep[e.g.,][]{lehnert1996}. At higher redshifts, however, studies of galactic winds often rely on spectral data tracing foreground gas absorbed against the light of background galaxies or quasars \citep[e.g.,][]{sato2009,weiner2009,steidel2010,rubin2010a,coil2011}. While absorption lines unambiguously probe gas between Earth and a more distant light source, emission lines can arise from either foreground or background gas due to scattering. Observations of emission lines associated with galactic winds can be used to map the spatial extent of circumgalactic gas \citep[e.g.,][]{rubin2010c} and measurements of emission lines accordingly comprise rich data sets complementing absorption-line studies. 

\begin{figure}
\centering
\includegraphics[width=3.5in]{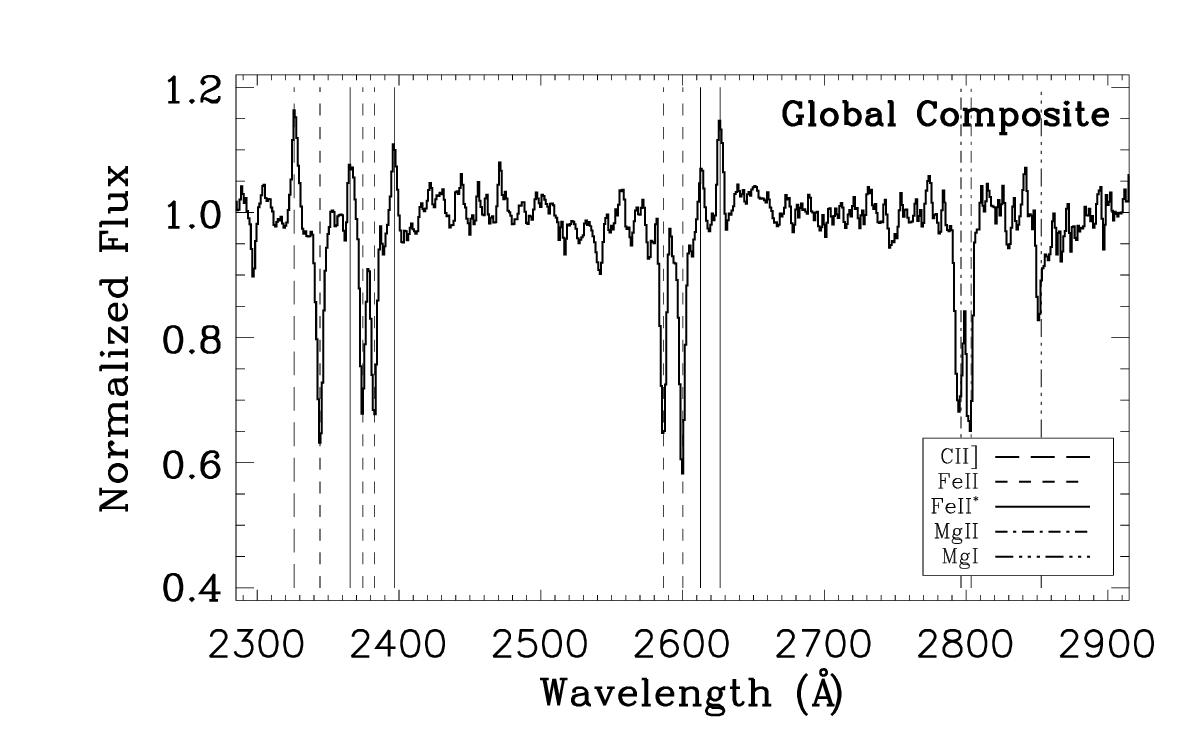} 
\caption{Composite spectrum of all the data in our sample. Prominent interstellar absorption lines (Fe~II, Mg~II, and Mg~I) and emission lines (C~II], fine-structure Fe~II$^*$) are labeled, where the C~II] line is a blend of several features and we mark a wavelength of 2326 \AA\ in the figure. The absorption line at 2297 \AA\ is a stellar C~III line. The S/N of this composite spectrum is $\sim$39 pixel$^{-1}$.}
\label{composite_blue_smooth}
\end{figure}

The H~I \lya line at 1216 \AA\ is an example of a well-studied resonant feature associated with galactic winds. This line has been observed in emission at \z3 in star-forming galaxies \citep{shapley2003}. At lower redshifts, Mg~II resonant emission at $\sim$2800 \AA\ or Na~I D resonant emission at $\sim$5900 \AA\ are typically used as emission-line probes of galactic winds. One of the first spectroscopic observations of resonant emission associated with outflowing gas was a Na~I P-Cygni profile in the local starburst galaxy NGC 1808 \citep{phillips1993}, although recombination features have also been used as tracers of galactic winds \citep[e.g.,][]{heckman1990,martin1998,genzel2011}. Fine-structure emission -- in which a photon is emitted following an electronic transition to an excited ground state -- is necessarily related to resonant emission lines as the upper electronic states of both kinds of features are populated by resonant absorption. Examples of fine-structure emission lines include Si~II$^*$ and Fe~II$^*$ features, where we adopt the convention of denoting fine-structure lines with an asterisk. In this paper, we present observations of Fe~II$^*$ emission at \z1 to study its prevalence and kinematics. Many authors have noted resonant and fine-structure emission lines in diverse samples of star-forming galaxies at 0.3 $<$ \emph{z} $<$ 4 hosting galactic winds \citep{shapley2003,martin2009,weiner2009,rubin2010a,rubin2010c,coil2011,kornei2012,martin2012,jones2012,talia2012,erb2012}. 

While kinematic measurements of emission lines should in principle be a useful diagnostic of the origin of the line-emitting gas, work by \citet{prochaska2011} has shown that it is difficult to determine if emission lines arise from galactic winds or star-forming regions. These authors propose that gas flows and stationary H~II regions can imprint similar kinematic signatures on line emission. As emission from an optically thin source will be visible over both its approaching and receding (i.e., blueshifted and redshifted) sides, the emission profile can remain centered at roughly 0 \kms\ while still tracing a gas flow. Likewise, an emission line arising from a stationary H~II region will exhibit a line profile largely at the systemic velocity. \citet{rubin2010c} investigated fine-structure Fe~II$^*$ emission in a starburst galaxy at \z0.7 and concluded that since the emission was redward or within 30 \kms\ of the systemic velocity, the velocity profile of fine-structure Fe~II$^*$ emission is significantly different from both absorption lines tracing galactic winds and nebular lines associated with H~II regions. \citet{coil2011} reported that the Fe~II$^*$ emission lines in a sample of 11 post-starburst and active galactic nucleus (AGN) host galaxies at 0.2 $<$ \emph{z} $<$ 0.8 are within 2$\sigma$ of the systemic velocity for all but two galaxies. At higher redshift, \citet{erb2012} studied 96 star-forming galaxies at 1 $\lesssim$ \emph{z} $\lesssim$ 2 and found that the measured velocities of fine-structure Fe~II$^*$ emission scattered around 0 \kms\ and that the similarity of the Fe~II$^*$ and Fe~II line profiles indicate that the lines arise from the same gas. At \z3, \citet{shapley2003} measured an average velocity of 100 $\pm$ 35 \kms\ for fine-structure Si~II$^*$ emission, although these authors cautioned that the presence of nearby absorption features may bias the emission centroids to more redshifted values. While these collective measurements have shown that fine-structure emission is generally observed at or near the systemic velocity, additional data obtained with higher resolution spectrographs are needed in order to more precisely investigate the kinematics of fine-structure emission. 

\begin{figure}
\begin{center}$
\begin{array}{c}
\includegraphics[width=3.5in]{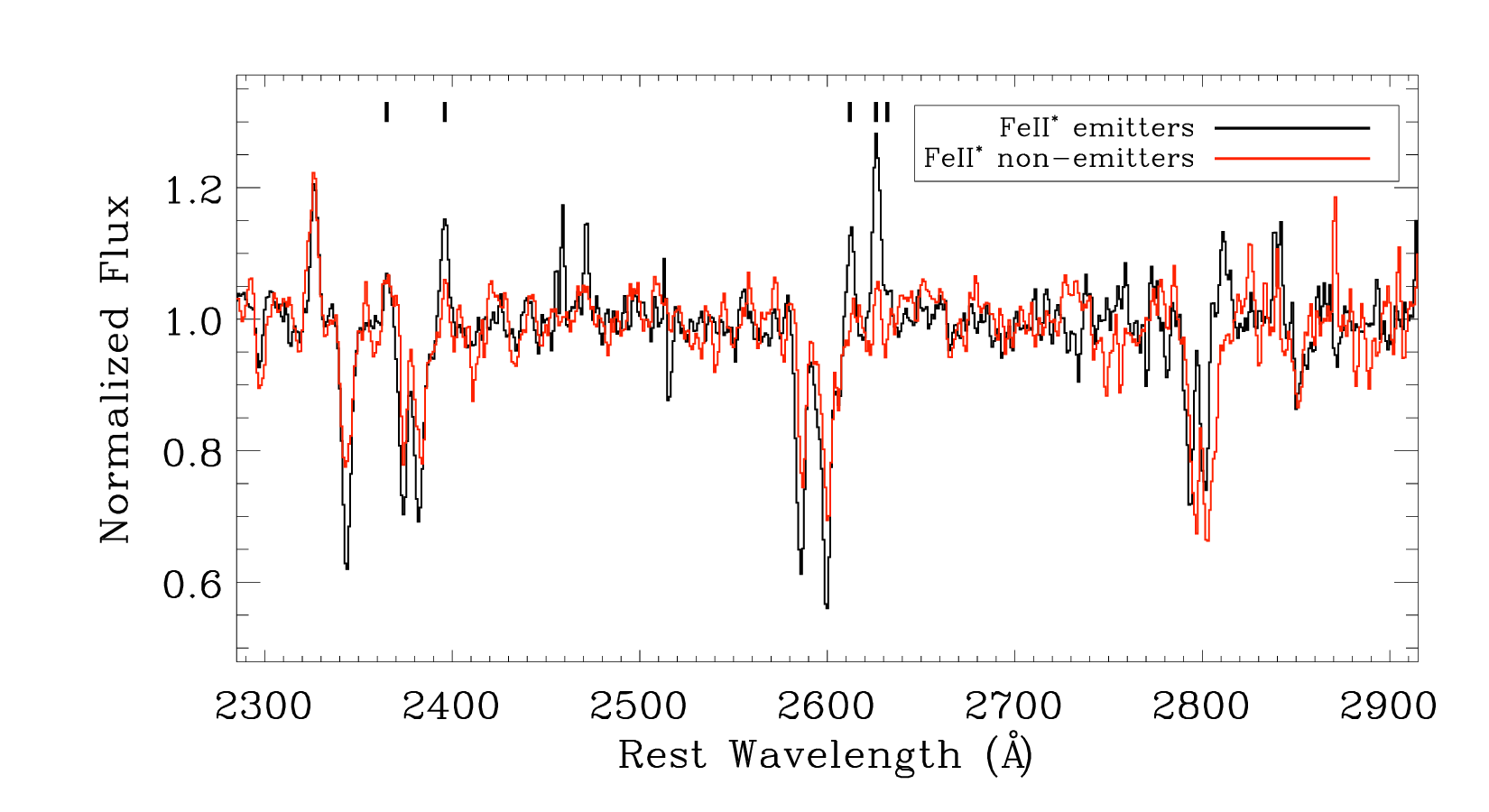} 
\end{array}$
\end{center}
\caption{A comparison of composite spectra assembled from Fe~II$^*$ emitters (black) and non-emitters (red). Short vertical lines delineate the locations of Fe~II$^*$ emission features at 2365, 2396, 2612, 2626, and 2632 \AA; the emission feature at 2326 \AA\ is a blend of several C~II] transitions. On average, Fe~II$^*$ emitters show deeper Fe~II absorption, weaker Mg~II absorption, and more blueshifted Mg~II absorption than Fe~II$^*$ non-emitters.}
\label{FeII_smooth}
\end{figure}

The prevalence of emission lines has been found to vary widely among different samples, with objects at higher redshifts more commonly exhibiting emission lines. Studies at 0.5 $<$ \emph{z} $<$ 4 have detected fine-structure Si~II$^*$, fine-structure Fe~II$^*$, and resonant Mg~II emission lines \citep[e.g.,][]{shapley2003,weiner2009,rubin2010a,kurk2012,jones2012}. In a study of 1406 star-forming galaxies at \z1.4, \citet{weiner2009} found that $\sim$4\% of objects showed excess Mg~II emission; these authors attributed the presence of Mg~II emission to either low-level AGNs\footnote{\citet{weiner2009} still observed Mg~II emission, however, in a sample where the AGN candidates had been removed.} or scattering off the backside of the galactic wind. \citet{rubin2010c} detected both Fe~II$^*$ and Mg~II emission in a starburst galaxy at \z0.7 and measured that the Mg~II emission was spatially extended to distances of $\sim$7 kpc. Local star-forming galaxies, on the other hand, do not show fine-structure emission lines \citep{leitherer2010}. \citet{giavalisco2011} and \citet{erb2012}, among others, suggest that slit losses are responsible for the lack of emission in nearby samples, given that spectroscopic observations in the local universe typically probe only the inner regions of galaxies where the emission may not originate.

While previous studies have collectively shown that emission lines are present in galaxies at \emph{z} $\gtrsim$ 0.5 exhibiting galactic winds, a systematic analysis of the prevalence and properties of emission lines as a function of host galaxy stellar populations, star-formation rate (SFR), SFR surface density, and outflow characteristics has thus far been absent from the literature. We present here an investigation of the frequency, strength, and kinematics of rest-frame ultraviolet fine-structure Fe~II$^*$ and resonant Mg~II emission lines in a sample of 212 galaxies at 0.2 $<$ \emph{z} $<$ 1.3 (all but four of which are at \emph{z} $>$ 0.4) for which stellar populations and outflow properties have been estimated. In Section \ref{sec: data}, we present the observations and in Section \ref{sec: determining_v} we discuss how outflow velocities were measured. Section \ref{sec: feIIstar} summarizes the Fe~II$^*$ emission features seen in the data while Section \ref{sec: mgII_introduction} is devoted to Mg~II emission observations. A discussion appears in Section \ref{sec: discussion} and conclusions are presented in Section \ref{sec: conclusions}. Throughout the paper, we assume a standard $\Lambda$CDM cosmology with \emph{H}$_{\rm 0}$ = 70 km s$^{-1}$ Mpc$^{-1}$, $\Omega_{\rm M}$ = 0.3, and $\Omega_{\Lambda}$ = 0.7. All wavelengths are measured in vacuum. At \emph{z} = 0.7 (1.3), an angular size of 1$''$ corresponds to 7.1 (8.4) kpc. 
 
\section{Observations} \label{sec: data}
 
We discuss the details of our observations in \citet{martin2012}. The 212 objects in our sample are drawn from the DEEP2 survey \citep{newman2012} utilizing the DEep Imaging Multi-Object Spectrograph (DEIMOS) on Keck \Rmnum{2}. While the DEIMOS spectra are generally dominated by nebular emission features, the majority of low- and high-ionization interstellar absorption features tracing outflows are in the rest-frame ultraviolet and are observed at shorter wavelengths than the blue edge of the typical DEIMOS spectra ($\sim$6500 \AA\ in the observed frame). In order to probe these outflow features (e.g., Fe~II $\lambda$2344, Fe~II $\lambda \lambda$2374,2382, Fe~II $\lambda \lambda$2587,2600, Mg~II $\lambda \lambda$2796,2803), we obtained spectroscopic data using the LRIS spectrograph on Keck \Rmnum{1}. 

\begin{figure}
\centering
\includegraphics[width=3.5in]{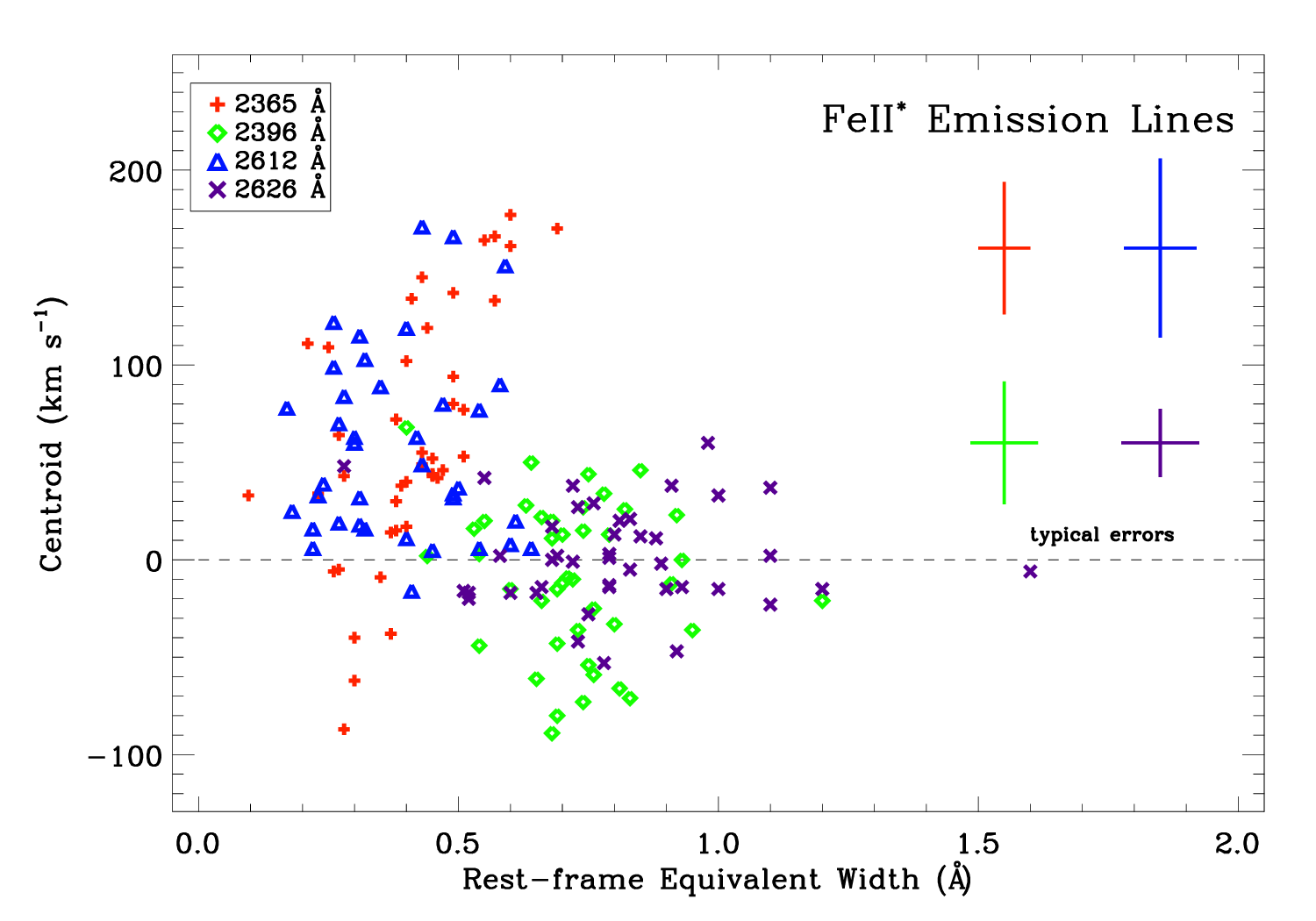} 
\caption{Fe~II$^*$ kinematics versus equivalent width for a set of high S/N composite spectra assembled according to different galaxy properties. Measurements of the four strongest Fe~II$^*$ emission lines are plotted and representative errors are shown in the lower right hand corner. The data points are not independent as there is substantial overlap of objects among the composite spectra. Stronger Fe~II$^*$ features at 2396 and 2626 \AA\ show kinematics scattering around 0 \kms\ while weaker Fe~II$^*$ lines at 2365 and 2612 \AA\ are on average redshifted by $\sim$100 \kms. The 2626 \AA\ feature is isolated from neighboring absorption lines and its centroid is therefore the most robust of the Fe~II$^*$ features. We accordingly conclude that the kinematics of Fe~II$^*$ emission are consistent with the systemic velocity.}
\label{IDLgaussian_plot}
\end{figure}

These LRIS data are described in detail in \citet{martin2012} and we provide here only a summary. The published \citet{martin2012} sample is inclusive of 208 galaxies; we include four additional low-redshift galaxies in this work to bring the sample total to 212 objects. The LRIS data were collected from 2007--2010 using 1$.''$2 slits on multi-object slitmasks targeting 20--28 objects each. We used two set-up configurations, both with the atmospheric dispersion corrector: the d680 dichroic with the 400 line mm$^{-1}$ grism and the 800 line mm$^{-1}$ grating (145 objects) and the d560 dichroic with the 600 line mm$^{-1}$ grism and the 600 line mm$^{-1}$ grating (67 objects). Integration times ranged from 3--9 hours per slitmask, where objects observed with the d560 dichroic had typically shorter exposures (3--5 hours) than objects observed with the d680 dichroic (5--9 hours). The slitmasks used with the d560 dichroic were reserved for brighter objects observed in poorer conditions. The resolutions of the 800, 600 and 400 line mm$^{-1}$ gratings/grisms are $R$ = 2000, 1100, and 700, respectively, and the reduction procedure -- flat-fielding, cosmic ray rejection, background subtraction, one-dimensional extraction, wavelength and flux calibration, and transformation to the vacuum wavelength frame -- was completed using {\tt IRAF} scripts \citep{martin2012}. The continuum signal-to-noise (S/N) ratios of the LRIS observations over the rest wavelength interval 2400--2500 \AA\ range from $\sim$1--25 pixel$^{-1}$ with a median of 6 pixel$^{-1}$. The average redshift of the sample is $\langle \emph{z} \rangle$ = 0.99 and the full range of redshifts is 0.2 $<$ $\emph{z}$ $<$ 1.3. 

The spectra were continuum normalized and composite spectra were assembled from stacks of mean-combined rest-frame spectra. No weights were applied in this stacking procedure. Since all data were continuum normalized prior to stacking, this procedure ensures that systems with differing luminosities contribute evenly to the stack and that the stack represents a mean in equivalent width. In assembling the composites, we smoothed the spectra of objects obtained with the 600 line mm$^{-1}$ grism or grating in order to account for the difference in resolution between those obtained with the 600 line mm$^{-1}$ and 400 line mm$^{-1}$ setups. In Figure \ref{composite_blue_smooth}, we show the composite spectrum assembled from all of the data in our sample. Fe~II, Mg~II, and Mg~I resonant absorption lines are significantly detected in the composite spectrum, as are Fe~II$^*$ and C~II] emission lines. Table \ref{abtable} summarizes the Fe~II absorption line and Fe~II$^*$ emission-line strengths of this composite spectrum. 

\begin{figure}
\centering
\includegraphics[trim = 0in 2.5in 2.7in 0in,clip,width=3.5in]{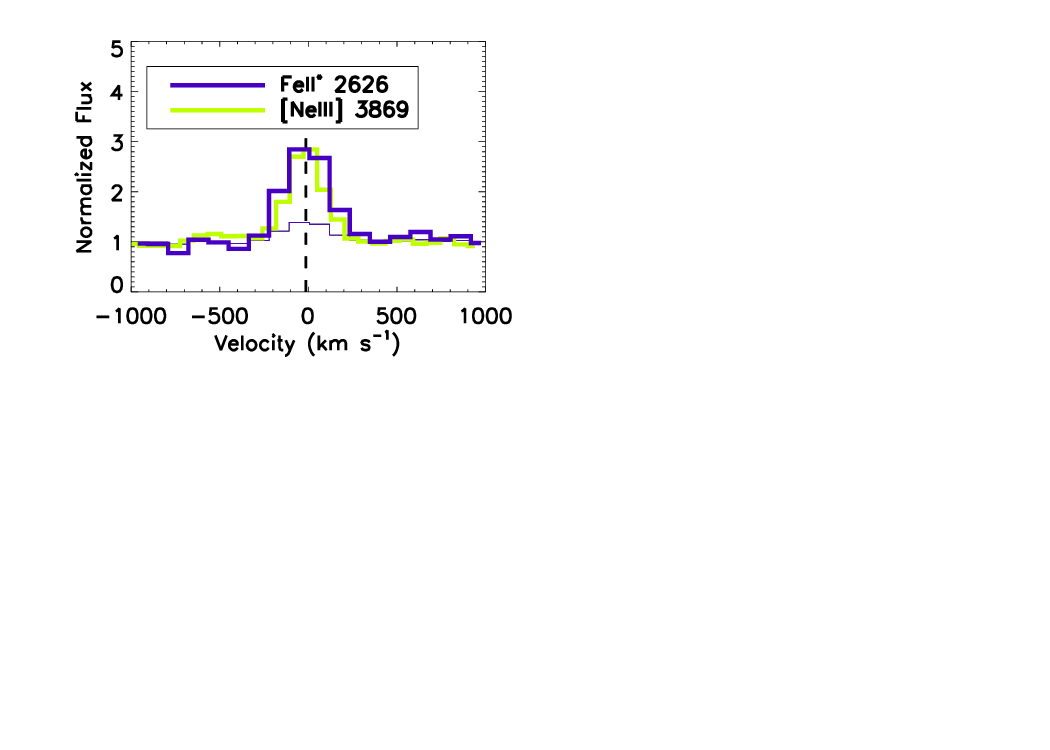} 
\caption{Comparison of scaled Fe~II$^*$ and [Ne~III] profiles for a spectral stack of the 12 objects showing $\ge$ 2$\sigma$ detections of Fe~II$^*$ 2626 \AA\ and $\ge$ 3$\sigma$ detections of the nebular emission line [Ne~III] 3869 \AA. The unscaled Fe~II$^*$ profile is shown as a thin line. The kinematics of Fe~II$^*$ and [Ne~III] are similar, with Fe~II$^*$ showing a velocity offset of --12 
\kms\ from systemic velocity and [Ne~III] shifted by --16 
\kms\ (overlapping vertical dashed lines). Given the uncertainty on the systemic redshift determination (Section \ref{sec: determining_v}), we conclude that the centroids of both the Fe~II$^*$ 2626 \AA\ and [Ne~III] 3869 \AA\ emission lines are consistent with 0 \kms.}
\label{composite_nebular}
\end{figure}

72 of the galaxies observed with LRIS fall in the Extended Groth Strip and accordingly have extensive multi-wavelength coverage from the All-Wavelength Extended Groth Strip International Survey \citep[AEGIS;][]{davis2007}. AEGIS observations cover a broad range in wavelength; we utilized \emph{Chandra} X-ray, \emph{Galaxy Evolution Explorer} (\emph{GALEX}) FUV and NUV imaging, \emph{Hubble Space Telescope} (\emph{HST}) ACS F606W ($V$) and F814W ($I$) imaging, \emph{BRI} Canada France Hawaii Telescope and Palomar/WIRC \emph{J} and $K_s$ imaging, and \emph{Spitzer} IRAC and MIPS pointings in our analyses. We specifically employed \emph{GALEX} observations to estimate dust-corrected SFRs and \emph{HST} imaging to estimate disk inclinations and galaxy areas. We calculated dust-corrected SFRs using UV measurements from \emph{GALEX}, where the dust correction was estimated based on the relationship between the spectral slope $\beta$ and dust extinction \citep{seibert2005}. $\beta$ parameterizes the slope of the flux over the rest-frame interval 1250--2500 \AA\ ($f_{\lambda}$ $\propto$ $\lambda^{\beta}$). The inclination of resolved galactic disks was estimated from axis ratios of rest-frame UV imaging, under the assumption that galaxy disks are intrinsically circular. We estimated galaxy areas using a methodology that flags only the brightest star-forming clumps satisfying a threshold star-formation rate surface density. In this paper, we also utilize stellar masses calculated from SED modeling with \emph{BRIK} photometry, assuming \citet{bruzual2003} spectral templates and a \citep{chabrier2003} initial mass function. For our study, modeling was done with \emph{BRI} photometry alone if objects lacked \emph{K}--band detections. We refer the reader to \citet{kornei2012} for a full description of these properties.

\section{The Determination of Systemic and Outflow Velocities} \label{sec: determining_v}

We discuss our procedure for determining outflow velocities in \citet{martin2012} and provide here only a summary of the methodology. As the kinematics of galactic winds are only meaningful when compared to a systemic redshift frame, $z_{\rm sys}$, we used nebular emission lines such as [O~II] $\lambda \lambda$3726,3729, [O~III] $\lambda \lambda$4959,5007, and the H~I Balmer series to define $z_{\rm sys}$ for each galaxy in our sample \citep{martin2012}. Comparing our measurements of $z_{\rm sys}$ determined from the LRIS data with those given for the DEIMOS data in the DEEP2 catalogs, we find a mean velocity discrepancy of --14 \kms\ with a standard deviation of 41 \kms. We accordingly assume a conservative systematic redshift uncertainty of 41 \kms\ on the systemic redshift measurements and require that secure detections of galactic winds have velocity offsets of at least 41 \kms. 

\begin{figure*}
\centering
\includegraphics[width=7in]{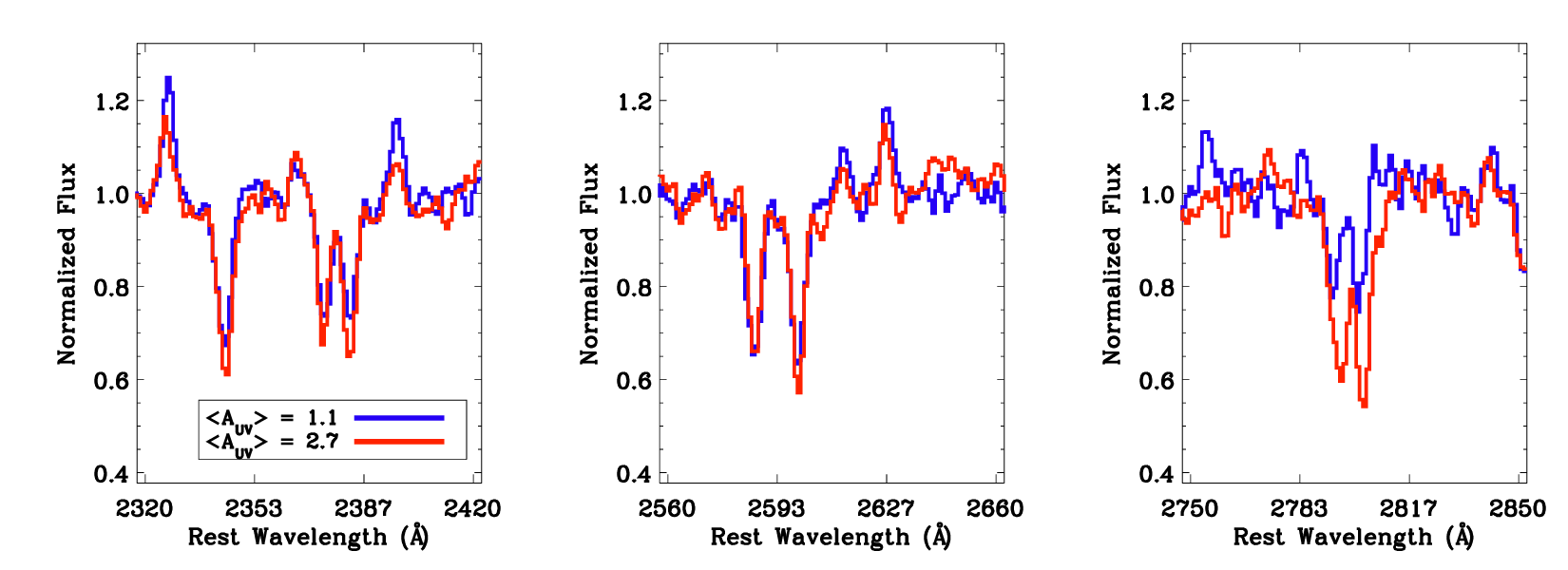} 
\caption{A comparison of composite spectra of different dust attenuation levels. Lower attenuation objects ($\langle$$A_{\rm UV}$$\rangle$ = 1.1) are shown in blue, while higher attenuation objects ($\langle$$A_{\rm UV}$$\rangle$ = 2.7) are plotted in red. Fe~II$^*$ emission lines are weaker in dustier systems, consistent with predictions by \citet{prochaska2011}. We find that objects with larger $A_{\rm UV}$ values show, on average, blueshifted 2626 \AA\ Fe~II$^*$ emission (--20 $\pm$ 42 \kms) while objects with lower $A_{\rm UV}$ values exhibit redshifted emission (37 $\pm$ 41 \kms). However, the difference between these two values is not statistically different from 0 \kms. We conclude that higher resolution observations are needed in order to test the \citet{prochaska2011} hypothesis that increased dust attenuation produces more blueshifted Fe~II$^*$ profiles.}
\label{AUV_smooth_FeII}
\end{figure*}

In \citet{martin2012}, we fit a single-component model simultaneously to five resonant Fe~II absorption lines tracing cool (T $<$ 10$^4$ K) gas at 2249.88, 2260.78, 2344.21, 2374.46, and 2586.65 \AA\ in the LRIS spectra\footnote{The Fe~II features at 2382.76 and 2600.17 \AA\ are purposefully omitted from fitting as these lines are more susceptible to filling from resonant emission; this ``emission filling" can shift the measured centroid of absorption lines to bluer wavelengths \citep[e.g.,][]{prochaska2011,martin2012}. We measure Fe~II absorption lines as opposed to Mg~II absorption lines as the latter suffer more emission filling.}. The model fit to the Fe~II lines has four free parameters: Doppler shift, optical depth at line center, Doppler width ($b$, where $b$ = $\sqrt{2}$$\sigma$ = FWHM/2$\sqrt{\rm ln2}$), and covering fraction. Due to the low spectral resolution and finite S/N of the observations, the Doppler shift is the best-constrained parameter and we will not discuss the other three parameters in this paper. We measured velocities for 172/212 objects, where 40 objects had no significant absorption lines and therefore could not be modeled, and find velocities ranging from $-302$ to $+401$ \kms\ with a mean of $-30$ \kms\ and a 1$\sigma$ sample standard deviation of 89 \kms. We define here the convention of employing ``$V_1$" to refer to the measured velocity shift of the deepest part of the Fe~II absorption line fit, relative to a systemic reference frame typically defined by nebular emission lines. Negative $V_1$ values refer to blueshifts (``outflows") while positive $V_1$ values correspond to redshifts (``inflows"). Fe~II velocity shifts significant at the 1$\sigma$ (3$\sigma$) level are observed in $\sim$67\% (27\%) of the sample. In this paper, we primarily utilize the $V_1$ measurements of composite spectra due to the high S/N of the composite spectra (15--45 pixel$^{-1}$). We also measured maximal outflow velocities from the blue wing of the 2796 \AA\ Mg~II feature (``$V_{\rm max}$(Mg~II)"; Martin et al. 2012). These maximal outflow velocities are not biased by the effects of Mg~II emission filling since these measurements are made from the wing of the absorption feature as opposed to the centroid. $V_{\rm max}$(Mg~II) ranges from $-1151$ \kms\ to a limit of $-435$ \kms in 400 line mm$^{-1}$ spectra and $-282$ \kms in 600 line mm$^{-1}$ spectra, with the limits set by the resolution of the spectra \citep{martin2012}.

\section{Fine-Structure Fe~II$^*$ Emission} \label{sec: feIIstar}

The Fe~II ion has many transitions in the rest-frame ultraviolet\footnote{We refer the reader to Figure 1 of \citet{erb2012} for an energy level diagram of the Fe~II ion.} and several authors have used Fe~II resonant absorption lines to trace the bulk motion of outflowing interstellar gas \citep{rubin2010a,coil2011,kornei2012,martin2012}. The absorption of a resonant photon can result in either the re-emission of another resonant photon to the ground state (scattering) or the emission of a photon to an excited ground state (fluorescence). We observe resonant Fe~II absorption in our data, but do not see resonant Fe~II emission. This lack of resonant emission may be due to the limited spectral resolution of our data (FWHM $\sim$435 \kms), as \citet{erb2012} observe resonant Fe~II emission at 2600 \AA\ in a subset of their higher resolution data. Here, we focus on the emission features of Fe~II resulting from fluorescence. In this work, we examine four Fe~II$^*$ lines at 2365.55, 2396.35, 2612.65, and 2626.45 \AA. An additional Fe~II$^*$ line within our data's wavelength coverage, at 2632.11 \AA, is absent in individual observations but plausibly detected in a stack of the strongest Fe~II$^*$-emitting galaxies. 

\subsection{Fe~II$^*$ Emitters and Non-emitters} \label{sec: feIIstaremitttersnonemitters}

We measured the equivalent widths of the four strongest Fe~II$^*$ features for each object in our sample with Fe~II$^*$ spectral coverage ($>$ 95\% of the sample). The equivalent width of each line was calculated over a fixed wavelength interval approximately 10 \AA\ wide, where the precise wavelength interval was derived from the 3$\sigma$ extent of each Fe~II$^*$ feature measured in a high S/N stack of all the spectral data\footnote{The exact wavelength intervals we use are: 2360.4--2371.4 \AA\ for Fe~II$^*$ 2365 \AA, 2390.9--2401.1 \AA\ for Fe~II$^*$ 2396 \AA, 2607.8--2617.2 \AA\ for Fe~II$^*$ 2612 \AA, and 2622.2--2630.3 \AA\ for Fe~II$^*$ 2626 \AA.}. We find typical rest-frame Fe~II$^*$ equivalent widths of several tenths of an angstrom, where 23, 37, 23, and 58 objects exhibit $\ge$ 2$\sigma$ detections in each of the four Fe~II$^*$ lines at 2365, 2396, 2612, and 2626 \AA, respectively. We observe that the 2626 \AA\ line is the most frequently detected of the Fe~II$^*$ lines. Since this feature is relatively isolated from neighboring absorption lines, the kinematics of the 2626 \AA\ line may furthermore be more robust than those of other Fe~II$^*$ features.

\begin{figure*}
\centering
\includegraphics[width=7in]{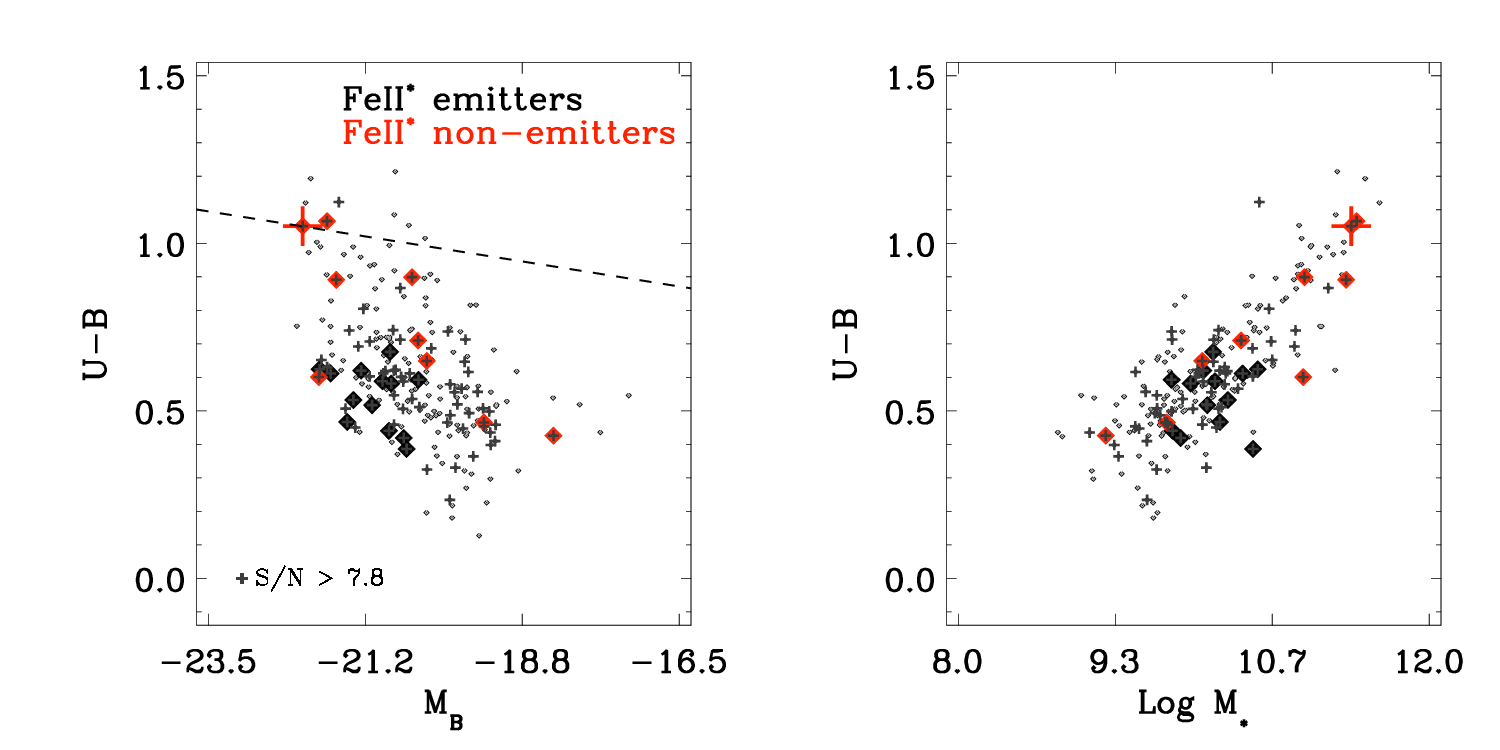} 
\caption{Left: color-magnitude diagram of the full sample, where the dashed line indicates the division between red sequence and blue cloud galaxies derived by \citet{willmer2006}. Objects with continuum S/N $>$ 7.8 are indicated with gray crosses. Within this subsample of objects with higher S/N, Fe~II$^*$ emitters are shown as black diamonds (13 objects) and Fe~II$^*$ non-emitters are shown as red diamonds (9 objects).  Object 12015320, a Fe~II$^*$ non-emitter, is differentiated with a red cross since it is likely an AGN based on its broad Ne~V emission and X-ray flux. Fe~II$^*$ emitters appear to be preferentially bright, blue galaxies, although Figure \ref{IDLgaussianEW} shows no statistically significant trend between Fe~II$^*$ strength and either luminosity or color, based on a binary division of the data. Right: color versus stellar mass plot, where the symbols are the same as in the left panel. Fe~II$^*$ emitters have lower stellar masses on average than the sample as a whole.}
\label{CMD_FeII}
\end{figure*}

With these measurements of Fe~II$^*$ equivalent widths, we isolated subsamples of objects exhibiting strong or weak Fe~II$^*$ emission and examined the structural, outflow, and star-forming properties of these galaxies. We focused our selection criteria on the 2396 \AA\ and 2626 \AA\ features given that the largest number of objects had significant ($\ge$ 2$\sigma$) detections in these lines. Furthermore, a high S/N spectral stack of all the data showed larger equivalent widths for the 2396 and 2626 \AA\ features (0.72 $\pm$ 0.07 and 0.80 $\pm$ 0.09 \AA, respectively) than for the 2365 and 2612 \AA\ lines (0.42 $\pm$ 0.05 and 0.41 $\pm$ 0.09 \AA, respectively). None of the Fe~II$^*$ lines are resolved in either our 400 line mm$^{-1}$ observations or our 600 line mm$^{-1}$ observations. 

We define a subsample of 13 ``Fe~II$^*$ emitters" based on objects having $>$ 4$\sigma$ equivalent width detections in either the 2396 \AA\ or 2626 \AA\ lines, where this detection threshold was motivated by visual inspection of objects showing obvious Fe~II$^*$ emission. We found that selecting objects having $>$ 3$\sigma$ equivalent width detections yielded a sample of 31 objects where several objects showed asymmetric Fe~II$^*$ profiles that did not appear robust. The 13 objects selected with the $>$ 4$\sigma$ criterion have continuum S/N ranging from 7.8--25.1 pixel$^{-1}$. We also collect a subsample of 9 ``Fe~II$^*$ non-emitters" selected to have $<$ 1$\sigma$ equivalent width detections in both the 2396 \AA\ and 2626 \AA\ lines and continuum S/N $>$ 7.8 pixel$^{-1}$. We imposed a S/N threshold for the Fe~II$^*$ non-emitter sample in order to ensure a fair comparison with the Fe~II$^*$ emitter sample. We adopted the simplified methodology of requiring the same S/N threshold for objects observed with either the 400 line mm$^{-1}$ grism or the 600 line mm$^{-1}$ grism. A fixed equivalent width sensitivity will translate to slightly different S/N thresholds depending on the resolution of the data \citep{martin2012}, so we acknowledge that our technique may be biased toward detecting weaker  lines in the higher-resolution 600 line mm$^{-1}$ data. Three of the Fe~II$^*$ non-emitters have colors typical of ``green valley" galaxies \citep[e.g.,][]{martin2007}; our conclusions remain unchanged if these objects are omitted from the sample.

In Figure \ref{FeII_smooth}, we compare the composite spectra assembled from the Fe~II$^*$ emitter and non-emitter samples. The 2365, 2396, 2612, and 2626 \AA\ Fe~II$^*$ lines are clearly detected in the composite of Fe~II$^*$ emitters, and the 2632 \AA\ Fe~II$^*$ line is plausibly seen as a shoulder on the red side of the 2626 \AA\ feature. The relatively low spectral resolution of our data (FWHM $\sim$435 \kms) makes it difficult to detect the 2632 \AA\ feature securely given its proximity to the Fe~II$^*$ line at 2626 \AA\ (--645 \kms). In a higher-resolution (FWHM $\sim$190 \kms) composite spectrum of 96 star-forming galaxies at 1 $\lesssim$ \emph{z} $\lesssim$ 2, \citet{erb2012} detect the 2632 \AA\ line at $\sim$4$\sigma$ significance and resolve this feature cleanly from the 2626 \AA\ line. The \citeauthor{erb2012} composite spectrum is inclusive of all these authors' data, while the corresponding stack of all of our data does not show any signature of 2632 \AA\ emission; we only observe evidence of 2632 \AA\ emission in the stack of Fe~II$^*$ emitters. We convolved the \citeauthor{erb2012} spectrum to our lower spectral resolution and found a blended 2626+2632 \AA\ complex consistent with the profile of our data. This similarity in profile shape supports our hypothesis that we detect the 2632 \AA\ line blended with the 2626 \AA\ feature in the stack of Fe~II$^*$ emitters. We fit two Gaussian profiles simultaneously to our data's 2626+2632 \AA\ complex and find that the strength of 2632 \AA\ emission is approximately half that of the 2612 \AA\ feature, as predicted by the ratio of the Einstein A coefficients of these lines. In the composite spectrum assembled from all of the data in our sample, however, we find that the 3$\sigma$ upper limit on the strength of 2632 \AA\ emission is 0.21 \AA, less than 50\% of the strength of the 2612 \AA\ feature (0.55 \AA). The relatively low spectral resolution of our data may be responsible for the weaker-than-expected 2632 \AA\ strength. 

The composite spectra assembled from the Fe~II$^*$ emitter and non-emitter samples exhibit different Fe~II$^*$ strengths, by construction. However, these spectra also show variation in the strength and kinematics of their Fe~II and Mg~II resonant absorption features. Evidently, the presence or absence of fine-structure Fe~II$^*$ emission is linked to the strength and velocity offsets of resonant absorption lines. Based on inspection of the composite spectra assembled from the Fe~II$^*$ emitter and non-emitter samples, two trends are visually apparent: 

1. Fe~II$^*$ emitters show stronger Fe~II resonant absorption lines and weaker Mg~II resonant absorption lines than Fe~II$^*$ non-emitters.

2. Fe~II$^*$ emitters show more blueshifted Mg~II resonant absorption lines than Fe~II$^*$ non-emitters.

The composite spectrum of Fe~II$^*$ emitters exhibits a $\sim$50\% larger Fe~II absorption equivalent width and a $\sim$50\% weaker Mg~II absorption equivalent width (integrated over both Mg~II absorption lines) than the composite spectrum of Fe~II$^*$ non-emitters. This anticorrelation between Fe~II and Mg~II absorption strengths is not typical in our sample; composite spectra assembled on the basis of other galaxy properties (Section \ref{sec: feIIstar_strength}) generally show Fe~II and Mg~II strengths varying in tandem, as \citet{coil2011} find. Furthermore, the changes in both Fe~II and Mg~II absorption strength seen in the Fe~II$^*$ emitter and non-emitter composite spectra are not as large as the changes observed in other composite spectra. As we show later, Fe~II$^*$ emitters are preferentially at larger redshifts than Fe~II$^*$ non-emitters. We propose that the redshift difference between Fe~II$^*$ emitters and non-emitters may be responsible for some of the trends we observe in Figure \ref{FeII_smooth}, as we discuss both in this section and later in the paper. In order to control against redshift evolution, we would ideally assemble a sample in which Fe~II$^*$ emitters and non-emitters are at similar redshifts. Future studies with more objects in each redshift bin will be instrumental in examining the properties of Fe~II$^*$ emitters and non-emitters independent of potential biases caused by redshift evolution.  

We explore the relationship between Fe~II$^*$ emission and Fe~II absorption more thoroughly in Martin et al. (in preparation) but note here that our finding of stronger Fe~II absorption in the Fe~II$^*$ emitter sample is contrary to the trends observed in most of the other composite spectra presented in this paper (Section \ref{sec: feIIstar_strength}). The majority of the composite spectra exhibit weaker Fe~II absorption paired with stronger Fe~II$^*$ emission, while Figure \ref{FeII_smooth} instead shows stronger Fe~II absorption with stronger Fe~II$^*$ emission. This observation that stronger Fe~II absorption accompanies stronger Fe~II$^*$ emission is furthermore inconsistent with the results of \citet{erb2012}. These authors examined 96 star-forming galaxies at 1 $\lesssim$ \emph{z} $\lesssim$ 2 and found a trend of decreasing Fe~II$^*$ emission with increasing Fe~II absorption. \citeauthor{erb2012} suggest that systems with stronger Fe~II absorption may be preferentially dusty and therefore show only weak Fe~II$^*$ emission. \citeauthor{erb2012} also propose that galaxy inclination modulates the observed ratio of emission and absorption equivalent widths, as an anisotropic (i.e., biconical) outflow will exhibit differing ratios of emission and absorption depending on viewing angle: a biconical wind viewed edge-on will show more emission than absorption while a face-on view of the same galactic wind will show more absorption than emission. These predictions make sense given that a wind seen face-on (down the barrel) shows material absorbed against the background light of the host galaxy while observations of a wind edge-on see the wind projected 90$^{\circ}$ and therefore record more scattered emission as opposed to absorption backlit by starlight. Furthermore, \citeauthor{erb2012} suggest that slit losses -- arising because the size of the Fe~II$^*$-emitting region is larger than the area encompassed by the spectroscopic slit -- may reduce the observed strength of Fe~II$^*$ emission in galaxies subtending larger angular sizes. In \z3 Lyman break galaxies, \citet{shapley2003} found that Si~II$^*$ emission lines were much weaker than the Si~II resonant absorption lines. These authors hypothesized that the narrow spectroscopic slit (1.$''$4) might be subtending only a small fraction of the area in which Si~II$^*$ emission was arising. \citet{steidel2010} showed that interstellar absorption persists at distances of several tens of kpc from galaxies; emission lines arising from similar galactocentric distances would fall beyond the extent of most spectroscopic slits. Findings by \citet{jones2012} also support the hypothesis that fine-structure emission may be spatially extended. These authors used a sample of 81 Lyman break galaxies at \z4 and found that objects had average Si~II$^*$/Si~II equivalent width ratios of $\sim$0.5, indicative of a loss of fine-structure photons. As the galaxies examined by \citet{jones2012} suffer only minimal dust attenuation (E(B$-V$) $\sim$0.10), this discrepancy in equivalent width is likely due to fine-structure emission being more spatially extended than resonant emission as opposed to emission being suppressed by dust. \citet{jones2012} also infer a change in the spatial extent of Si~II$^*$ emission between \z3 and \z4 based on the different Si~II$^*$/Si~II equivalent width ratios seen in their data and a \z3 Lyman break galaxy sample \citep{shapley2003}. As objects at \z4 show a larger average Si~II$^*$/Si~II equivalent width ratio ($\sim$0.5) than objects at \z3 ($\sim$0.2), \citet{jones2012} suggest that the characteristic size of the Si~II$^*$-emitting region decreases from \z3 to \z4. A similar redshift dependency of the Fe~II$^*$/Fe~II equivalent width ratio is not observed, however, between our current data set ($\langle$\emph{z}$\rangle$ = 1.0) and the \citet{erb2012} sample ($\langle$\emph{z}$\rangle$ = 1.6); the composite spectra of both studies show Fe~II$^*$/Fe~II equivalent width ratios of $\sim$0.3 $\pm$ 0.02. Since the slit widths employed in both studies are comparable and the difference in angular diameter distance at the average redshift of each sample is only $\sim$6\%, we conclude that the size of the Fe~II$^*$-emitting region does not evolve substantially in the 1.8 Gyr intervening between \z1 and \z1.6. 

\begin{figure}
\centering
\includegraphics[width=3.5in]{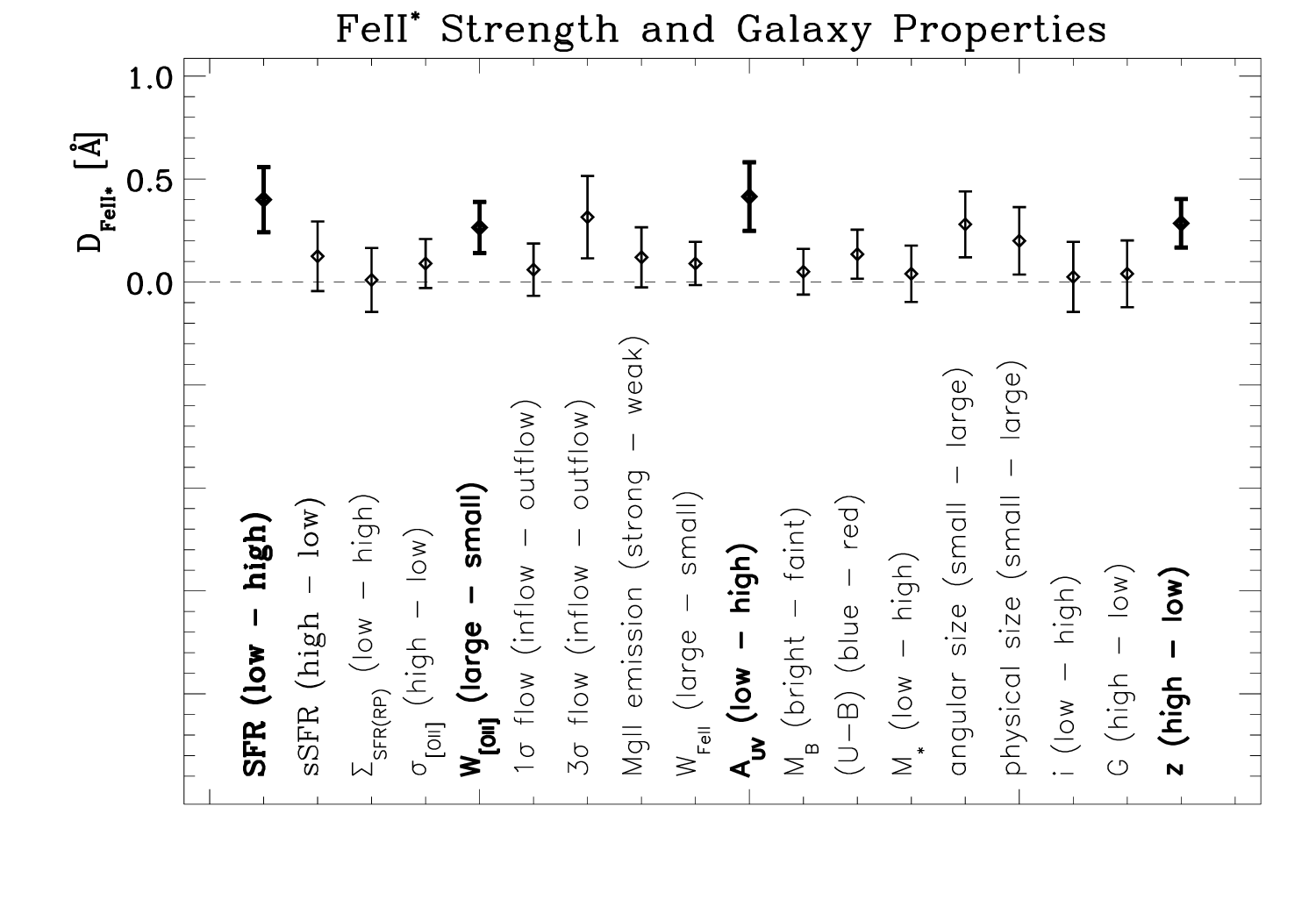} 
\caption{Variation of Fe~II$^*$ emission strength with galaxy properties. D$_{\rm FeII^*}$, a parameterization of how Fe~II$^*$ strength changes between two spectra (Section \ref{sec: feIIstar_strength}), is shown for pairs of composite spectra assembled according to 18 different galaxy properties. The largest D$_{\rm FeII^*}$ values, significant at $>$ 2$\sigma$, are observed for three galaxy properties: SFR, $A_{\rm UV}$, and $W_{\rm [OII]}$, as well as \emph{z} (bold text). Stronger Fe~II$^*$ emission is seen for lower SFR, larger $W_{\rm [OII]}$, lower $A_{\rm UV}$, and higher \emph{z} objects. D$_{\rm FeII^*}$ values are significant at $>$ 1$\sigma$ for the following additional parameters: 3$\sigma$ gas flow, $U-B$ color, angular size, and physical size, such that Fe~II$^*$ emission is more pronounced in systems with 3$\sigma$ inflows (as opposed to 3$\sigma$ outflows), bluer $U-B$ colors, smaller angular sizes, and smaller physical sizes. Our finding that objects with smaller angular sizes show stronger Fe~II$^*$ emission is consistent with the hypothesis that slit losses modulate Fe~II$^*$ emission strength.}
\label{IDLgaussianEW}
\end{figure}

\begin{deluxetable*}{llllll}
\tablewidth{0pt}
\tablecaption{Fe~II and Fe~II$^*$ Lines in the Global Composite Spectrum}
\tablehead{
\multicolumn{1}{c}{$\lambda$\tablenotemark{a}} 
& \multicolumn{1}{c}{J$_{\rm lower}$\tablenotemark{b}} 
& \multicolumn{1}{c}{J$_{\rm upper}$\tablenotemark{b}} 
& \multicolumn{1}{c}{$A_{\rm 21}$\tablenotemark{c}} 
& \multicolumn{1}{c}{Feature Type} 
& \multicolumn{1}{c}{Equivalent Width\tablenotemark{d}} \\
\colhead{(\AA)}
& \multicolumn{1}{c}{}
& \multicolumn{1}{c}{}
& \multicolumn{1}{c}{(s$^{-1}$)}
& \multicolumn{1}{c}{}
& \multicolumn{1}{c}{(\AA)}
}
\startdata
2344.21 & 9/2  & 7/2 & 1.7 $\times$ 10$^8$ & resonant abs. & 1.94 $\pm$ 0.06 \\
2365.55 & 7/2 & 7/2 & 5.9 $\times$ 10$^7$ & fine-structure em. & --0.42 $\pm$ 0.05 \\
2380.76 & 5/2 & 7/2 & 3.1 $\times$ 10$^7$ & fine-structure em. & \nodata\tablenotemark{e} \\
& & \\
2374.46 & 9/2 & 9/2 & 4.3 $\times$ 10$^7$ & resonant abs. & 1.39 $\pm$ 0.08 \\
2396.35 & 7/2 & 9/2 & 2.6 $\times$ 10$^8$ & fine-structure em. & --0.72 $\pm$ 0.07\\ 
& & \\
2382.77 & 9/2 & 11/2 & 3.1 $\times$ 10$^8$ & resonant abs. & 1.53\tablenotemark{e} $\pm$ 0.06 \\
& & \\
2586.65 & 9/2 & 7/2 & 9.0 $\times$ 10$^7$ & resonant abs. & 1.72 $\pm$ 0.06 \\
2612.65 & 7/2 & 7/2 & 1.2 $\times$ 10$^8$ & fine-structure em. & --0.41 $\pm$ 0.09 \\
2632.11 & 5/2 & 7/2 & 6.3 $\times$ 10$^7$ & fine-structure em. & --0.21\tablenotemark{f}\\
& & \\
2600.17 & 9/2 & 9/2 & 2.4 $\times$ 10$^8$ & resonant abs. & 2.35 $\pm$ 0.09 \\
2626.45 & 7/2 & 9/2 & 3.5 $\times$ 10$^7$ & fine-structure em. & --0.80 $\pm$ 0.09 \\
\enddata
\tablenotetext{a}{Vacuum wavelengths from \citet{ralchenko2005}.}
\tablenotetext{b}{Orbital angular momenta.}
\tablenotetext{c}{Einstein A coefficients.}
\tablenotetext{d}{Equivalent width of feature measured in the global composite spectrum (Figure \ref{composite_blue_smooth}). Negative equivalent widths correspond to features in emission.}
\tablenotetext{e}{Fe~II$^*$ 2380.76 \AA\ is blended with Fe~II 2382.77 \AA.}
\tablenotetext{f}{3$\sigma$ upper limit.}
\label{abtable}
\end{deluxetable*}

Fe~II$^*$ emitters are marked by stronger Fe~II absorption, and weaker Mg~II absorption, than Fe~II$^*$ non-emitters. These spectral results are surprising given that the majority of composite spectra used in our study show Fe~II and Mg~II strengths varying together. Furthermore, most composites that show stronger Fe~II$^*$emission also exhibit comparable, or weaker, Fe~II absorption (Section \ref{sec: feIIstar_strength}). We propose that the weaker Mg~II absorption seen in the composite spectrum of Fe~II$^*$ emitters is primarily caused by Mg~II emission filling. Emission filling is a likely explanation given that Fe~II$^*$ emitters are bluer than Fe~II$^*$ non-emitters (Section \ref{sec: feIIstar_strength}) and therefore suffer less dust attenuation. Furthermore, the composite spectrum of Fe~II$^*$ emitters shows emission redwards of the Mg~II 2803 \AA\ absorption feature; that this emission is present supports the theory of emission filling. Additionally, the kinematic shifts observed in the Mg~II profiles of the Fe~II$^*$ emitter and non-emitter composites are consistent with the effects of emission filling. Fe~II$^*$ emitters show, on average, significantly blueshifted Mg~II absorption, as expected from emission filling \citep[e.g.,][]{prochaska2011,martin2012}. The centroids of both features of the Mg~II doublet are blueshifted in the Fe~II$^*$ emitter composite relative to the Fe~II$^*$ non-emitter composite. The Fe~II$^*$ emitter composite has Mg~II lines with centroids of --319 $\pm$ 28 \kms\ and --247 $\pm$ 46 \kms\ for the 2796 and 2803 \AA\ features, respectively. The Fe~II absorption lines are not significantly blueshifted in the Fe~II$^*$ emitter spectrum. The Mg~II doublet is optically thick at lower column densities than the majority of the Fe~II transitions studied here. Mg~II is therefore a better probe of rarefied gas, which may have a high-velocity component \citep{martin2012}. The Fe~II$^*$ non-emitter composite has Mg~II centroids of 0 $\pm$ 65 \kms\ and --58 $\pm$ 44 \kms, respectively. These results suggest that the conditions enabling observations of high-velocity Mg~II gas may be the same conditions conducive to seeing Fe~II$^*$ emission. This hypothesis is consistent with predictions by \citet{prochaska2011} that stronger Fe~II$^*$ emission is seen when viewing a galactic wind face-on (i.e., in conditions favorable for detecting blueshifted absorption lines; Kornei et al. 2012), but contradictory with the interpretation from \citet{erb2012} that a galactic wind seen face-on will show more absorption than emission. 

While emission filling may be responsible for shifting the centroids of Mg~II, we also observe that the blue wings of the Mg~II doublets are offset between the Fe~II$^*$ emitter and non-emitter composite spectra. The different velocities seen in the blue wings of the features -- where emission filling is negligible -- indicate an intrinsic discrepancy in the observed Mg~II gas kinematics between the Fe~II$^*$ emitter and non-emitter populations. However, the $V_{\rm max}$(Mg~II) values for the Fe~II$^*$ emitter and non-emitter composite spectra are consistent within their errors given the similar line profiles of the composite spectra where $V_{\rm max}$(Mg~II) is measured; we emphasize the overall difference in the line profiles here as opposed to relying on the single parametrization of $V_{\rm max}$(Mg~II). The distinct blue wing profiles of the Fe~II$^*$ emitter and non-emitter composite spectra suggest that the line profiles differ more substantially than can be explained solely by emission filling. 

Although these composite spectra show a direct link between Mg~II gas kinematics and the strength of Fe~II$^*$ emission lines such that galaxies with blueshifted Mg~II absorption lines are more likely to have stronger Fe~II$^*$ emission, not all of our findings are consistent with the trend that observations of gas flows are most often observed when stronger Fe~II$^*$ emission is observed. We assembled composite spectra on the basis of both $V_1$ and $V_{\rm max}$(Mg~II) and found no statistically significant difference in Fe~II$^*$ strength when the data were divided by gas flow velocity.  Furthermore, we examined distributions of $V_1$ and $V_{\rm max}$(Mg~II) and found no evidence that objects with either strong or weak Fe~II$^*$ emission were clustered in velocity space. 

\begin{figure}
\begin{center}$
\begin{array}{c}
\includegraphics[width=3.5in]{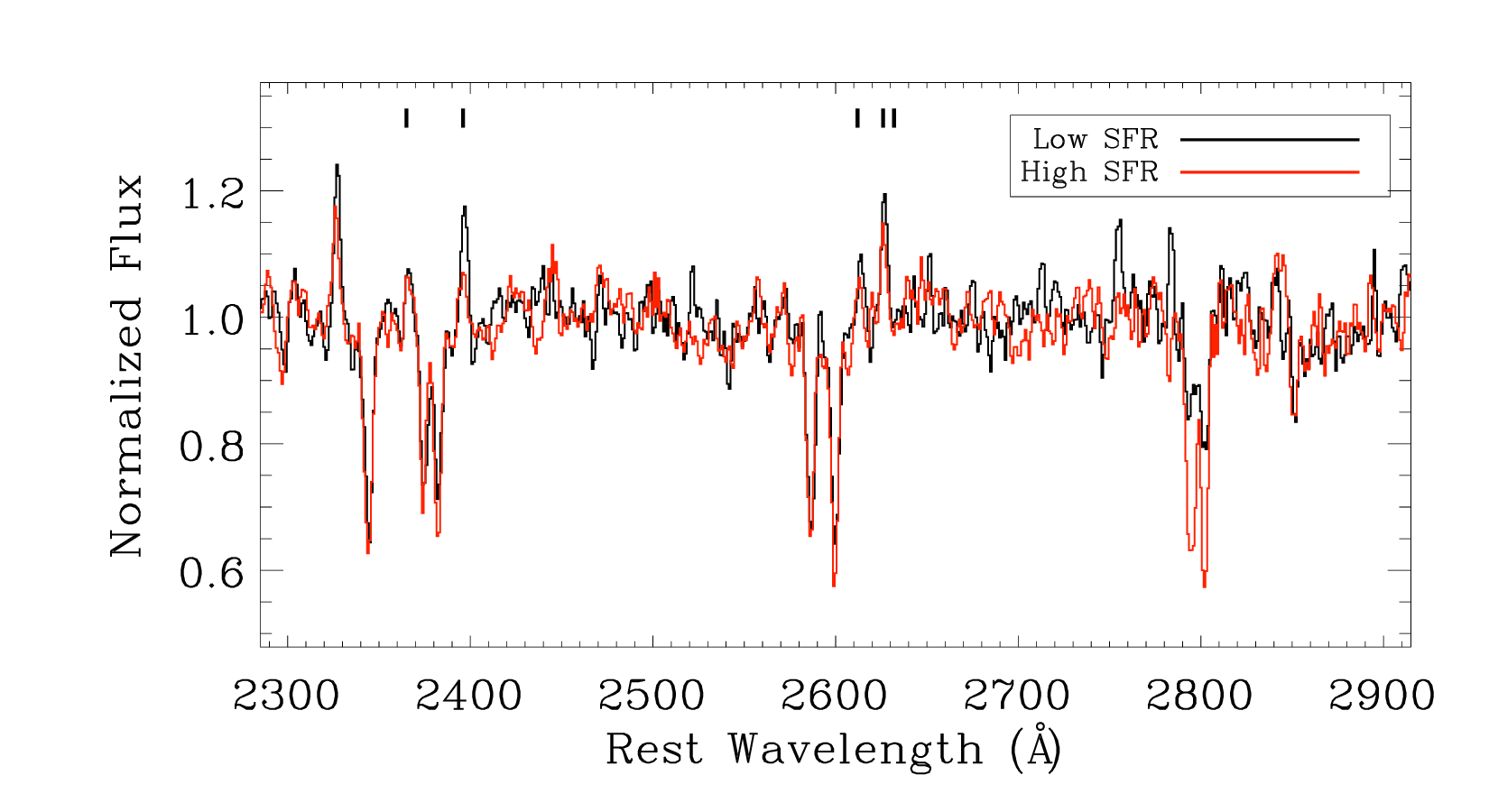} \\ 
\includegraphics[width=3.5in]{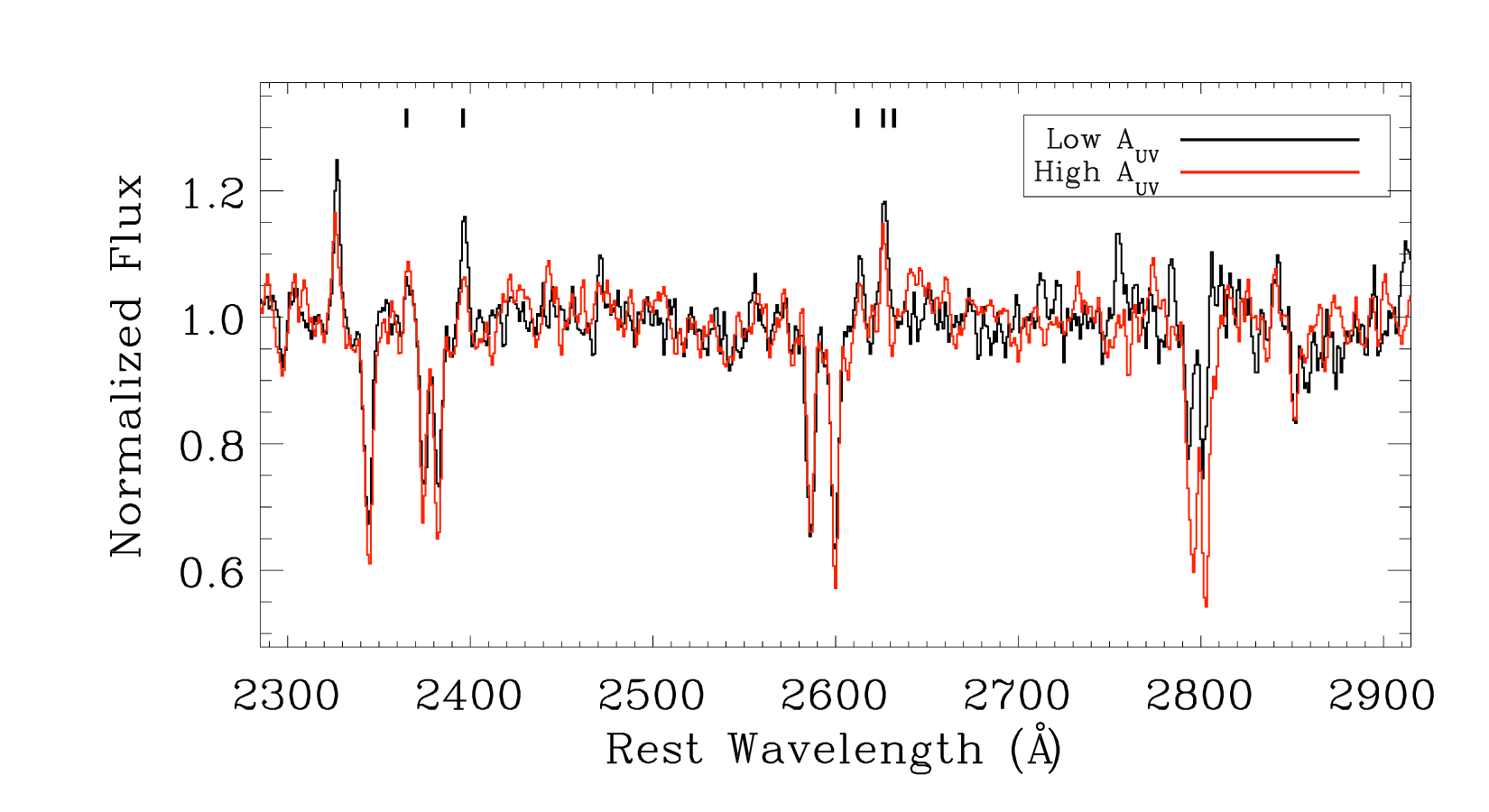} \\ 
\includegraphics[width=3.5in]{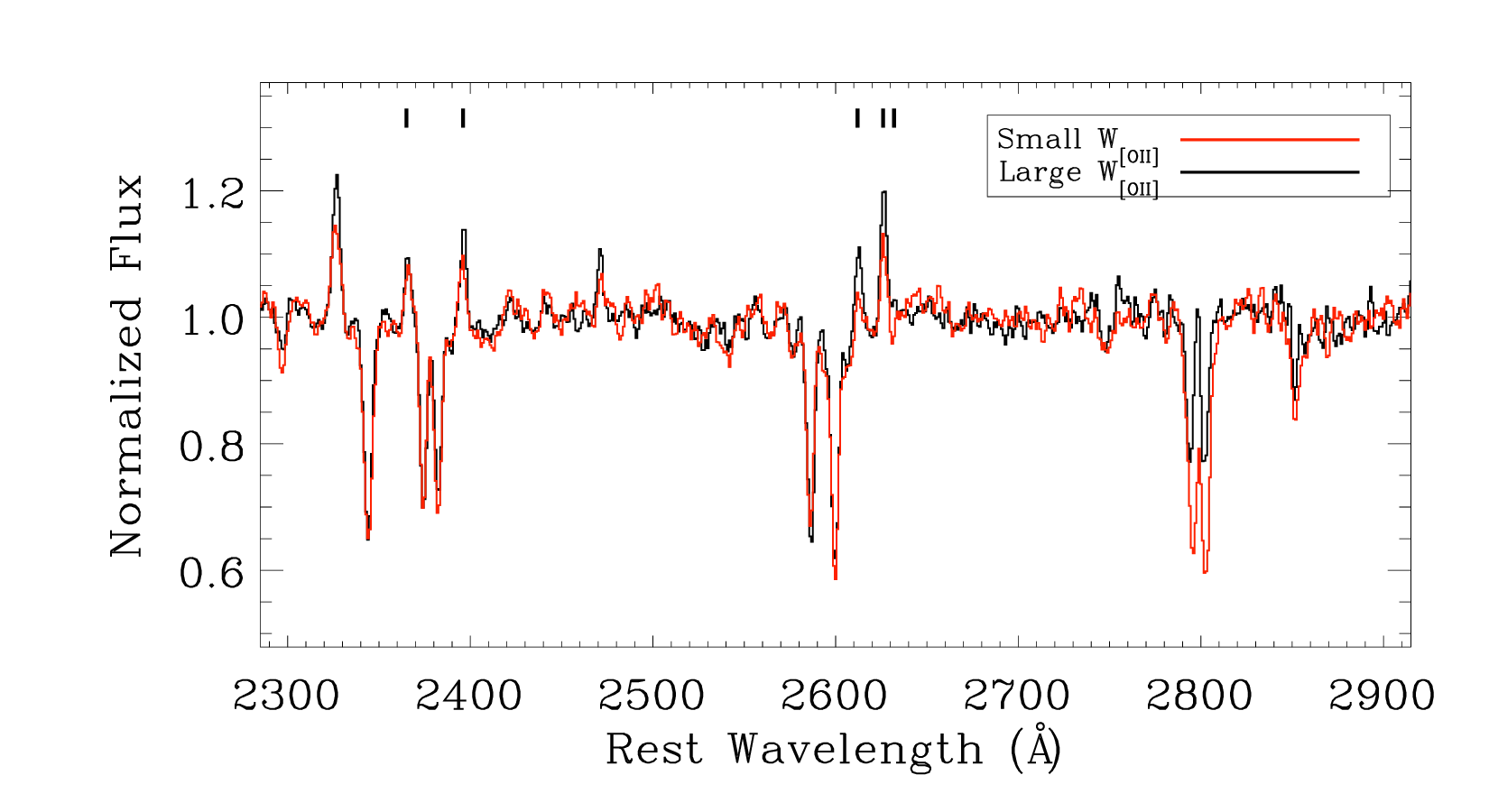} \\ 
\includegraphics[width=3.5in]{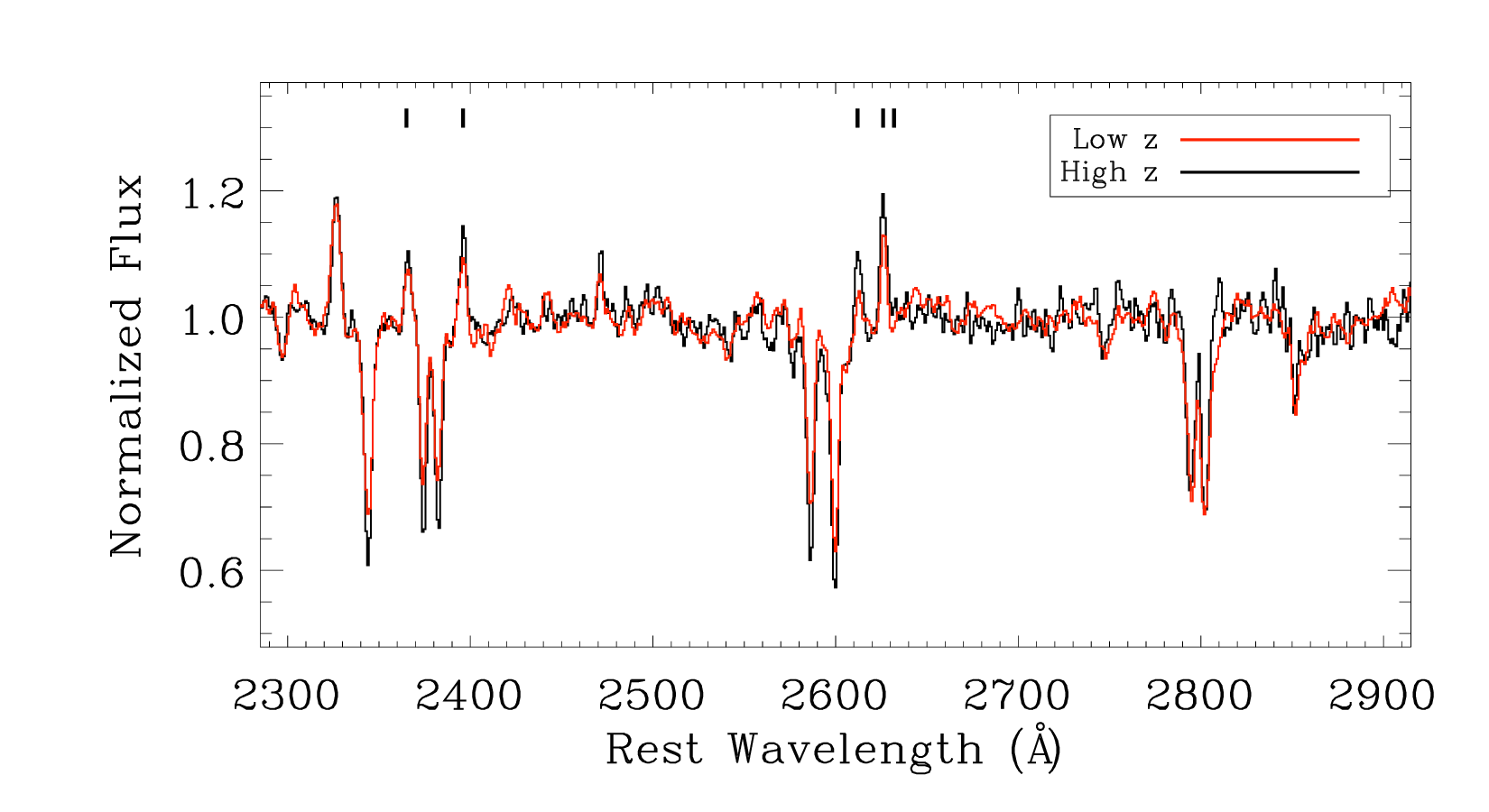} 
\end{array}$
\end{center}
\caption{Composite spectra of galaxy parameters that strongly modulate Fe~II$^*$ emission strength. From top to bottom, composite spectra are shown comparing low and high SFR, low and high $A_{\rm UV}$, small and large $W_{\rm [OII]}$, and low and high \emph{z}. In each case, the composite spectrum with stronger Fe~II$^*$ emission is plotted in black. The composite spectra have S/N $\sim$30 pixel$^{-1}$.}
\label{SFR_smooth_FeII}
\end{figure}

Modeling of galactic winds is important for understanding the link between the observability of Fe~II$^*$ emission and the frequency of observed prevalence of gas flows. In particular, models of biconical galactic winds -- a geometry prevalent in both local and \z1 samples \citep[e.g.,][]{heckman1990,martin2012} -- will be critical to investigating how the measured frequency of fine-structure emission varies as a function of observed gas flow properties. \citet{prochaska2011} studied the prevalence of emission in a model of a hemispherical wind, but thus far accurate modeling of more collimated outflows observed at different viewing angles has been lacking from the literature. 

\subsection{Fe~II$^*$ Kinematics} \label{sec: kinematics}

\begin{figure*}
\begin{center}$
\begin{array}{c}
\includegraphics[width=2.3in]{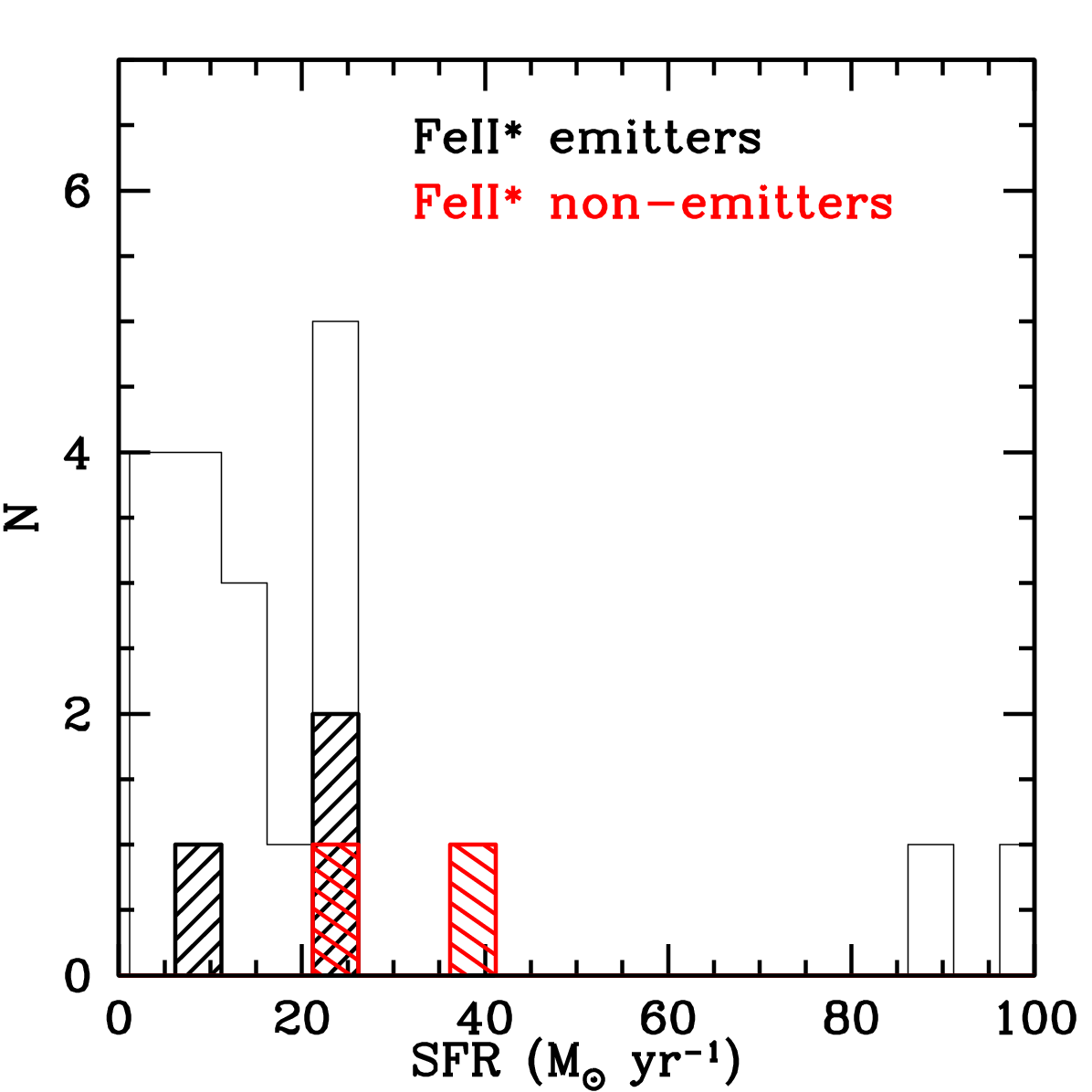} 
\includegraphics[width=2.3in]{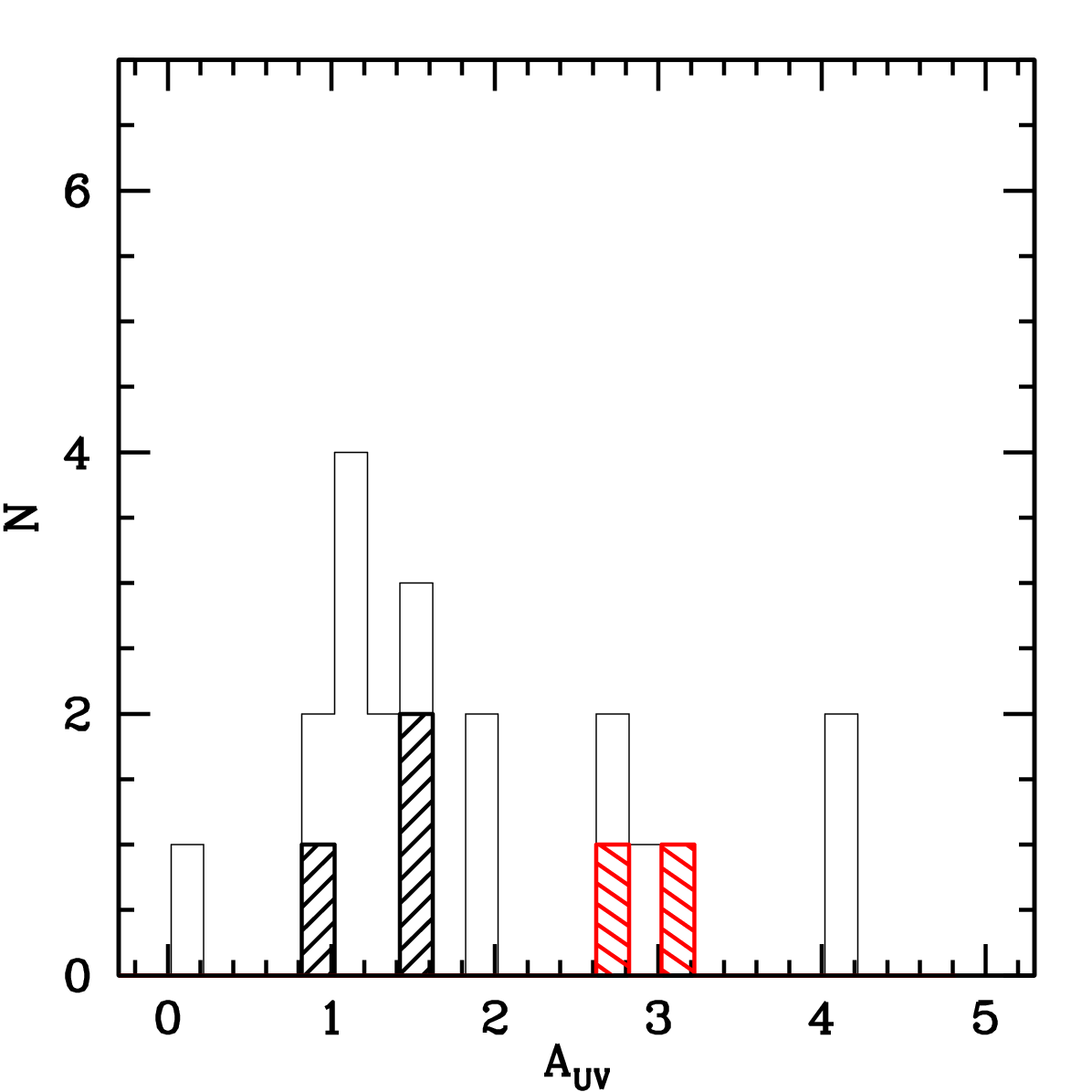} \\ 
\includegraphics[width=2.3in]{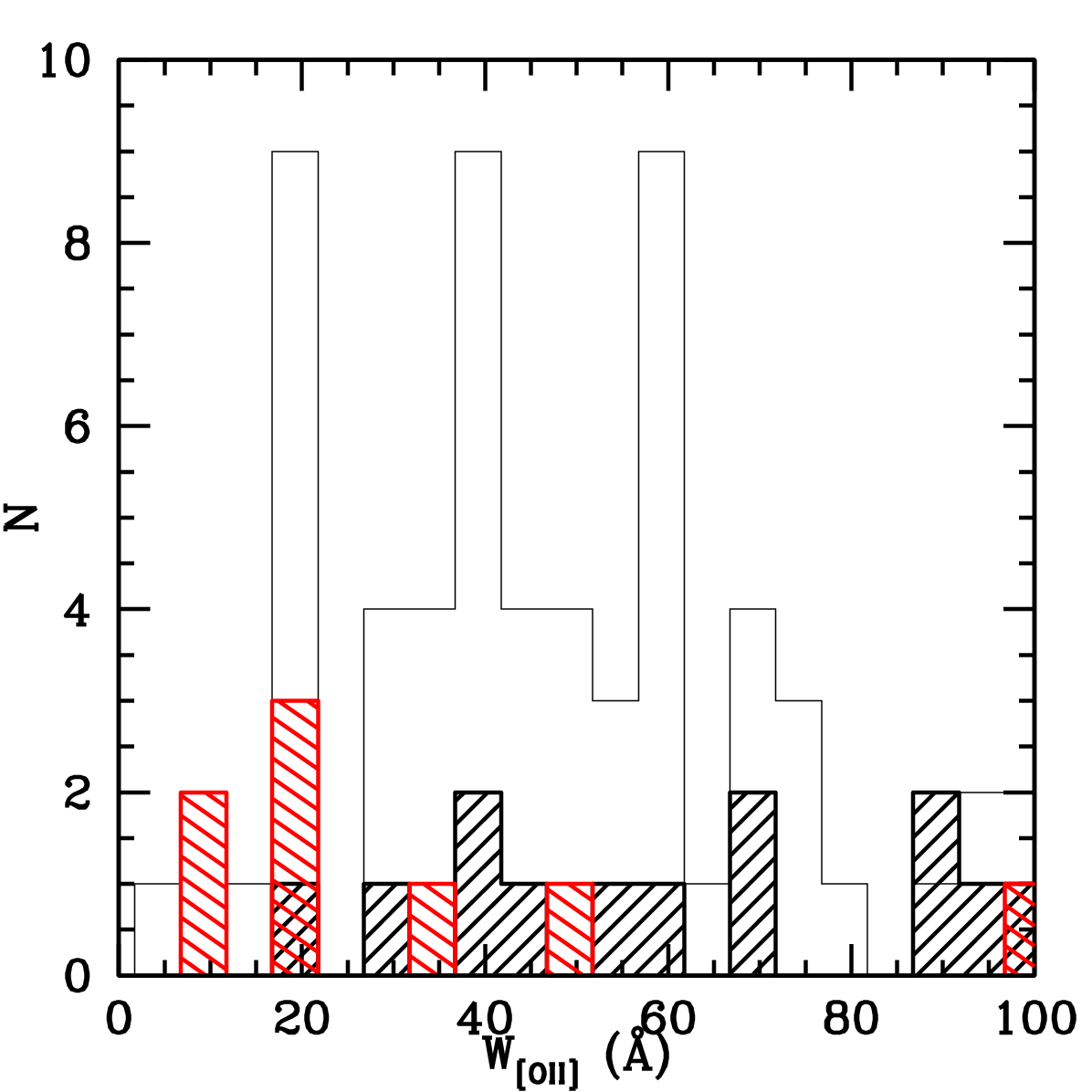} 
\includegraphics[width=2.3in]{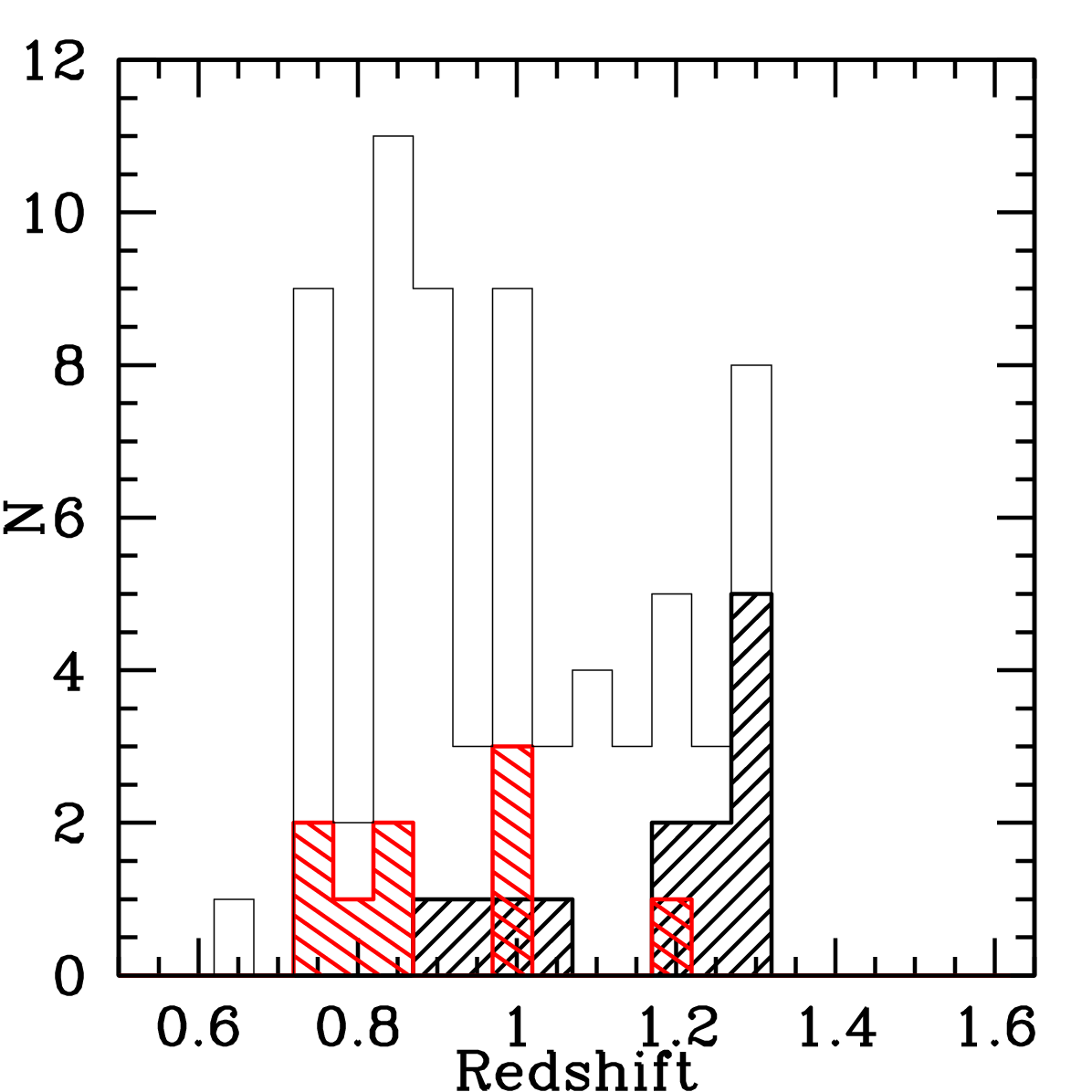} 
\end{array}$
\end{center}
\caption{Histograms of the four parameters most strongly modifying Fe~II$^*$ emission strength: SFR, $A_{\rm UV}$, $W_{\rm [OII]}$, and \emph{z}. In each panel, the open histogram shows the parent sample of objects with S/N $>$ 7.8. Fe~II$^*$ emitters are shown in the black shaded histogram and Fe~II$^*$ non-emitters are indicated with the red shaded histogram. Fe~II$^*$ emitters are characterized on average by smaller SFRs, lower dust attenuations, larger $W_{\rm [OII]}$, and larger redshifts than Fe~II$^*$ non-emitters.}
\label{auv_FeII}
\end{figure*}

While blueshifted or redshifted Fe~II$^*$ emission is consistent with moving gas, \citet{prochaska2011} showed that emission profiles arising in the presence of gas flows can still be centered at roughly 0 \kms. Therefore, the absence of a net kinematic offset does not necessarily imply that the associated gas is at rest with respect to a galaxy's stars. We measured the centroids of the Fe~II$^*$ lines in the Fe~II$^*$ emitter composite spectrum using Gaussian fits from the {\tt IDL} routine {\tt gaussfit}. We find that the 2365, 2396, 2612, and 2626 \AA\ lines have centroids of --40 $\pm$ 111, --36 $\pm$ 28, 7 $\pm$ 33, and --6 $\pm$ 21 \kms, respectively, where errors were estimated from Monte Carlo simulations. The large uncertainty on the centroid of the 2365 \AA\ line is due to the low equivalent width of this line. We conclude that the Fe~II$^*$ centroids are consistent with the systemic velocity, given both the measured uncertainties on the velocities and the uncertainty on the systemic redshift determination (Section \ref{sec: data}). These measurements are accordingly consistent with Fe~II ions tracing either gas flows or stationary H~II regions. In a sample of star-forming galaxies at 1 $\lesssim$ \emph{z} $\lesssim$ 2, \citet{erb2012} found that [O~II]/Fe~II$^*$ line ratios were too large to be consistent with simple photoionization models. These authors suggest that Fe~II$^*$ emission arises from photon scattering in the galactic outflow.

We also measured the centroids of the Fe~II$^*$ lines in other composite spectra assembled on the basis of an array of morphological (e.g., size) and stellar population (e.g., SFR) parameters and we show the results in Figure \ref{IDLgaussian_plot}. In the ensemble of composite spectra, which are not independent due to substantial overlap of objects among the composites, we find that the strongest Fe~II$^*$ lines at 2396 and 2626 \AA\ have centroids scattering around 0 \kms\ while the weaker Fe~II$^*$ lines at 2365 and 2612 \AA\ are predominantly redshifted. All of the Fe~II$^*$ lines of all of the composite spectra are within 3$\sigma$ of 0 \kms, however, with the sole exception of one line (the 2365 \AA centroid of the composite spectrum assembled from objects with large angular sizes). We propose that the systematic redshift observed in the 2365 and 2612 \AA\  line centroids are due to the weaker strength of these lines relative to the other Fe~II$^*$ features and the proximity of these lines to Fe~II absorption features. Since all the Fe~II$^*$ emission lines trace the same underlying population of gas, we conclude that the kinematics of Fe~II$^*$ emission are consistent with 0 \kms\ based on the measurements of the strongest Fe~II$^*$ features. Given the relatively low S/N of our data, it is difficult to robustly measure the kinematics of Fe~II$^*$ emission in individual objects. 

\begin{figure*}
\centering
\includegraphics[width=6in]{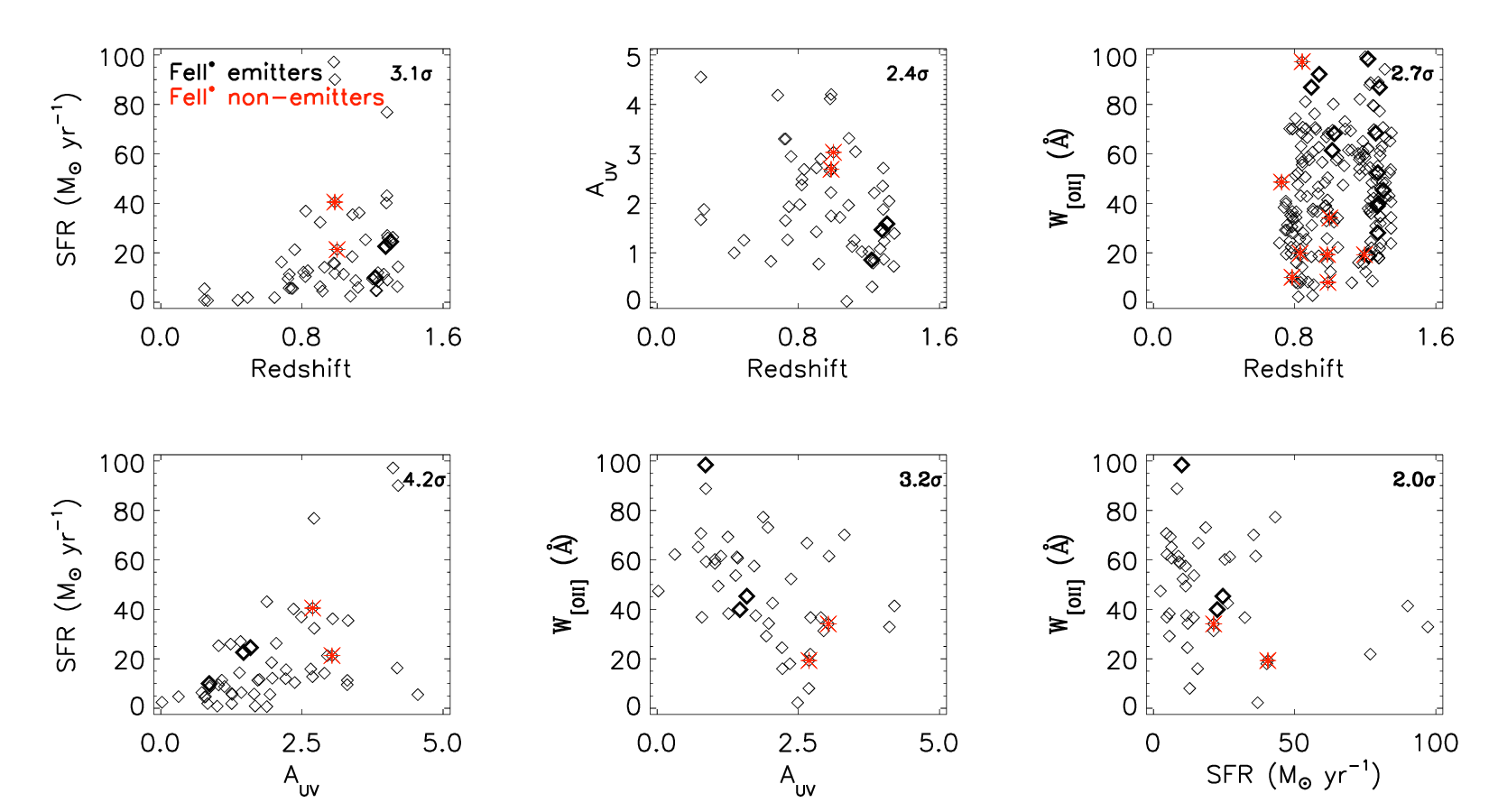} 
\caption{Top panels: SFR, $A_{\rm UV}$, and $W_{\rm [OII]}$ (from left to right) versus redshift, where Fe~II$^*$ emitters are indicated with thick black diamonds and Fe~II$^*$ non-emitters are shown as red stars. Higher redshift objects have lower $A_{\rm UV}$ values, on average, than objects at lower redshifts; this trend favors galaxies at higher redshifts exhibiting stronger Fe~II$^*$ emission than lower-redshift galaxies. Bottom panels: correlations among $A_{\rm UV}$, SFR, and $W_{\rm [OII]}$, using the same notation as above. SFR and $A_{\rm UV}$ are positively correlated, while both $W_{\rm [OII]}$ and $A_{\rm UV}$ and $W_{\rm [OII]}$ and SFR are inversely correlated. These correlations are consistent with $A_{\rm UV}$ being the primary modulator of Fe~II$^*$ emission, given that Fe~II$^*$ emitters are characterized by smaller SFRs and larger $W_{\rm [OII]}$ values than Fe~II$^*$ non-emitters.}
\label{sfr_z}
\end{figure*}

A complementary investigation of Fe~II$^*$ kinematics is to compare the velocity offsets of a strong Fe~II$^*$ line and a nebular emission line \citep[e.g.,][]{rubin2010c}. Nebular emission lines such as [Ne~III] trace star-forming regions and therefore probe stationary gas. In Figure \ref{composite_nebular}, we show the Fe~II$^*$ 2626 \AA\ and [Ne~III] 3869 \AA\ features of a composite spectrum assembled from the spectra of 12 objects with strong detections of both Fe~II$^*$ 2626 \AA\ and [Ne~III] 3869 \AA. The kinematics of these two lines are similar; the Fe~II$^*$ line has a centroid of --12 
\kms\ while the [Ne~III] line has a centroid of --16 
\kms. In light of the uncertainty on our determination of the systemic redshift, we conclude that both Fe~II$^*$ and [Ne~III] have kinematics consistent with being centered at 0 \kms. 

\citet{rubin2010c} measured the kinematics of Fe~II$^*$ lines in the spectrum of a starburst galaxy at \emph{z} $\sim$ 0.69  and found that Fe~II$^*$ lines trace gas that is redward or within 30 \kms\ of the systemic velocity. These authors note that the kinematics of Fe~II$^*$ features are inconsistent with the kinematics of both Fe\Rmnum{2} resonant absorption lines ($\Delta$V $\sim$ --200 km s$^{-1}$, where $\Delta$V is the offset from systemic velocity) and features tracing galactic H~II regions ([Ne\Rmnum{3}], H$\delta$, H$\gamma$; $\Delta$V $\sim$ 0 km s$^{-1}$). Consistent with \citet{rubin2010c}, \citet{coil2011} found that Fe~II$^*$ lines are typically within 2$\sigma$ of the systemic redshift in a sample of post-starburst and AGN host galaxies at 0.2 $<$ \emph{z} $<$ 0.8. These authors propose that Fe~II$^*$ emission arises in galactic winds as opposed to star-forming regions, since Fe~II$^*$ emission is observed in post-starburst galaxies not currently experiencing star formation. In composite spectra of star-forming galaxies at 1 $\lesssim$ \emph{z} $\lesssim$ 2, \citet{erb2012} find that Fe~II$^*$ 2626 \AA\ exhibits blueshifted velocity centroids of about --50 \kms, although these authors report that individual spectra show Fe~II$^*$ 2626 \AA\ centroids scattering around 0 \kms. \citeauthor{erb2012} conclude that Fe~II$^*$ 2626 \AA\ is on average observed at velocities closer to systemic than the resonant Fe~II absorption lines. While the high S/N composite spectra of \citeauthor{erb2012} show Fe~II$^*$ kinematics near the systemic velocity, as we find in our own sample, higher resolution observations of a larger number of individual objects are needed in order to definitively measure the kinematics of fine-structure emission lines on a per-object basis. The composite spectra used here for measuring Fe~II$^*$ kinematics have significantly higher S/N (15--45 pixel$^{-1}$) but lower resolution (FWHM $\sim$435 \kms) than the spectra of individual objects employed by both \citet{rubin2010c} and \citet{coil2011}. Our data also have higher S/N than the composite spectra employed by \citet{erb2012}, although our observations are lower resolution than the \citet{erb2012} data (FWHM $\sim$190 \kms).

Numerous studies have developed models of line emission associated with galactic winds \citep[e.g.,][]{verhamme2006,steidel2010,rubin2010c,prochaska2011}. In particular, \citet{prochaska2011} present modeling of Fe~II$^*$ emission arising from photons scattered in galactic winds. These authors predict Fe~II$^*$ emission at velocities close to the systemic velocity, since an optically thin galaxy will transmit the Fe~II$^*$ emission scattered from both the backside and frontside of the wind. However, \citet{prochaska2011} also note that increased dust opacity may produce more blueshifted Fe~II$^*$ profiles as the redshifted photons scattering off the backside of the wind will be preferentially absorbed by dust due to their longer path lengths. In Figure \ref{AUV_smooth_FeII}, we compare composite spectra assembled on the basis of $A_{\rm UV}$. While we find that objects with larger $A_{\rm UV}$ values show, on average, more blueshifted 2626 \AA\ Fe~II$^*$ emission (--20 $\pm$ 42 \kms) than objects with smaller $A_{\rm UV}$ values (37 $\pm$ 41 \kms), the kinematic differences are small and not statistically significant given the systematic uncertainties in systemic redshift of our data. We note that the C~II] emission line at 2326 \AA\ is more blueshifted in objects with larger $A_{\rm UV}$ values, but as this feature is a blend of several C~II] lines the precise rest-frame centroid of this line is uncertain. Higher resolution data are needed in order to test the hypothesis of \citet{prochaska2011} that increased dust attenuation produces more blueshifted emission line profiles. 

\subsection{Fe~II$^*$ Emission Strength and Galaxy Properties} \label{sec: feIIstar_strength}

In light of the diversity of Fe~II$^*$ emission strengths observed in our sample, it is important to determine which galaxy properties modulate Fe~II$^*$ emission. Figure \ref{CMD_FeII} shows a color-magnitude diagram with the 13 Fe~II$^*$ emitters and 9 Fe~II$^*$ non-emitters highlighted. Objects showing strong Fe~II$^*$ emission are brighter ($\langle M_B \rangle$ = --21.0 $\pm$ 0.1), bluer ($\langle U-B \rangle$ = 0.54 $\pm$ 0.02), and lower stellar mass ($M_*$ = 1.6 $\pm$ 0.2 $\times$ 10$^{10}$ $M_{\odot}$) than both Fe~II$^*$ non-emitters ($\langle M_B \rangle$ = --20.7 $\pm$ 0.4, $\langle U-B \rangle$ = 0.75 $\pm$ 0.07, $\langle M_{\odot} \rangle$ = 9.7 $\pm$ 3.1 $\times$ 10$^{10}$ $M_{\odot}$) and the global sample of objects with comparable S/N in their LRIS spectra ($\langle M_B \rangle$ = --20.5 $\pm$ 0.1, $\langle U-B \rangle$ = 0.59 $\pm$ 0.02, $\langle M_{\odot} \rangle$ = 2.8 $\pm$ 0.5 $\times$ 10$^{10}$ $M_{\odot}$). This result is consistent with the interpretation by \citet{prochaska2011} and \citet{erb2012} that Fe~II$^*$ emitters likely have little dust attenuation. Below, we use our extensive multi-wavelength data set to investigate how Fe~II$^*$ strength depends on a variety of galaxy properties. 

We measured the strength of Fe~II$^*$ emission in 18 pairs of composite spectra assembled based on star-forming, gas flow, interstellar gas (i.e., Fe~II and Mg~II), stellar population, size, morphological, and redshift parameters. The star-forming properties include SFR, specific SFR (sSFR = SFR/stellar mass), SFR surface density assuming a Petrosian galaxy area ($\Sigma_{\rm SFR}$(R$_{\rm P}$)), [O~II] emission linewidth corrected for instrumental resolution ($\sigma_{\rm [OII]}$), and [O~II] emission line equivalent width ($W_{\rm [OII]}$). The gas flow properties are described by the presence of outflows or inflows detected at or above the 1$\sigma$ or 3$\sigma$ levels. Interstellar gas properties include the presence of Mg~II emission and the Fe~II 2344 \AA\ absorption equivalent width ($W_{\rm FeII}$). Dust attenuation ($A_{\rm UV}$), $B$-band luminosity ($M_B$), $U-B$ color, and stellar mass ($M_*$) are the measured stellar population parameters. Size and morphological properties include angular size, physical size, disk inclination (\emph{i}), and Gini coefficient \citep[\emph{G;}][]{lotz2004}. Redshift, measured spectroscopically, is an additional parameter. We summarize the composite spectra used in this paper in Table \ref{sumtable}.

For each property, we assembled composite spectra based on a binary division of the data according to that parameter (i.e., low SFR and high SFR, etc.). Fe~II$^*$ emission is prominent in all the composite spectra, although its strength varies as a function of galaxy properties. We investigated how the strength of Fe~II$^*$ changed between each pair of composite spectra using equivalent width measurements of the strongest Fe~II$^*$ features at 2396 and 2626 \AA. We measured local continua around each Fe~II$^*$ line and defined the extent of each feature as the region where the flux was greater than the continuum. A Gaussian profile was fit to each Fe~II$^*$ line and an equivalent width was estimated by integrating the Gaussian fit and dividing the summed flux by the local continuum. We define the quantity D$_{\rm FeII^*}$ as the average of the equivalent width differences between each pair of composite spectra: 

\begin{equation} \rm D_{\rm FeII^*} = \frac{\Delta EW_{2396} + \Delta EW_{2626}}{2} \label{eqn: D1} \end{equation} 

\noindent where $\Delta$EW$_{2396}$ is the difference in Fe~II$^*$ 2396 \AA\ equivalent widths between the composite spectra and $\Delta$EW$_{2626}$ is the analogous difference for the 2626 \AA\ line. D$_{\rm FeII^*}$ accordingly has units of \AA. In Figure \ref{IDLgaussianEW}, we show the values of D$_{\rm FeII^*}$ for the 18 parameters. We observe D$_{\rm FeII^*}$ ranging from 0.0--0.4 \AA. The uncertainty on D$_{\rm FeII^*}$, $\delta$D$_{\rm FeII^*}$, was estimated through Monte Carlo realizations: for each composite spectrum, we constructed 1000 simulated spectra by perturbing the data at each wavelength by a value drawn from a Gaussian distribution of 1$\sigma$ width equal to the data's error spectrum at that wavelength. We measured the Fe~II$^*$ equivalent widths for each ensemble of 1000 simulated spectra and adopted the standard deviation of the equivalent width distribution as the error. We estimated $\delta$D$_{\rm FeII^*}$ for each parameter by propagating errors through Equation \ref{eqn: D1}. The average $\delta$D$_{\rm FeII^*}$ of the sample is 0.15 \AA\ and $\delta$D$_{\rm FeII^*}$ ranges from 0.11--0.20 \AA.

D$_{\rm FeII^*}$ is significant at $\ge$ 2$\sigma$ for four parameters: \emph{z}, SFR, $A_{\rm UV}$, and $W_{\rm [OII]}$. Stronger Fe~II$^*$ emission is seen in higher redshift objects, those with lower SFRs, lower $A_{\rm UV}$ values, and larger $W_{\rm [OII]}$ measurements. \citet{erb2012} also find stronger Fe~II$^*$ emission in objects with lower SFRs; their analyses on the basis of $A_{\rm UV}$ were inconclusive and these authors do not assemble composite spectra according to $W_{\rm [OII]}$ or redshift. In Figure \ref{SFR_smooth_FeII}, we show the four pairs of \emph{z}, SFR, $A_{\rm UV}$, and $W_{\rm [OII]}$ composite spectra. It is important to highlight that the composite spectra assembled on the basis of SFR, $A_{\rm UV}$, and $W_{\rm [OII]}$ exhibit different properties from those in the composite spectra assembled according to \emph{z}. The stronger Fe~II$^*$ emission observed in the lower SFR, lower $A_{\rm UV}$, and larger $W_{\rm [OII]}$ composite spectra is accompanied by weaker Mg~II absorption and similar, if not weaker, Fe~II absorption. However, the higher \emph{z} composite shows stronger Fe~II$^*$ emission and yet comparable Mg~II absorption and stronger Fe~II absorption than the lower \emph{z} composite; these spectral trends echo those seen in the composite spectra divided according to Fe~II$^*$ emission strength (Figure \ref{FeII_smooth}). It appears that the division of objects on the basis of redshift yields significantly different spectral trends from those based on the division of the sample according to SFR, $A_{\rm UV}$, or $W_{\rm [OII]}$. We propose that the different redshift distributions of Fe~II$^*$ emitters and non-emitters -- where emitters are preferentially at higher redshift -- and the changing make-up of our sample properties with redshift contribute to the trends seen in the composite spectra assembled in bins by redshift, since objects in the DEEP2 survey at higher redshifts are preferentially bluer. In summary, Fe~II$^*$ emission strength increases at higher redshifts, in contrast to the lack of evolution in the Fe~II$^*$/Fe~II equivalent width ratio described in Section \ref{sec: feIIstaremitttersnonemitters}. We attribute this evolution in Fe~II$^*$ emission strength to changes in global galaxy properties with redshift. 

While the above analyses relied on a binary division of the data according to galaxy properties, we also employed a complementary investigation of the links between Fe~II$^*$ emission and \emph{z}, SFR, $A_{\rm UV}$, and $W_{\rm [OII]}$. Specifically, we utilized our samples of Fe~II$^*$ emitters and non-emitters (Section \ref{sec: feIIstaremitttersnonemitters}) to study if and how these two samples separate in \emph{z}, SFR, $A_{\rm UV}$, and $W_{\rm [OII]}$ parameter space (Figure \ref{auv_FeII}). Each of these four parameters shows a separation of Fe~II$^*$ emitters and non-emitters, although the sample size is small in the cases of SFR and $A_{\rm UV}$. The larger sample of objects with redshifts and $W_{\rm [OII]}$ measurements permits a Kolmogorov-Smirnov (KS) test of the probability that emitters and non-emitters are drawn from the same parent population. We find a probability of 0.003 that the redshifts of the Fe~II$^*$ emitters and non-emitters arise from the same distribution and a 0.03 probability that the $W_{\rm [OII]}$ values of the emitters and non-emitters are from the same parent population. While this result indicates in particular the high significance with which Fe~II$^*$ emitters and non-emitters separate in redshift space, such that Fe~II$^*$ emitters are preferentially found at larger redshifts, \emph{z} is not an intrinsic property of a galaxy. It is therefore important to examine how SFR, $A_{\rm UV}$, and $W_{\rm [OII]}$ themselves are correlated with \emph{z} in order to investigate if the observed trend between Fe~II$^*$ emission strength and \emph{z} is only a secondary correlation or a function of redshift correlating with intrinsic galaxy propeties.

In Figure \ref{sfr_z}, we show how SFR, $A_{\rm UV}$, and $W_{\rm [OII]}$ are themselves interrelated: systems with larger SFRs have larger $A_{\rm UV}$ values, systems with stronger $W_{\rm [OII]}$ have lower $A_{\rm UV}$ values, and systems with stronger $W_{\rm [OII]}$ have lower SFRs. SFR, $A_{\rm UV}$, and $W_{\rm [OII]}$ are accordingly correlated such that the stronger Fe~II$^*$ emission in objects with lower SFRs, lower $A_{\rm UV}$ values, and larger $W_{\rm [OII]}$ measurements can be explained as arising primarily due to the effect of a single parameter: $A_{\rm UV}$. Furthermore, the correlation between redshift and $A_{\rm UV}$ is likely responsible for the observed correlation between Fe~II$^*$ emission strength and \emph{z}. Figure \ref{sfr_z} also shows the correlations of SFR, $A_{\rm UV}$, and $W_{\rm [OII]}$ versus redshift. Redshift and SFR are correlated at the 3.1$\sigma$ level ($r_S$ = 0.44, where $r_S$ is the Spearman rank-order correlation coefficient) and higher redshift systems have higher SFRs. This positive correlation between redshift and SFR is in the opposite sense of the trends we find with Fe~II$^*$ emission strength, where both low-SFR and high-redshift systems show stronger Fe~II$^*$ emission. We find a correlation at the 2.4$\sigma$ level ($r_S$ = --0.33) between redshift and $A_{\rm UV}$, such that higher redshift systems are less attenuated. That higher redshift objects have both higher SFR and lower $A_{\rm UV}$ appears at first contradictory, given that SFR and $A_{\rm UV}$ are themselves positively correlated at a given redshift \citep[this work;][]{brinchmann2004}. These trends can be explained by the results of \citet{adelberger2000}, where these authors noted that at a given SFR, objects at higher redshifts are less attenuated than their lower-redshift counterparts. Additionally, the \emph{R}-band selection limits of the DEEP2 survey dictate that objects at higher redshifts must be either brighter or bluer to still fall in the selection window. Therefore, it would be expected that objects at higher redshifts are preferentially bluer than lower-redshift galaxies. Finally, we find a 2.7$\sigma$ correlation ($r_S$ = 0.20) between redshift and $W_{\rm [OII]}$, in the sense that higher-redshift objects have larger $W_{\rm [OII]}$ measurements. The correlation between redshift and $A_{\rm UV}$ is suggestive that the trend between redshift and Fe~II$^*$ arises simply as a secondary correlation and that $A_{\rm UV}$ is the primary property modulating Fe~II$^*$ emission strength (Section \ref{sec: modulatedbydust}). Furthermore, the correlation observed between $A_{\rm UV}$ and Fe~II$^*$ emission strength cannot simply be an artifact of the manner in which $A_{\rm UV}$ is estimated and Fe~II$^*$ emission equivalent width measured, even though both quantities depend on the properties of the UV continuum. Given that galaxies with lower $A_{\rm UV}$ values have slightly higher UV continuum luminosity densities on average than those with higher $A_{\rm UV}$ values, one expects a \emph{lower} Fe~II$^*$ emission equivalent width at lower $A_{\rm UV}$ for a fixed Fe~II$^*$ emission-line flux (contrary to what we observe). 

Other parameters besides SFR, $A_{\rm UV}$, $W_{\rm [OII]}$, and \emph{z} may modulate Fe~II$^*$ emission strength. We find that 3$\sigma$ gas flow, $U-B$ color, the presence of Mg~II emission, angular size, and physical size have D$_{\rm FeII^*}$ values significant at $\ge$ 1$\sigma$, such that stronger Fe~II$^*$ emission is seen in systems with 3$\sigma$ inflows (as opposed to 3$\sigma$ outflows), bluer $U-B$ colors, stronger Mg~II emission, smaller angular sizes, and smaller physical sizes. The stronger Fe~II$^*$ emission observed in systems with smaller angular sizes is consistent with the theory that slit losses may be responsible for the lack of Fe~II$^*$ emission in local galaxy samples \citep[e.g.,][]{giavalisco2011,erb2012}. Curiously, we do not find a significant correlation between Fe~II$^*$ emission strength and Fe~II equivalent width in the composite spectra assembled according to $W_{\rm FeII}$. This result is striking given that the Fe~II$^*$ emitter and non-emitter composite spectra in Figure \ref{FeII_smooth} show pronounced differences in Fe~II absorption strength. However, the fact that Fe~II$^*$ emitters are preferentially at higher redshifts than Fe~II$^*$ non-emitters may contribute to the stark differences in Fe~II absorption strength seen in Figure \ref{FeII_smooth}. If the properties of interstellar gas evolve with redshift such that systems at higher redshifts show stronger resonant Fe~II absorption, then one would expect that objects with strong Fe~II$^*$ emission also exhibit deeper Fe~II absorption profiles. We tested this hypothesis by assembling composite spectra holding $A_{\rm UV}$ constant and varying redshift; we find that objects at higher redshifts do exhibit preferentially stronger resonant Fe~II absorption than systems at lower redshifts. This result explains the trends seen in the Fe~II$^*$ emitter and non-emitter composite spectra (Figure \ref{FeII_smooth}) as the Fe~II$^*$ emitters are both at higher redshifts and also exhibit stronger Fe~II resonant absorption. In order to probe how Fe~II$^*$ emission and Fe~II resonant absorption relate, we must therefore look at samples at fixed redshift. 

\section{Resonant Mg~II Emission} \label{sec: mgII_introduction}

Mg~II, a low ionization state of a cosmically-abundant element \citep{savage1996}, is a useful tracer of interstellar gas. Mg~II transitions in the rest-frame ultraviolet are commonly used as probes of galactic winds at intermediate redshifts due to their placement above the atmospheric cut-off for samples at \emph{z} $\gtrsim$ 0.2. In unsaturated systems, the bluer line of the Mg~II doublet at $\lambda \lambda$ 2796,2803 \AA\ is twice as strong as the redder line. However, this 2:1 line ratio is not always seen in our data. Rather, we find some objects with inverted Mg~II ratios (i.e., stronger absorption in the 2803 line than in the 2796 \AA\ line) and a majority of objects with roughly equal absorption-line strengths. The former case may be correlated with dense galactic winds, as discussed below, while the latter case indicates line saturation. 

The Mg~II doublet, while primarily seen in absorption, is also observed with a P-Cygni emission profile in objects as varied as Seyfert 1s \citep{wu1983}, ultraluminous infrared galaxies \citep{martin2009}, local star-forming spiral galaxies \citep{kinney1993}, and high-redshift starburst galaxies \citep{weiner2009,rubin2010c,giavalisco2011,erb2012,martin2012}. The physical origin of Mg~II emission has been attributed to several processes, including photon scattering off the backside of a galactic wind \citep{phillips1993,weiner2009,rubin2010c}, where this process has also been seen in Ly$\alpha$ emission in Lyman break galaxies \citep{pettini2001,shapley2003}. Both features of the Mg~II doublet can be strongly affected by emission filling due to the lack of excited ground states available for fluorescence. We examine here the diversity of Mg~II profiles in our sample, focusing specifically on the incidence of Mg~II emission. Even in our relatively low-resolution data, the Mg~II doublet is resolved given its wide velocity separation ($\sim$770 \kms). 

\subsection{Mg~II Emitters and Non-emitters} \label{sec: mgII}

\begin{figure}
\centering
\includegraphics[width=3.5in]{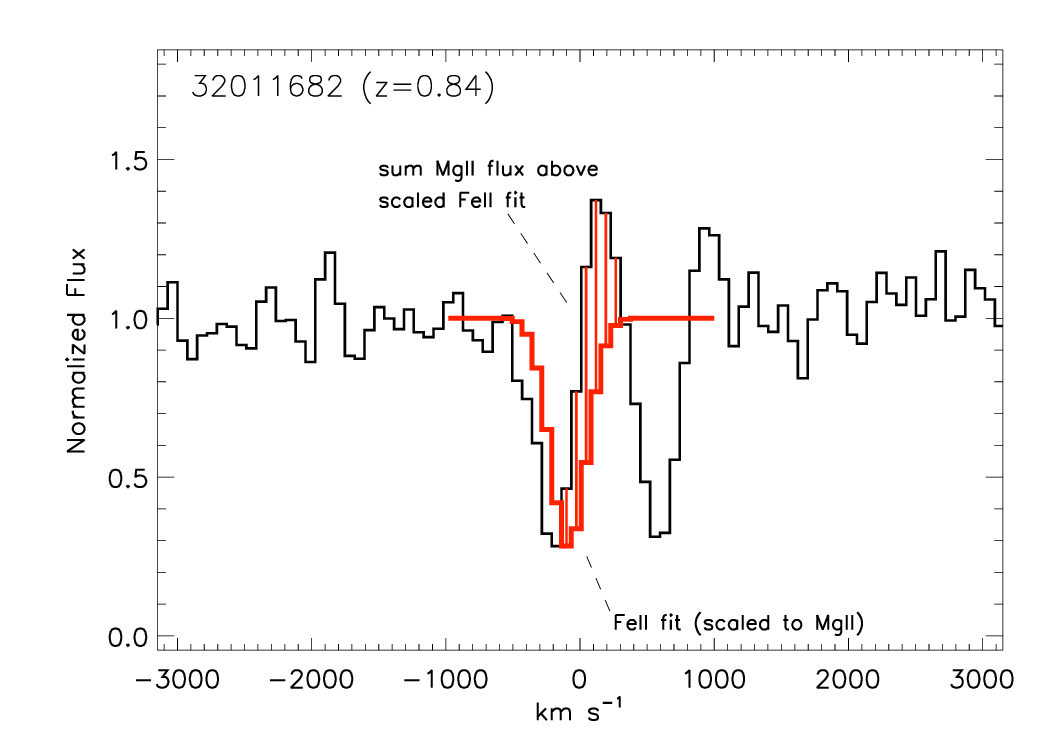} 
\caption{Methodology for identifying Mg~II emission. We assume that the Fe~II profile represents a fiducial absorption profile largely unaffected by emission filling. After normalizing the depth of the Fe~II profile to that of the Mg~II profile, we sum the normalized flux in the Mg~II feature above the scaled Fe~II profile over a range of 500 \kms, beginning at the minimum of the Fe~II absorption profile. The summed flux is representative of the emission contribution of Mg~II to the overall Mg~II profile. This methodology is repeated for both the 2796 and 2803 \AA\ Mg~II lines and the emission strengths of each line are added to produce a total estimate of the Mg~II emission strength.}
\label{MgII_32011682_norm_min_paperfig}
\end{figure}

Several authors have investigated the frequency of Mg~II emission in samples of star-forming galaxies. \citet{weiner2009} report Mg~II emission in 4\% of their sample of \z1.4 star-forming galaxies. \citet{rubin2010a} find that 4/468 ($<$ 1\%) star-forming galaxies at a similar epoch exhibit Mg~II emission. Both of these authors identified the presence of Mg~II emission in continuum-normalized spectra by comparing the fluxes above the continuum level in two wavelength windows tracing the continuum and the region immediately redward of the 2796 \AA\ line, respectively (Figure 3 in Weiner et al. 2009). \citet{weiner2009} find Mg~II emission preferentially in blue, luminous galaxies and propose that the emission may be due to low-level AGNs, although these authors fail to find any emission lines characteristic of active galaxies (e.g., Ne~V 3425 \AA) in a co-added spectrum of all their data. \citet{weiner2009} also assemble a composite spectrum excluding objects with individual detections of Mg~II emission and still find Mg~II emission in the co-added stack. These authors argue that stellar chromospheric Mg~II emission is unlikely to account for the observed emission; recombination in H~II regions or scattering in galactic winds are more likely causes of Mg~II emission. 

While the technique of \citet{weiner2009} and \citet{rubin2010a} robustly identifies strong Mg~II emitters, it may miss objects with only weak Mg~II emission. In light of the increased S/N of our data ($\sim$11 per resolution element) relative to that of the \citeauthor{weiner2009} data ($\sim$1 per resolution element) and the \citeauthor{rubin2010a} data ($\sim$2 per resolution element), we developed a new method to systematically isolate objects with Mg~II emission. Instead of relying on measurements of flux above the continuum within prescribed wavelength windows, we employed the fits to the resonant Fe~II absorption features in the data \citep{martin2012}. Since the Fe~II lines we fit in our absorption-line measurements are not as susceptible to emission filling as Mg~II, we used the Fe~II profile for this subset of lines as a proxy for the intrinsic Mg~II absorption profile free of the effects of emission filling. In Figure \ref{MgII_32011682_norm_min_paperfig}, we schematically illustrate how we compare the Mg~II and Fe~II profiles to estimate the contribution of Mg~II emission to the overall Mg~II profile. For each of the objects with a $V_1$ measurement and spectral coverage of Mg~II, we plotted the Fe~II absorption fit and the Mg~II profile together in velocity space. We normalized the Fe~II fit to the lowest pixel of the Mg~II absorption trough and estimated the Mg~II emission component by summing the positive flux between the Mg~II profile and the normalized Fe~II fit. We summed the flux only over a limited velocity range -- from the minimum of the Fe~II profile to 500 \kms\ redward -- to ensure that the 2796 and 2803 \AA\ lines are not artificially extended blueward or redward of the true line profile. The resulting flux measurements have units of \AA, as these measurements represent areas integrating a continuum-normalized flux (unitless) with wavelength (units of \AA). This method is sensitive both to strong Mg~II emitters -- such as those found using the technique of \citet{weiner2009} and \citet{rubin2010a} -- and also to objects in which Mg~II emission is not visually apparent but the kinematic profile of Fe~II differs from that of Mg~II. The ability of this technique to isolate objects with different Fe~II and Mg~II kinematics results in a more complete sample of Mg~II emitters than was found by either \citet{weiner2009} or \citet{rubin2010a}. Our methodology furthermore selects objects where the Mg~II emission may not obviously extend above the local continuum; both \citet{weiner2009} or \citet{rubin2010a} used a technique in which only Mg~II emission above the continuum was considered. 

\begin{figure*}
\centering
\includegraphics[width=7in]{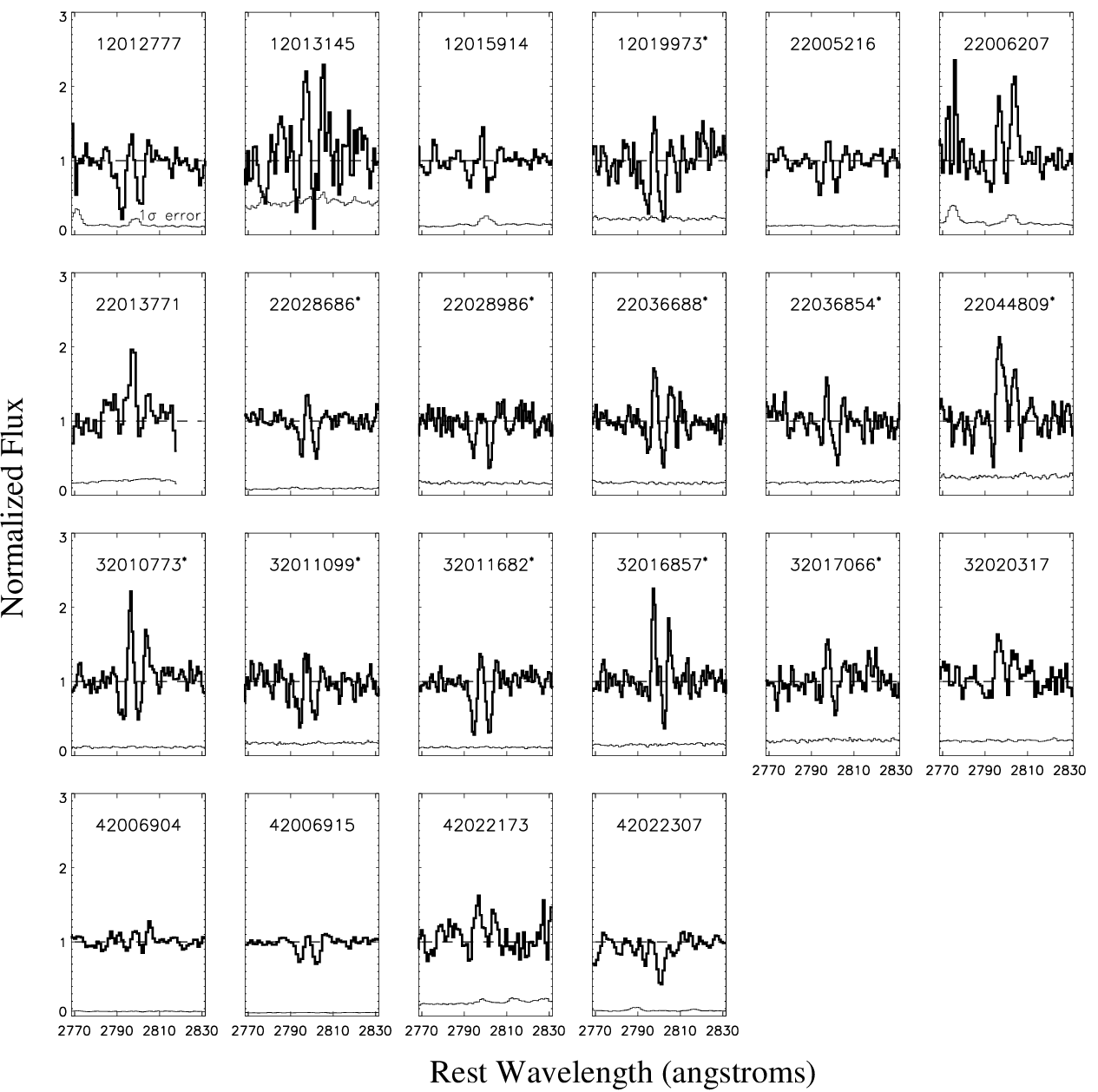} 
\caption{Thumbnails of the 22 objects with significant Mg~II emission, selected using the method described in Section \ref{sec: mgII}. The majority of objects show pronounced Mg~II emission above the continuum, where the continuum is indicated by the dashed horizontal line. Object ID numbers followed by an asterisk indicate that the object was observed with the d560 dichroic and the 600 line mm$^{-1}$ grism and the 600 line mm$^{-1}$ grating while object numbers lacking an asterisk correspond to galaxies observed with the d680 dichroic and the 400 line mm$^{-1}$ grism and the 800 line mm$^{-1}$ grating. These spectra are unsmoothed. The 1$\sigma$ error spectra are shown below the data spectra; the average S/N of the Mg~II emitters is 7.5 (12.0) pixel$^{-1}$ for the d560 (d680) samples.}
\label{IDLzoomMgIIem}
\end{figure*}

We searched for emission in the 2796 and 2803 \AA\ lines separately and we assigned an emission significance to each line using the error spectrum of the data. We then combined the emission significances of the 2796 and 2803 \AA\ features for each object, adding their associated errors in quadrature, to obtain a total Mg~II emission strength and uncertainty for each object. Of the 145 objects in our sample with a $V_1$ measurement and spectral coverage of both features of the Mg~II doublet, 22 ($\sim$15\%) show a combined 2796/2803 emission significance at or above the 6$\sigma$ level while also exhibiting at least a 3$\sigma$ emission significance in each individual line (Figure \ref{IDLzoomMgIIem}). We call this subsample of objects ``Mg~II emitters"\footnote{While 18/22 Mg~II emitters show stronger emission in the 2796 \AA\ line than in the 2803 \AA\ line -- consistent with expectations based on oscillator strengths -- four objects show more emission in the 2803 \AA\ line. These line ratios, which deviate from what is expected based on oscillator strengths ($f_{\rm 12}$ = 0.608 for the 2796 \AA\ line and $f_{\rm 12}$ = 0.303 for the 2803 \AA\ line), are predicted to occur more frequently in cases where a higher density of particles populates the galactic wind \citep{prochaska2011}. Higher quality data are necessary in order to determine if a population of objects with inverted emission ratios exists, as suggested by \citet{prochaska2011}, or if our finding is simply the result of the limited S/N of our data set.}. In \citet{martin2013}, we analyze the 2D spectra and find that one of these Mg~II emitters exhibits spatially-extended emission. If we instead require that Mg~II emitters exhibit a 3$\sigma$ detection in only at least one of the Mg~II lines, we find that 57 objects ($\sim$39\%) meet the Mg~II emitter criterion. There are clearly many possible ways of isolating Mg~II emitters, but the criteria we use primarily isolate objects with visually striking emission and fully utilize the information provided by our coverage of both Fe~II and Mg~II features.

Whereas we find that $\sim$15\% of objects in our sample show strong Mg~II emission, \citet{erb2012} observe that 33/96 star-forming galaxies at 1 $\lesssim$ \emph{z} $\lesssim$ 2 ($\sim$30\%) exhibit strong Mg~II emission. \citet{erb2012} flagged Mg~II-emitting galaxies by searching for two adjacent pixels at least 1.5$\sigma$ above the continuum in either the 2796 or 2803 \AA\ lines. This technique clearly depends on the S/N of the spectra. Our study finds a higher fraction of Mg~II-emitters galaxies than both \citet{weiner2009} and \citet{rubin2010a} and we attribute this discrepancy to the increased sensitivity of our method to objects with different Fe~II and Mg~II kinematics. We assembled a spectral stack of the 22 objects in our sample exhibiting Mg~II emission and find no evidence for Ne~V AGN emission at 3425 \AA\ in the 19 objects with spectral coverage of Ne~V. Based on Chandra X-ray flux catalogs available for the Extended Groth Strip, only one object (out of 72 objects in this field with LRIS spectroscopy) is likely an AGN due to its X-ray flux. This object does not have spectral coverage of Mg~II and therefore is not included in the above analyses. 

\begin{figure*}
\centering
\includegraphics[width=5.5in]{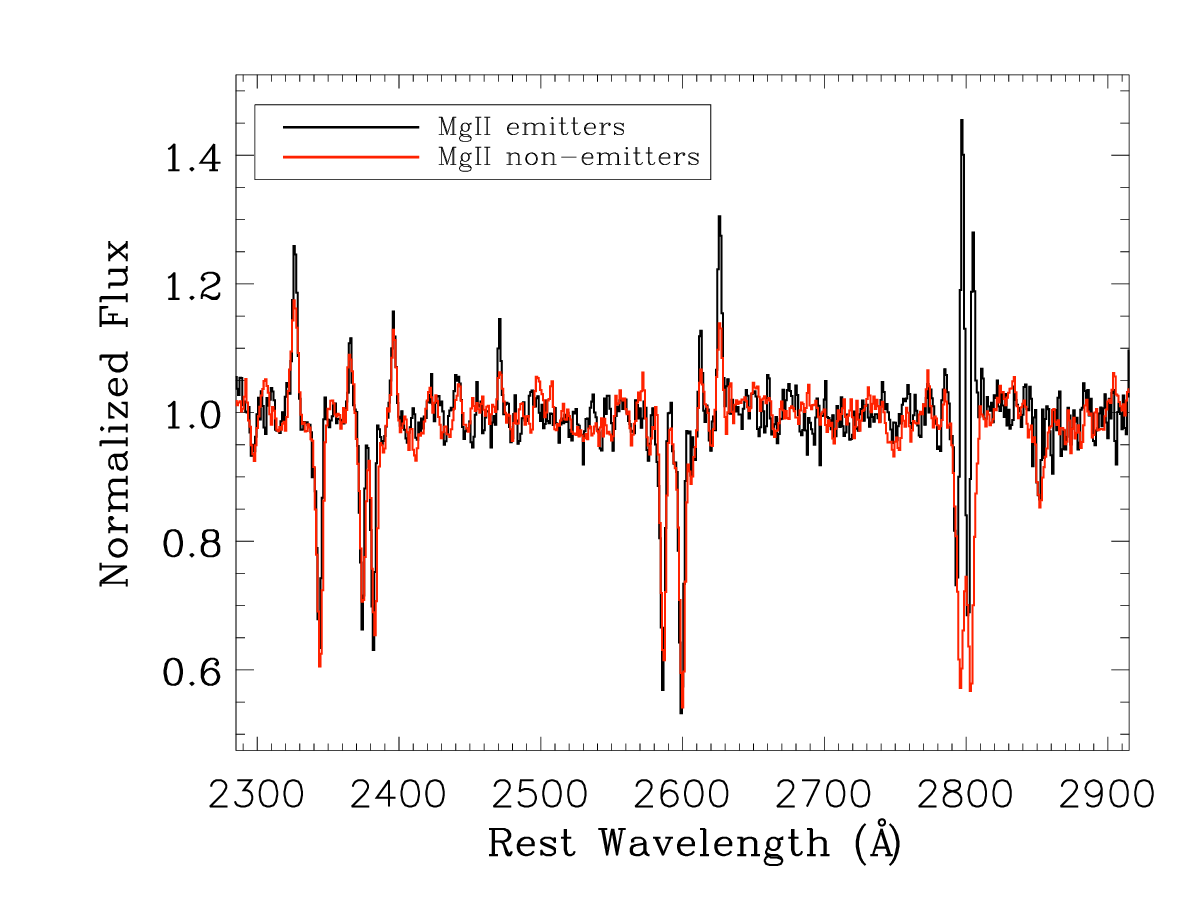} 
\caption{Comparison of composite spectra assembled from stacks of the 22 Mg~II emitters (black) and 34 Mg~II non-emitters (red). On average, Mg~II emitters show stronger Fe~II$^*$ emission than objects with weaker Mg~II emission. Typical errors on the Mg~II emitter (non-emitter) composite spectra are 0.03 (0.02) in units of normalized flux.}
\label{MgII_composites}
\end{figure*}

We also isolated a sample of 34 ``Mg~II non-emitters" with $<$ 2$\sigma$ Mg~II emission detections, where we required that each object have a continuum S/N greater than the lowest S/N observed in the Mg~II emitter sample (4.66 pixel$^{-1}$). This methodology of requiring comparable continuum S/N in the Mg~II emitter and non-emitter samples ensures that objects do not scatter into the Mg~II non-emitter sample purely because of noise. Three Mg~II non-emitters have colors indicative of ``green valley" galaxies; our conclusions remain unchanged if these objects are removed from the non-emitter sample. In Figure \ref{MgII_composites}, we plot composite spectra assembled from Mg~II emitters and non-emitters, respectively. While Fe~II$^*$ emission is stronger in the stack of Mg~II emitters, the composite spectra are otherwise comparable in terms of Fe~II kinematics and strength.

According to radiative transfer modeling by \citet{prochaska2011}, the geometry of galactic winds affects the strength of Mg~II emission. Specifically, the opening angle of the wind, assuming a biconical outflow geometry, modulates the strength of Mg~II emission. More collimated outflows show weaker Mg~II emission while more isotropic outflows show stronger Mg~II emission. The 22 objects in our sample that show significant Mg~II emission are therefore expected to have gas flows opening into large solid angles. As objects with wide-angle gas flows should show kinematic signatures of galactic winds more frequently than systems with more collimated flows, we checked the frequency of 1$\sigma$ and 3$\sigma$ detections of gas flows in the sample of Mg~II emitters. We find that 59\% (50\%) of Mg~II emitters show 1$\sigma$ (3$\sigma$) gas flows, compared with 67\% (27\%) of the sample as a whole. These results show that the frequency of objects exhibiting strong ($\ge$ 3$\sigma$) gas flows is higher in the Mg~II emitter sample than in the parent sample. Therefore, our data do support the hypothesis of \citet{prochaska2011} that more isotropic outflows are associated with stronger Mg~II emission. Observations with upcoming integral field units on large telescopes (e.g., the Keck Cosmic Web Imager \citep[KCWI;][]{martin2010} and the Multi Unit Spectroscopic Explorer \citep[MUSE;][]{bacon2010} on the VLT) will be important for determining the geometry of Mg~II emission. 

We measured the kinematics of the Mg~II emission peaks in the composite spectrum assembled from the 22 objects showing significant Mg~II emission. We fit a Gaussian profile to the 2796 and 2803 \AA\ features separately and find that the emission peaks are located at 111 $\pm$ 16 and 168 $\pm$ 30 \kms, respectively. 
\citet{weiner2009} and \citet{rubin2010a} also note Mg~II emission peaks redshifted by approximately $\sim$100 \kms\ in their composite spectra of star-forming galaxies at \z1.4 and 0.7 $<$ \emph{z} $<$ 1.5, respectively.  

\subsection{Mg~II Emission Strength and Galaxy Properties} \label{sec: mgiiandproperties}

\begin{figure*}
\centering
\includegraphics[width=7in]{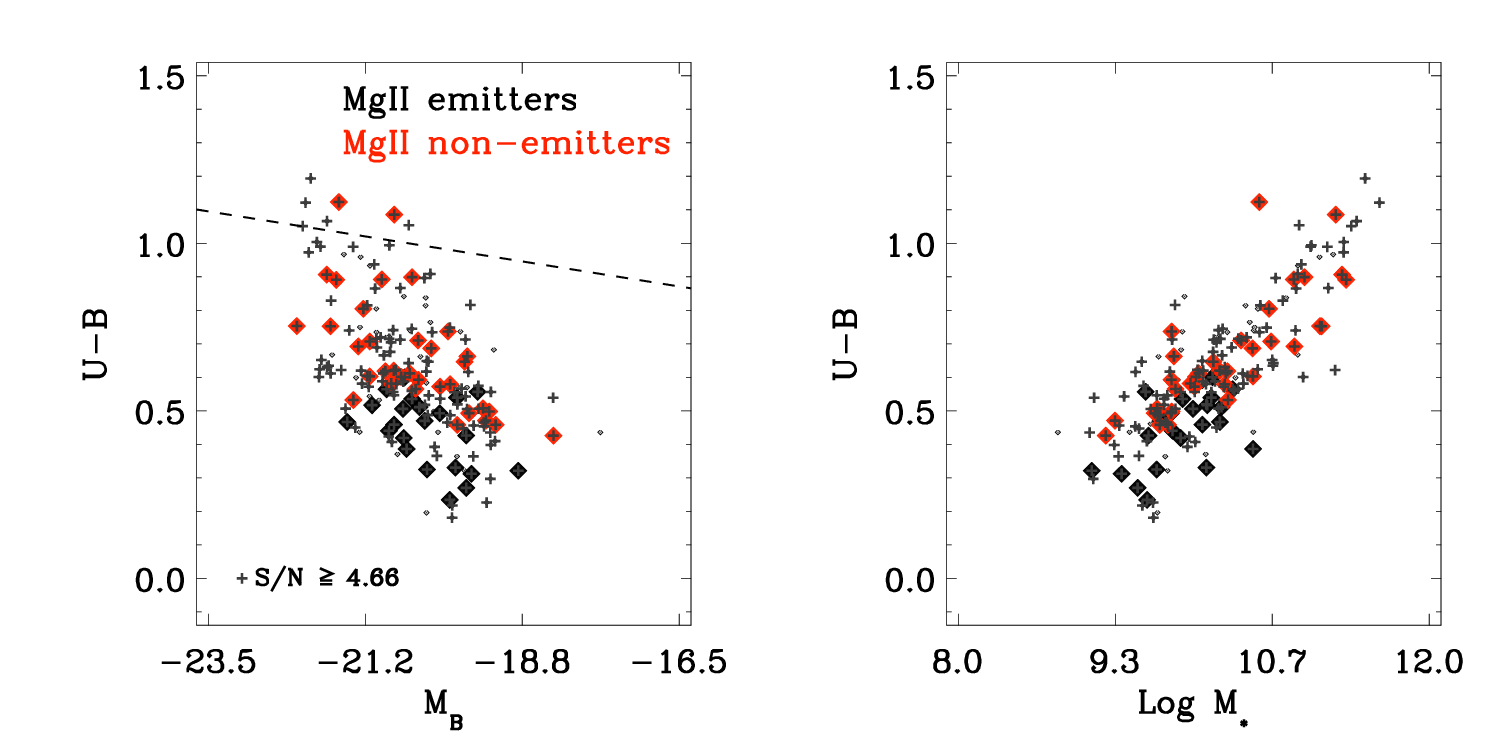} 
\caption{Left: color-magnitude diagram with the 22 Mg~II emitters shown as filled black diamonds and the 34 Mg~II non-emitters shown as filled red diamonds. Gray crosses indicate objects with S/N $>$ 4.66 pixel$^{-1}$, the minimum S/N of the Mg~II emitter sample. Consistent with the results of \citet{weiner2009} and \citet{martin2012}, we find that objects showing Mg~II in emission are bluer than the general galaxy poulation: the Mg~II emitters have an average $U-B$ color of 0.44 $\pm$ 0.02 compared with an average $U-B$ color of 0.67 $\pm$ 0.03 for the Mg~II non-emitters. The global sample of objects is characterized by an average $U-B$ color of 0.60 $\pm$ 0.02, where these errors represent the standard deviation of the mean. Right: color versus stellar mass plot, where the symbols are the same as in the left panel. Mg~II emitters have characteristically lower stellar masses ($\langle$$M_*$$\rangle$ = 1.0 $\pm$ 0.2 $\times$ 10$^{10}$ $M_{\odot}$) than both objects with weaker Mg~II emission ($\langle$$M_*$$\rangle$ = 4.1 $\pm$ 0.9 $\times$ 10$^{10}$ $M_{\odot}$) and the population as a whole ($\langle$$M_*$$\rangle$ = 3.3 $\pm$ 0.4 $\times$ 10$^{10}$ $M_{\odot}$), consistent with the results of \citet{martin2012}.} 
\label{CMD_MgII}
\end{figure*}

As strong Mg~II emission is only seen in a subset of our sample, we investigate here the galaxy properties modulating Mg~II emission. In the left-hand panel of Figure \ref{CMD_MgII}, we show a color-magnitude diagram with the 22 Mg~II emitters and the 34 Mg~II non-emitters highlighted. We find that the galaxies showing strong Mg~II emission are preferentially bluer ($\langle U-B \rangle$ = 0.44 $\pm$ 0.02) and lower stellar mass ($\langle$$M_*$$\rangle$ = 1.0 $\pm$ 0.2 $\times$ 10$^{10}$ $M_{\odot}$) than both Mg~II non-emitters ($\langle U-B \rangle$ = 0.67 $\pm$ 0.03, $\langle$$M_*$$\rangle$ = 4.1 $\pm$ 0.9 $\times$ 10$^{10}$ $M_{\odot}$) and the global sample of objects with comparable S/N in their LRIS spectra ($\langle U-B \rangle$ = 0.60 $\pm$ 0.02, $\langle$$M_*$$\rangle$ = 3.3 $\pm$ 0.4 $\times$ 10$^{10}$ $M_{\odot}$); 
the errors on these colors represent the standard deviation of the mean. \citet{martin2013} discuss one Mg~II emitter with spatially-extended Mg~II emission and remark that this object has a notably blue $U-B$ color and a stellar mass falling in the lowest third of the sample. 

\citet{weiner2009} find that objects at \z1.4 showing Mg~II in emission are typically drawn from a bluer and more luminous population. To test if our own data also show Mg~II emitters being more luminous than the sample as a whole, we conducted a KS test comparing the distributions of the \emph{B}-band absolute luminosities of the 22 objects with significant Mg~II emission and the entire sample of objects with $V_1$ measurements and spectral coverage of Mg~II. We find a probability of $\sim$30\% that the two distributions are drawn from the same parent population. Our data accordingly do not suggest an intrinsic luminosity difference between objects showing Mg~II in emission and the general population, although we acknowledge that our sample is substantially smaller than that of \citet{weiner2009}. 

\begin{figure}
\centering
\includegraphics[width=3.5in]{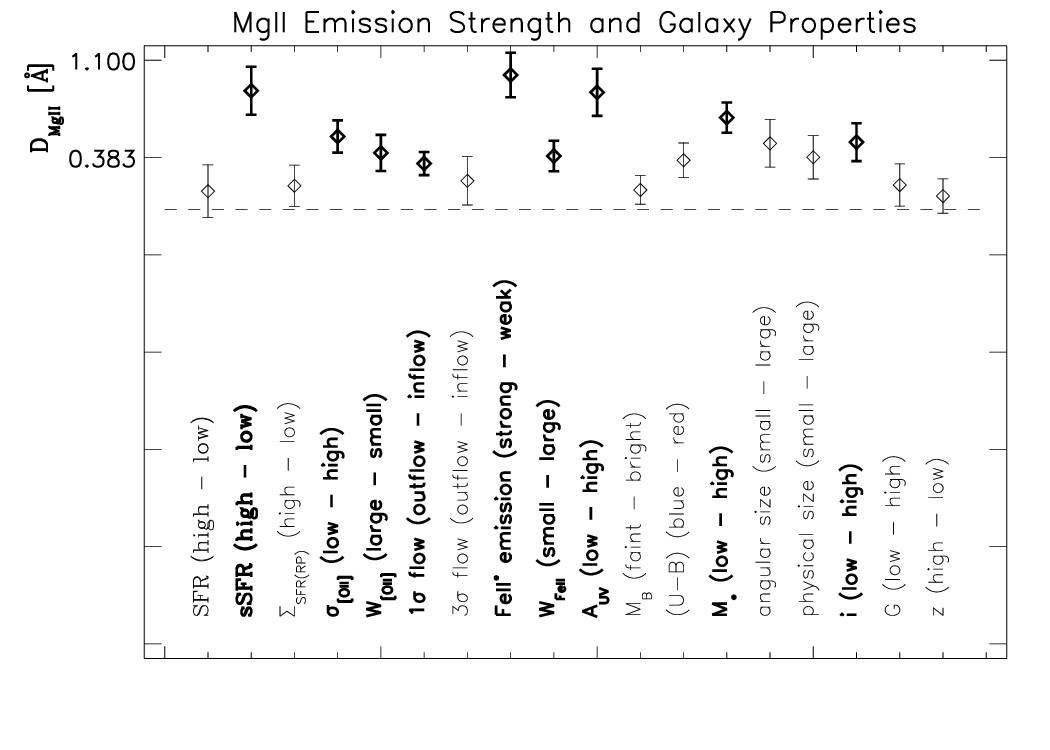} 
\caption{Variation of Mg~II emission strength with galaxy properties. D$_{\rm MgII}$, a parameterization of the change in Mg~II emission strength between two spectra, is shown for pairs of composite spectra assembled according to 18 different galaxy properties. The largest D$_{\rm MgII}$ values, significant at $>$ 3$\sigma$, are observed for nine intrinsic properties: sSFR, $\sigma_{\rm [OII]}$, $W_{\rm [OII]}$, 1$\sigma$ gas flows, Fe~II$^*$ emission strength, $W_{\rm FeII}$, $A_{\rm UV}$, $M_*$, and disk inclination \emph{i} (bold text). Stronger Mg~II emission is observed in objects with higher sSFR, lower $\sigma_{\rm [OII]}$, larger $W_{\rm [OII]}$, 1$\sigma$ outflows, stronger Fe~II$^*$ emission, smaller $W_{\rm FeII}$, lower $A_{\rm UV}$, lower $M_*$, and lower $i$.}
\label{MgIIemitters_figure}
\end{figure}

As a corollary to the analyses in Section \ref{sec: feIIstar_strength} investigating the links between Fe~II$^*$ emission and galaxy properties, we now turn to systematically analyzing how a variety of galaxy properties are correlated with Mg~II emission strength. We showed above that individual objects with strong Mg~II emission have bluer colors and lower stellar masses than the sample as a whole (Figure \ref{CMD_MgII}), consistent with the results of \citet{martin2012} based on both composite and individual Äspectra. Now, we employ composite spectra to investigate how the strength of Mg~II emission varies as a function of different galaxy properties. We included 17 of the galaxy properties used for the Fe~II$^*$ analysis is Section \ref{sec: feIIstar_strength}, where we omitted the Mg~II emission strength property. Instead of Mg~II emission strength, we used Fe~II$^*$ emission strength. In the 18 pairs of composite spectra assembled according to these galaxy properties, we measured the strength of Mg~II emission using the same technique employed for measuring Mg~II emission in the individual spectra (Section \ref{sec: mgII}). We then calculated the change in Mg~II emission strength  between each pair of composite spectra using a method analogous to that employed for Fe~II$^*$ in Section \ref{sec: feIIstar_strength}. The quantity D$_{\rm MgII}$ is defined as the difference in Mg~II emission flux between each pair of composite spectra: 

\begin{equation} \rm D_{\rm MgII} = \Delta (f_{2796} + f_{2803}) \label{eqn: D2} \end{equation} 

\noindent where f$_{2796}$ and f$_{2803}$ are the emission fluxes in each Mg~II line. In Figure \ref{MgIIemitters_figure}, we show the values of D$_{\rm MgII}$ for the 18 galaxy parameters. We observe D$_{\rm MgII}$ ranging from 0.1--1.0 \AA. The uncertainty on D$_{\rm MgII}$, $\delta$D$_{\rm MgII}$, was calculated from the individual errors on the Mg~II flux measurements. Propagating errors through Equation \ref{eqn: D2}, we find that $\delta$D$_{\rm MgII}$ ranges from 0.09--0.19 \AA, with a sample average of 0.15 \AA. 

Nine galaxy properties modulate Mg~II emission at or above the 3$\sigma$ level: sSFR, $\sigma_{\rm [OII]}$, $W_{\rm [OII]}$, 1$\sigma$ gas flows, Fe~II$^*$ emission strength, $W_{\rm FeII}$, $A_{\rm UV}$, $M_*$, and disk inclination $i$. Stronger Mg~II emission is observed in objects with higher sSFR, lower $\sigma_{\rm [OII]}$, larger $W_{\rm [OII]}$, 1$\sigma$ outflows (as opposed to 1$\sigma$ inflows), stronger Fe~II$^*$ emission, smaller $W_{\rm FeII}$, lower $A_{\rm UV}$, lower $M_*$, and lower inclination. 

\begin{figure*}
\begin{center}$
\begin{array}{c}
\includegraphics[width=3.5in]{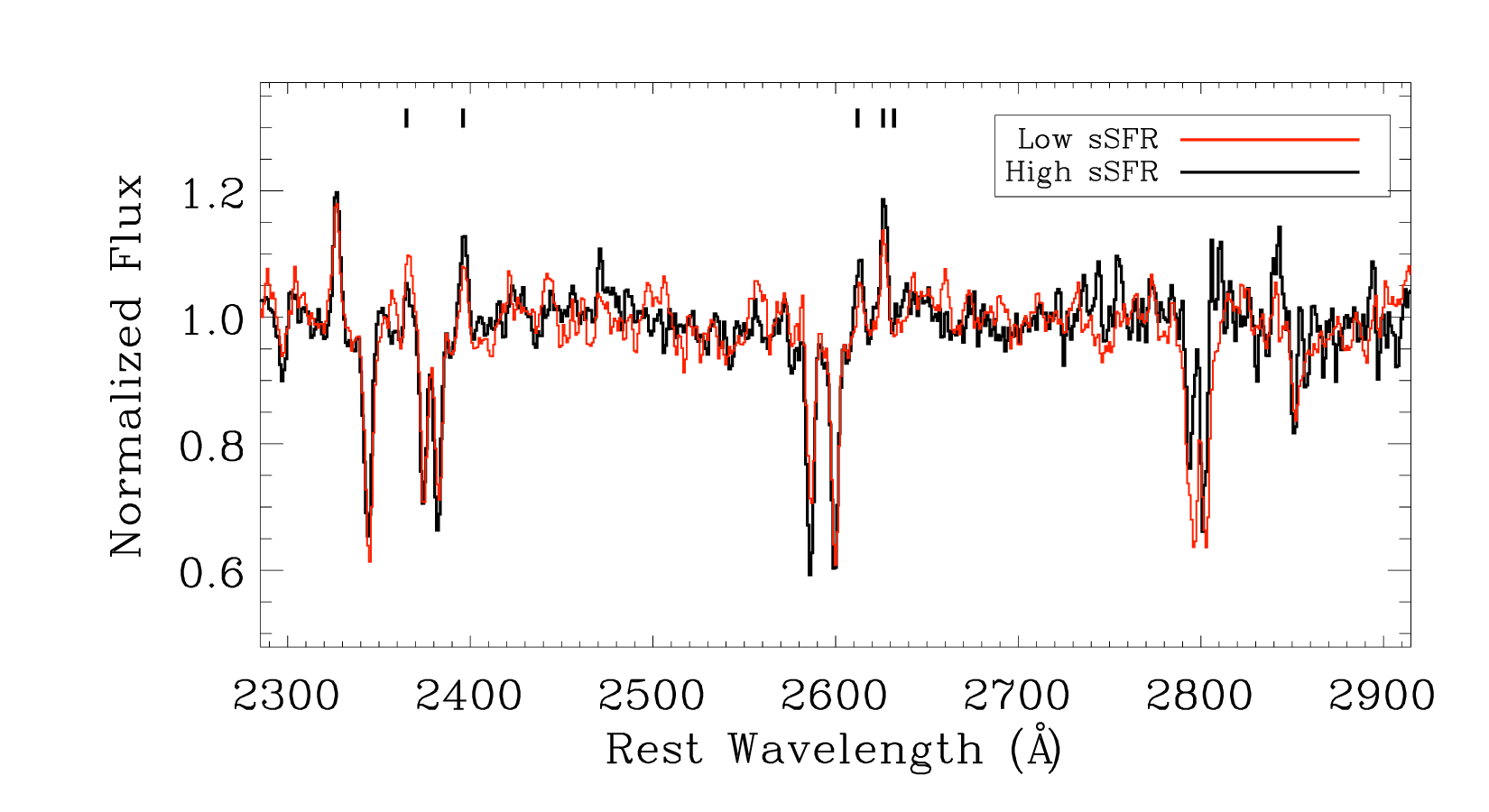} 
\includegraphics[width=3.5in]{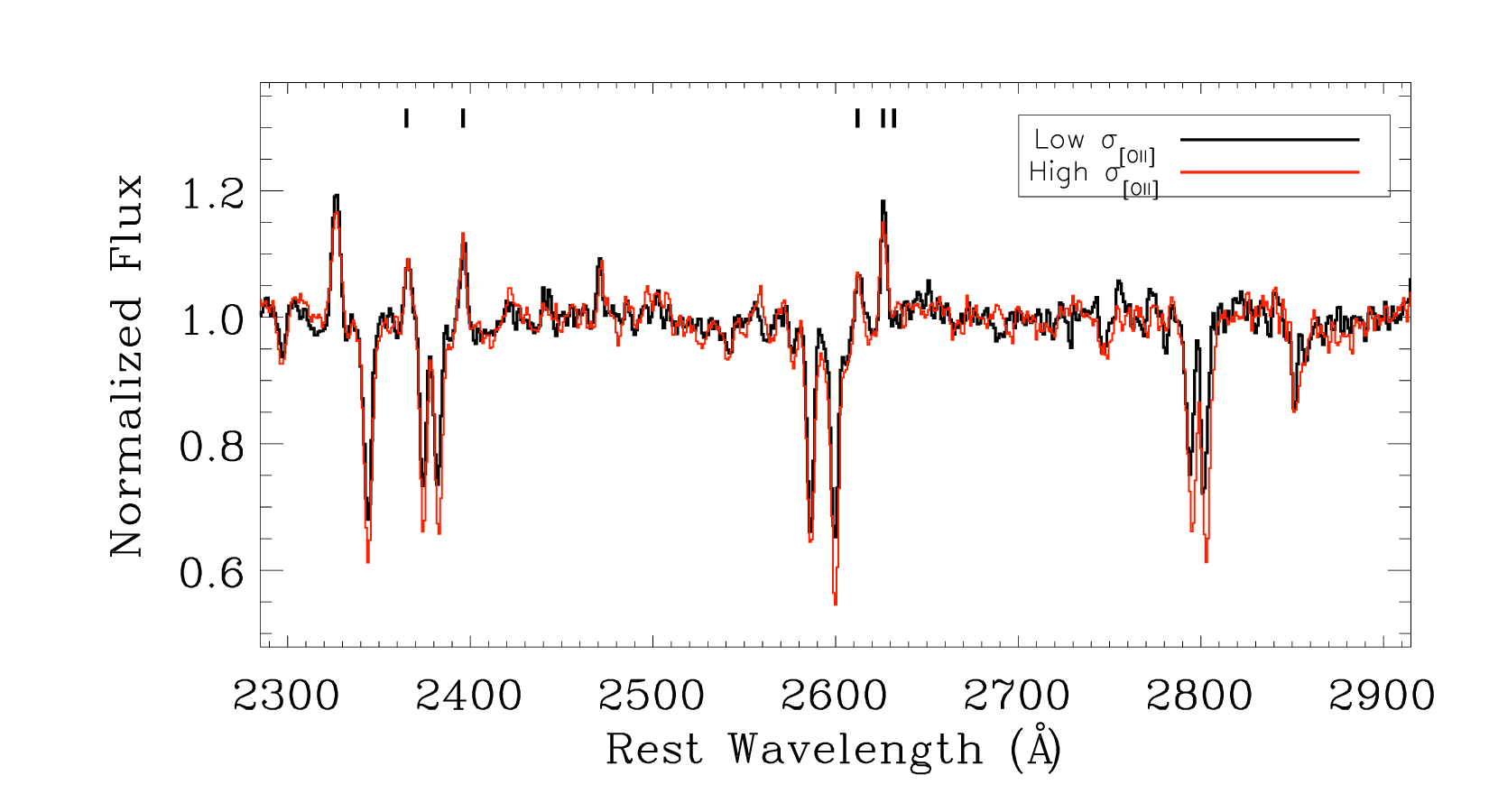} \\ 
\includegraphics[width=3.5in]{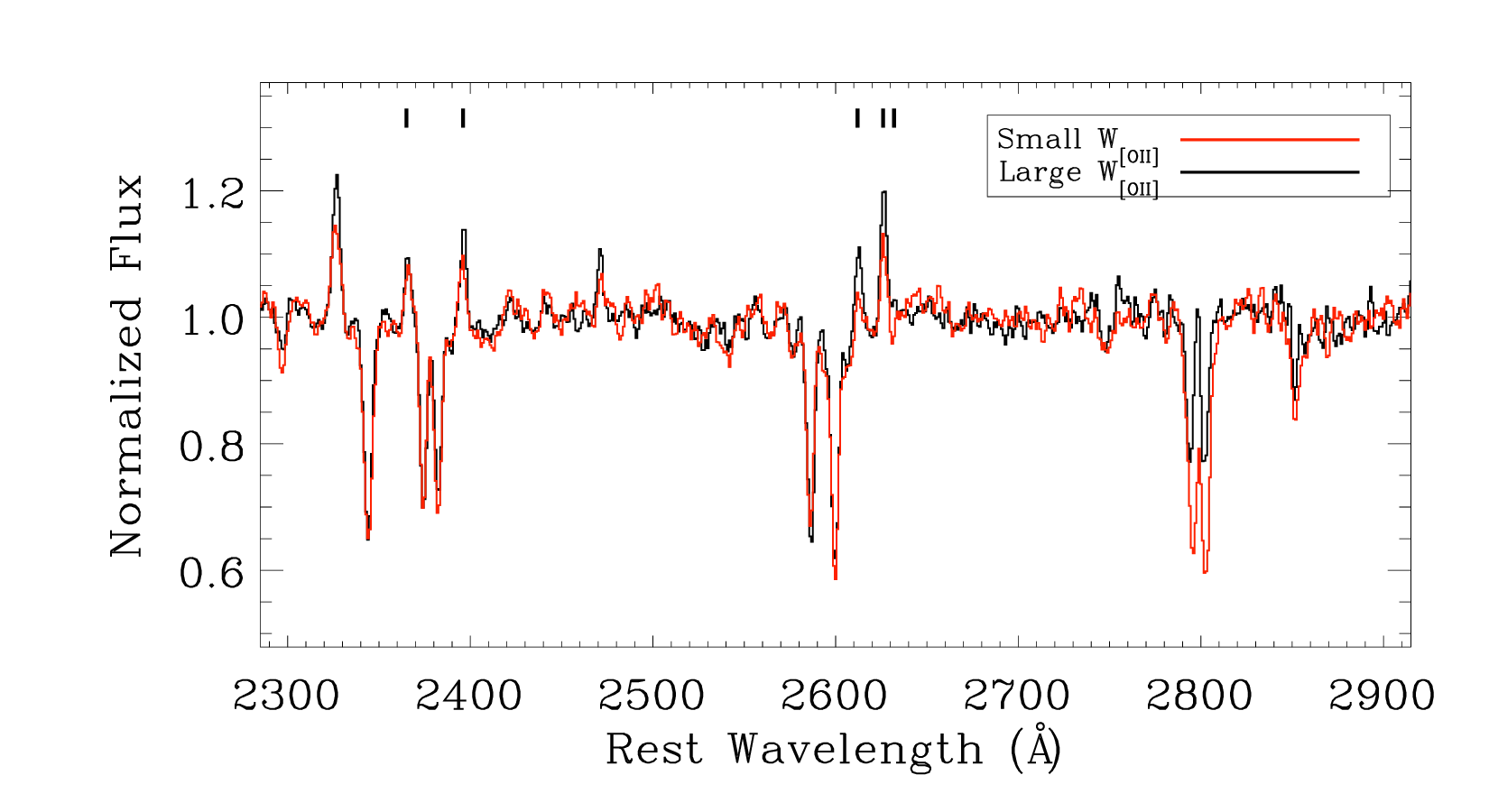} 
\includegraphics[width=3.5in]{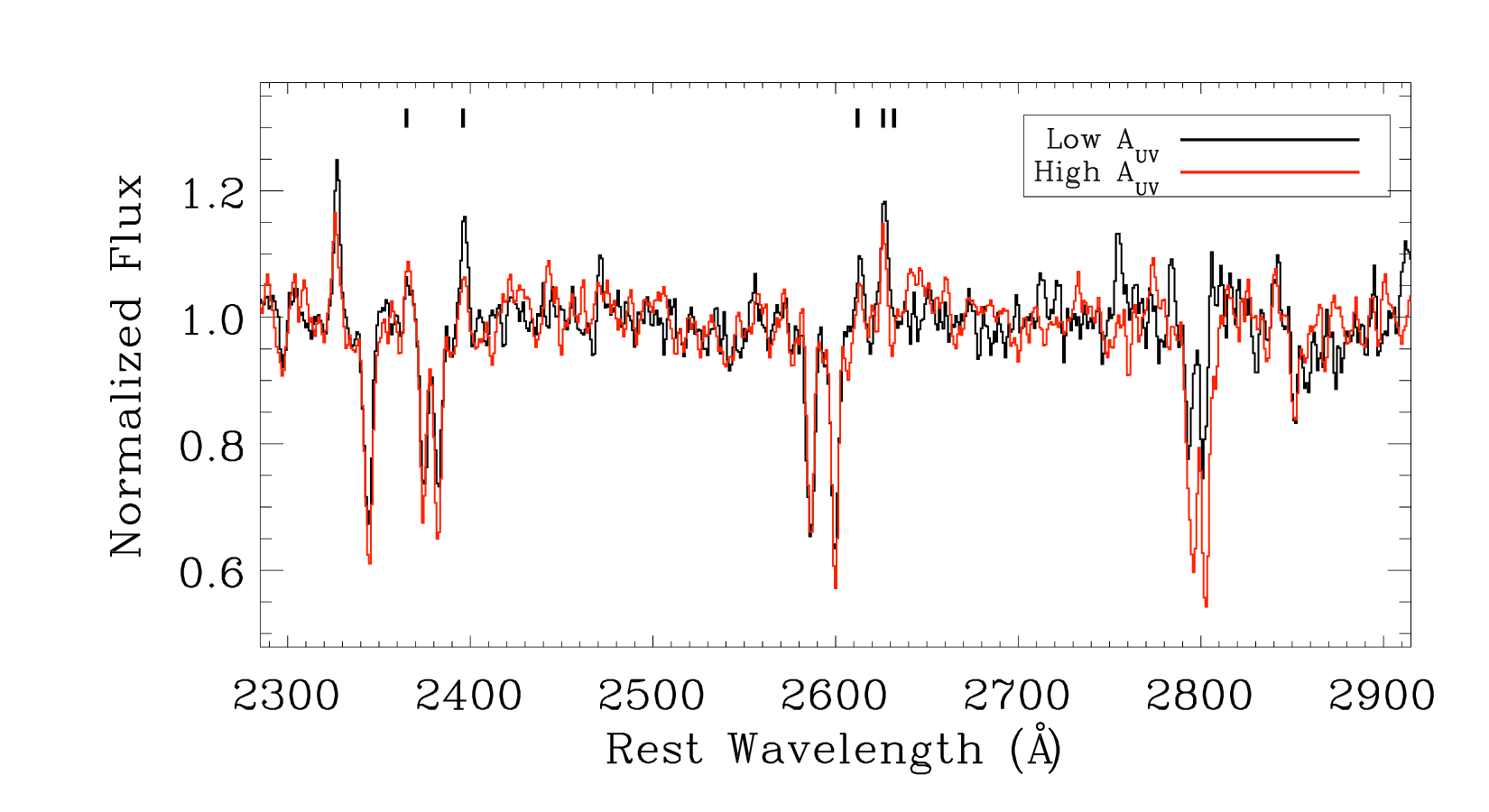} \\ 
\includegraphics[width=3.5in]{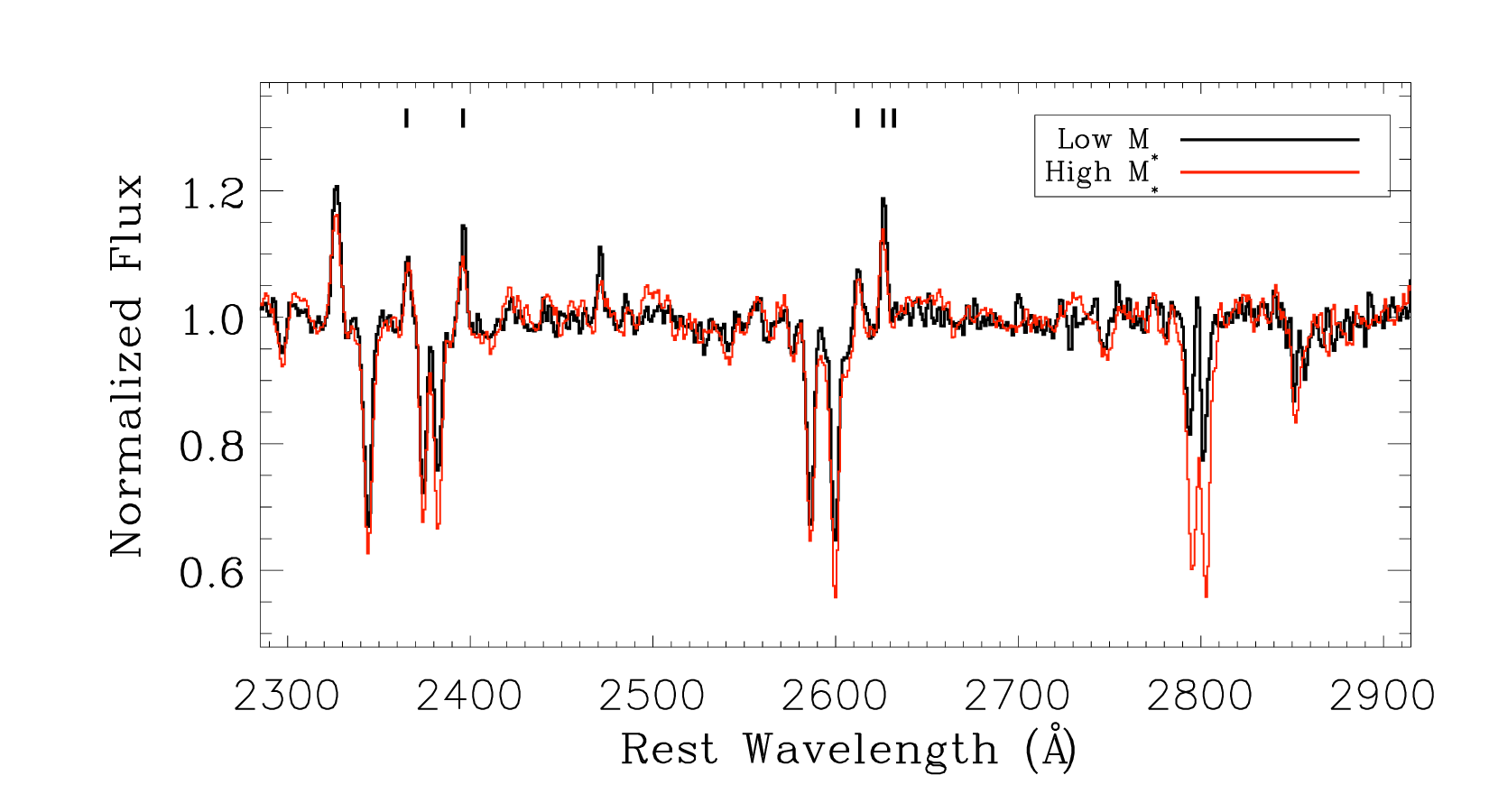} 
\end{array}$
\end{center}
\caption{Composite spectra of key galaxy parameters that strongly modulate Mg~II emission strength. Spectra are shown assembled according to sSFR, $\sigma_{\rm [OII]}$, $W_{\rm [OII]}$, $A_{\rm UV}$, and $M_*$. In each case, the composite spectrum with stronger Mg~II emission is plotted in black.}
\label{sSFR_smooth_onewindow}
\end{figure*}

Five of the above properties -- sSFR, $\sigma_{\rm [OII]}$, $W_{\rm [OII]}$, $A_{\rm UV}$, and $M_*$ -- describe the stellar and H~II region properties of galaxies and we focus on these five properties for the following analyses. In Figure \ref{sSFR_smooth_onewindow}, we show composite spectra assembled according to these five properties. In each case, stronger Mg~II emission is accompanied by stronger Fe~II$^*$ emission (unsurprising since stronger Fe~II$^*$ emission is found to be statistically correlated with stronger Mg~II emission). We show in Figure \ref{MgIIemission_stellarmass} individual measurements of Mg~II emission versus measurements of sSFR, $\sigma_{\rm [OII]}$, $W_{\rm [OII]}$, $A_{\rm UV}$, and $M_*$, respectively. The correlation significances and Spearman rank-order correlation coefficients are as follows: [3.0$\sigma$, 0.50], [3.6$\sigma$, --0.30], [4.4$\sigma$, 0.38], [1.2$\sigma$, --0.21], and [3.5$\sigma$, --0.30]. The strongest correlation (4.4$\sigma$) is observed between Mg~II emission strength and $W_{\rm [OII]}$. As the sample sizes of objects with sSFR and $A_{\rm UV}$ information are roughly one fourth those of the samples with $\sigma_{\rm [OII]}$, $W_{\rm [OII]}$, or $M_*$ measurements, is important to remember that the correlation significances of Mg~II emission with either sSFR or $A_{\rm UV}$ will necessarily be lower simply due to smaller number statistics. We re-calculated the correlation significances of Mg~II emission strength and $\sigma_{\rm [OII]}$, $W_{\rm [OII]}$, and $M_*$, respectively, including only objects also having sSFR and $A_{\rm UV}$ measurements. The correlation significances and Spearman rank-order correlation coefficients are as follows: [2.0$\sigma$, --0.34], [2.3$\sigma$, 0.42], and [1.9$\sigma$, --0.32]. We find that the strongest correlation, when equal sample sizes are compared, is between Mg~II emission strength and sSFR (3.0$\sigma$). In Figure \ref{MgIIhistograms}, we show histograms of sSFR, $\sigma_{\rm [OII]}$, $W_{\rm [OII]}$, $A_{\rm UV}$, and $M_*$, highlighting Mg~II emitters and non-emitters. A clear distinction between Mg~II emitters and non-emitters is observed in $\sigma_{\rm [OII]}$, $W_{\rm [OII]}$, and $M_*$ parameter space, but the smaller sample size of objects with sSFR and $A_{\rm UV}$ information makes it difficult to definitively discern a difference in the sSFR and $A_{\rm UV}$ properties of Mg~II emitters and non-emitters. In Figure \ref{MgIIintercorrelations}, we show the intercorrelations of sSFR, $\sigma_{\rm [OII]}$, $W_{\rm [OII]}$, $A_{\rm UV}$, and $M_*$. As in the case of SFR, $A_{\rm UV}$, and $W_{\rm [OII]}$ modulating Fe~II$^*$ emission strength, all the significant correlations are in consistent senses. In other words, there is overlap between objects with high sSFR, low $\sigma_{\rm [OII]}$, high $W_{\rm [OII]}$, low $A_{\rm UV}$, and low $M_*$ such that a single property could be modulating these intercorrelations  and the trends of sSFR, $\sigma_{\rm [OII]}$, $W_{\rm [OII]}$, $A_{\rm UV}$, and $M_*$ with Mg~II emission strength. In the next section, we show that sSFR appears to drive variations in Mg~II emission. We note that the correlation observed between sSFR and Mg~II emission cannot arise simply as a result of the methods used for estimating these two quantities. As sSFR is defined as the dust-corrected SFR normalized by stellar mass, it turns out to be weakly correlated with the observed UV-continuum luminosity density. Mg~II emission strength is based on the Mg~II emission-line flux divided by the UV-continuum luminosity density. Accordingly, for fixed Mg~II emission-line flux, one expects lower Mg~II emission strengths for objects with higher sSFR (contrary to what we observe). 

While we have focused here on the relationships between the stellar and H~II region properties of galaxies and the strength of Mg~II emission, it is also important to understand how the strength of interstellar absorption -- in this case, parameterized by $W_{\rm FeII}$ -- is correlated with the strength of interstellar emission. Other authors have studied how the strengths of resonant emission and absorption lines are linked; we continue this investigation here motivated by the fact that our data include several strong resonant lines of both Fe~II and Mg~II. In a sample of Lyman break galaxies at \z3, \citet{shapley2003} noted that objects showing weaker \lya emission had larger interstellar Si~II, C~II, Fe~II, and Al~II absorption equivalent widths while objects marked by stronger \lya emission had correspondingly weaker interstellar absorption lines. These authors attributed these trends to different covering fractions of dusty clouds, where galaxies with higher cloud covering fractions typically suffer more extinction (reducing the \lya emission strength and increasing the interstellar absorption equivalent width). In a lower-redshift sample ($\langle z \rangle$ = 2.3), \citet{erb2006} found that higher-mass galaxies showed stronger interstellar absorption lines and weaker \lya emission, consistent with the results from \citet{shapley2003}. \citet{martin2012} furthermore observed that higher-mass galaxies exhibit characteristically stronger Mg~II absorption lines. We also find here -- in the same data set used by \citet{martin2012} -- that objects with stronger Mg~II emission show weaker Fe~II absorption (Figure \ref{MgII_composites}). Additionally, we recover a statistically significant D$_{\rm MgII}$ value (0.40 $\pm$ 0.11 \AA) when we divide objects on the basis of $W_{\rm FeII}$ and measure their Mg~II emission strengths; objects with larger Fe~II absorption equivalent widths have weaker Mg~II emission. These findings support the results of both \citet{shapley2003} and \citet{erb2006} and suggest that Mg~II emission is stronger when the covering fraction of interstellar gas is lower.

\section{Discussion} \label{sec: discussion}

We have shown that fine-structure Fe~II$^*$ and resonant Mg~II emission are characteristic of star-forming galaxies at \z1. In this section, we build on our previous analyses of the correlations of Fe~II$^*$ and Mg~II emission strength with galaxy properties. We propose physical explanations for these observed correlations and suggest that galaxies with strong Fe~II$^*$ or Mg~II emission have typically lower $A_{\rm UV}$, higher sSFR, and lower $M_*$ than the galaxy population as a whole. We conclude this section with a discussion of the striking absence of Fe~II$^*$ emission in local galaxies and argue that slit losses may be largely responsible for the lack of Fe~II$^*$ emission in nearby objects.

\subsection{Fe~II$^*$ Emission is Modulated by Dust} \label{sec: modulatedbydust}

Several authors have proposed explanations for the variety of Fe~II$^*$ emission strengths observed in star-forming galaxies, including slit losses \citep{giavalisco2011,erb2012}, viewing angle effects of observing a non-spherical wind \citep[e.g.,][]{erb2012}, and dust attenuation \citep{prochaska2011}. We address each of these explanations in turn and examine how our results -- that systems at higher redshifts and those with lower SFRs, lower $A_{\rm UV}$ values, and larger $W_{\rm [OII]}$ measurements show stronger Fe~II$^*$ emission -- support or do not support these hypotheses. 

As discussed later in this section, the striking absence of Fe~II$^*$ emission in local galaxies has been attributed to slit losses \citep{giavalisco2011,erb2012}. \citet{erb2012} measured the spatial extent of Fe~II$^*$ emission in a sample of star-forming galaxies at 1 $\lesssim$ \emph{z} $\lesssim$ 2 and found that Fe~II$^*$ emission may be more spatially extended than the continuum, consistent with the hypothesis that Fe~II$^*$ emission may be missed by narrow spectroscopic slits. In Figure \ref{angular_size}, we show that dividing objects on the basis of angular size (i.e., simulating the effects of slit losses) yields a difference in Fe~II$^*$ strength at the 1.8$\sigma$ level such that smaller objects show stronger Fe~II$^*$ emission. However, the trend of Fe~II$^*$ emission strength with angular size is less significant than the trends of Fe~II$^*$ emission strength with SFR, $A_{\rm UV}$, $W_{\rm [OII]}$, and \emph{z}. We accordingly conclude that slit losses may be partially responsible for the variety of Fe~II$^*$ emission strengths observed in our sample, but that other galaxy properties play a larger role in the regulation of Fe~II$^*$ emission. It is important to emphasize that the effects of slit losses will be more pronounced over a larger redshift baseline (i.e., between local samples and \z1) than is spanned by our current data set.

Fe~II$^*$ emission can also be modulated by viewing angle effects of observing a non-spherical galactic wind. A biconical outflow will show variations in emission strength depending on the viewing geometry of the observer with respect to the wind \citep{erb2012}. For a wind arising perpendicular to a galaxy disk, these authors propose that observations of the wind face-on will yield stronger absorption signatures and weaker emission measurements while the wind viewed edge-on will predominantly show emission. The model predictions make sense given that a wind seen face-on means that the observer is looking down the barrel and therefore seeing material absorbed against the background light of the host galaxy. Conversely, observations of a wind edge-on see the wind projected 90$^{\circ}$ to the line of sight and accordingly observe more scattered emission as opposed to absorption backlit by starlight. 

In our sample, we find that objects showing stronger Fe~II$^*$ emission also show more blueshifted Mg~II absorption (Figure \ref{FeII_smooth}). This result is contrary to the model presented by \citet{erb2012}, in which stronger emission lines would be more prevalently seen in edge-on systems not expected to show large blueshifts in their interstellar absorption lines. As \citet{martin2012} estimated that the geometry of galactic winds at \z1 is roughly biconical with a wind opening angle of $\sim$40$^{\circ}$, disk inclination and the observability of interstellar blueshifts should be correlated. Disk inclination estimates are available for 46 objects in our sample and we assemble composite spectra from samples of both high- and low-inclination objects, where we divide the sample at \emph{i} = 45$^{\circ}$. The average inclination of the two samples are $\langle i \rangle$ = 58$^{\circ}$ and 38$^{\circ}$, respectively. We find no significant change in Fe~II$^*$ emission strength in the samples divided on the basis of disk inclination, although our sample is small. While we do not find a trend between disk inclination and Fe~II$^*$ emission strength, we do find a trend at the $\sim$3.6$\sigma$ level between disk inclination and Mg~II emission strength such that systems with low disk inclinations (i.e., more face on) show stronger Mg~II emission. The $\sim$10$^{\circ}$ uncertainties on our inclinations -- where inclination was estimated from rest-frame ultraviolet imaging sensitive only to current episodes of star formation -- make it difficult to construct subsamples that are precisely divided according to viewing angle. 

The absorption of photons by dust additionally modulates Fe~II$^*$ emission. \citet{prochaska2011} modeled Fe~II$^*$ emission lines arising from galactic winds and found that increasing dust attenuation suppressed Fe~II$^*$ emission (although resonant Fe~II absorption was minimally affected by changes in attenuation). These authors proposed that emission-line fluxes are reduced by a factor of order (1 + $\tau_{\rm dust}$), where $\tau_{\rm dust}$ = $A_{\rm V}$/1.086 $\sim$ $A_{\rm UV}$/1.9 assuming a \citet{calzetti2000} dust attenuation law. The 54 objects in our sample with dust attenuation measurements have $A_{\rm UV}$ values ranging from 0.02--4.55, with a median $A_{\rm UV}$ of 1.8 and a corresponding median $\tau_{\rm dust}$ of $\sim$0.9. When we divide our data on the basis of $A_{\rm UV}$, we find that objects with stronger dust attenuation show $\sim$40\% weaker Fe~II$^*$ emission (Figure \ref{AUV_smooth_FeII}), in rough agreement with predictions by \citet{prochaska2011}. The other three galaxy properties that are significantly correlated with Fe~II$^*$ emission strength -- SFR, $W_{\rm [OII]}$, and \emph{z} -- are themselves correlated with $A_{\rm UV}$ (Figure \ref{sfr_z}). As galaxies at larger redshifts with lower SFRs and larger $W_{\rm [OII]}$ measurements have lower $A_{\rm UV}$ values, our results are consistent with a single parameter -- $A_{\rm UV}$ -- being the primary driver of Fe~II$^*$ emission strength. 

\begin{deluxetable*}{llll}
\tablewidth{0pt}
\tablecaption{Summary of Composite Spectra}
\tablehead{
\multicolumn{1}{c}{Name} 
& \multicolumn{1}{c}{Number of Objects} 
& \multicolumn{1}{c}{Fraction of Global Sample}
& \multicolumn{1}{c}{Figure} \\
\colhead{}
& \multicolumn{1}{c}{}
& \multicolumn{1}{c}{}
& \multicolumn{1}{c}{}
}
\startdata
Global sample & 212 & 100\% & \ref{composite_blue_smooth}, \ref{FeII_compare} \\
Fe~II$^*$ emitters & 13 & 6\% & \ref{FeII_smooth} \\
Fe~II$^*$ non-emitters & 9 & 4\% & \ref{FeII_smooth} \\
Mg~II emitters & 22 & 10\% & \ref{IDLzoomMgIIem}, \ref{MgII_composites} \\
Mg~II non-emitters & 34 & 16\% & \ref{MgII_composites} \\
SFR\tablenotemark{$\dagger$} & 54 & 25\% & \ref{SFR_smooth_FeII} \\
sSFR\tablenotemark{$\dagger$}  &  54 & 25\% & \ref{sSFR_smooth_onewindow} \\
$\Sigma_{\rm SFR}$(R$_{\rm P}$)\tablenotemark{$\dagger$} & 40 & 19\% & \nodata \\
$\sigma_{\rm [OII]}$\tablenotemark{$\dagger$} & 206 & 97\% & \ref{sSFR_smooth_onewindow} \\
$W_{\rm [OII]}$\tablenotemark{$\dagger$} & 189 & 89\% & \ref{SFR_smooth_FeII}, \ref{sSFR_smooth_onewindow} \\
Inflows at 1$\sigma$ & 33 &16\% & \nodata \\
Outflows at 1$\sigma$ & 84 &40\% & \nodata \\
Inflows at 3$\sigma$ & 11 &5\% & \nodata \\ 
Outflows at 3$\sigma$ & 35 & 17\% & \nodata \\
Mg~II emission\tablenotemark{$\dagger$} & 165 & 78\% & \nodata \\
$W_{\rm FeII}$\tablenotemark{$\dagger$} & 203 & 96\% & \nodata \\
$A_{\rm UV}$\tablenotemark{$\dagger$} & 54 & 25\% & \ref{AUV_smooth_FeII}, \ref{SFR_smooth_FeII}, \ref{sSFR_smooth_onewindow} \\
$M_B$\tablenotemark{$\dagger$} & 212 & 100\% & \nodata \\
$U-B$\tablenotemark{$\dagger$} & 212 & 100\% & \nodata \\
$M_*$\tablenotemark{$\dagger$} & 212 & 100\% & \ref{sSFR_smooth_onewindow} \\
Angular size\tablenotemark{$\dagger$} & 51 & 24\% & \ref{angular_size} \\
Physical size\tablenotemark{$\dagger$} & 51 & 24\% & \nodata \\
Inclination\tablenotemark{$\dagger$}  & 54 & 25\% & \nodata \\
\emph{G}\tablenotemark{$\dagger$}  & 53 & 25\% & \nodata \\
\emph{z}\tablenotemark{$\dagger$}  & 212 & 100\% & \ref{SFR_smooth_FeII} \\
\enddata
\tablenotetext{$\dagger$}{The sample for this parameter was divided in half at the median value in order to produce the composite spectra.}
\label{sumtable}
\end{deluxetable*}

In order to further investigate if $A_{\rm UV}$ is most strongly driving the variation in Fe~II$^*$ emission, we assembled composite spectra holding $A_{\rm UV}$ constant and varying SFR, $W_{\rm [OII]}$, and \emph{z} respectively. If $A_{\rm UV}$ is indeed primarily responsible for modulating Fe~II$^*$ emission, then we expect these spectra to show weaker changes in Fe~II$^*$ emission compared with composite spectra holding SFR, $W_{\rm [OII]}$, and \emph{z} constant, respectively, and varying $A_{\rm UV}$. We do find that the spectra holding $A_{\rm UV}$ constant show weaker changes in Fe~II$^*$ emission strength (D$_{\rm FeII^*}$ = 0.12--0.28 \AA) than the spectra holding SFR, $W_{\rm [OII]}$, or \emph{z} constant and varying $A_{\rm UV}$ (D$_{\rm FeII^*}$ = 0.42--0.72 \AA). These results are consistent with $A_{\rm UV}$ being the primary driver of Fe~II$^*$ emission variation. 

As Fe~II$^*$ emission is thought to both originate in galaxy halos and also be strongly modulated by dust attenuation, it is important understand the spatial distribution of dust in galaxy halos. In Section \ref{sec: kinematics}, we showed that our kinematic measurements of Fe~II$^*$ are consistent with Fe~II$^*$ emission arising in either stationary H~II regions or extended galactic winds \citep{prochaska2011}. Given the spatial extent of Fe~II$^*$ emission reported by \citet{erb2012}, we choose to focus on the physical picture of emission presented by \citet{rubin2010c} in which Fe~II$^*$ emission is created by photon scattering at spatial distances of $\sim$kpc from the galactic disk. Assuming that Fe~II$^*$ emission arises at large galactocentric distances, it is surprising that we find a strong correlation between Fe~II$^*$ emission strength and $A_{\rm UV}$. Our measurements of dust attenuation are based on observations of star-forming regions and these values are accordingly indicative of attenuation only near galactic disks. 

Several authors have reported that dust is present at significant distances from galaxy disks. \citet{nelson1998} stacked \emph{Infrared Astronomical Satellite} profiles of local galaxies and found that 100$\mu$m emission tracing dust extended $\sim$20--30 kpc. \citet{holwerda2009} studied a pair of occulting galaxies in which light from the background galaxy at \emph{z} = 0.06 was used to probe the extended halo of the foreground object. These authors found dust extinction in the foreground galaxy at $\sim$6$R_{\rm 50}$, where $R_{\rm 50}$ is the galaxy's effective radius. In a large sample of background quasars and foreground galaxies, \citet{menard2010} inferred the presence of dust on scales from 20 kpc to a few Mpc around galaxies. These authors find that the dust mass in galaxy halos is comparable to the dust mass in galaxy disks. Dust is clearly present at significant distances from galaxies. As the $A_{\rm UV}$ values estimated for our sample are based on observations of star-forming regions, it is remarkable that we observe a strong correlation between $A_{\rm UV}$ and Fe~II$^*$ emission strength. There is no \emph{a priori} reason that the dust attenuation around star-forming regions has to be representative of the dust attenuation in the extended halo, although winds removing gas from galaxies may well entrain dust as well \citep[e.g.,][]{heckman2000}. Future observations of spatially-resolved emission around galaxies will be instrumental for estimating the dust attenuation at large galactocentric distances. 

We have shown that attenuation by dust is a leading candidate to explain the diversity of Fe~II$^*$ emission strengths in star-forming galaxies at \z1. However, it is important to acknowledge that our estimates of both galaxy angular size and disk inclination (i.e., quantities used to test the alternate hypotheses of slit losses and viewing angle effects, respectively) may be uncertain given the rest-frame ultraviolet data from which these values were calculated. As ultraviolet emission traces only high-mass star formation, galaxies typically appear more clumpy in ultraviolet passbands although this effect is strongest at \emph{z} $\gtrsim$ 2 \citep[e.g.,][]{law2007}. Future analyses of angular size and disk inclination using longer wavelength data, paired with a larger sample size of objects with $A_{\rm UV}$ measurements, will be nonetheless important in verifying the role of dust and other properties in modulating Fe~II$^*$ emission at \z1. 

\begin{figure*}
\begin{center}$
\begin{array}{c}
\includegraphics[width=3in]{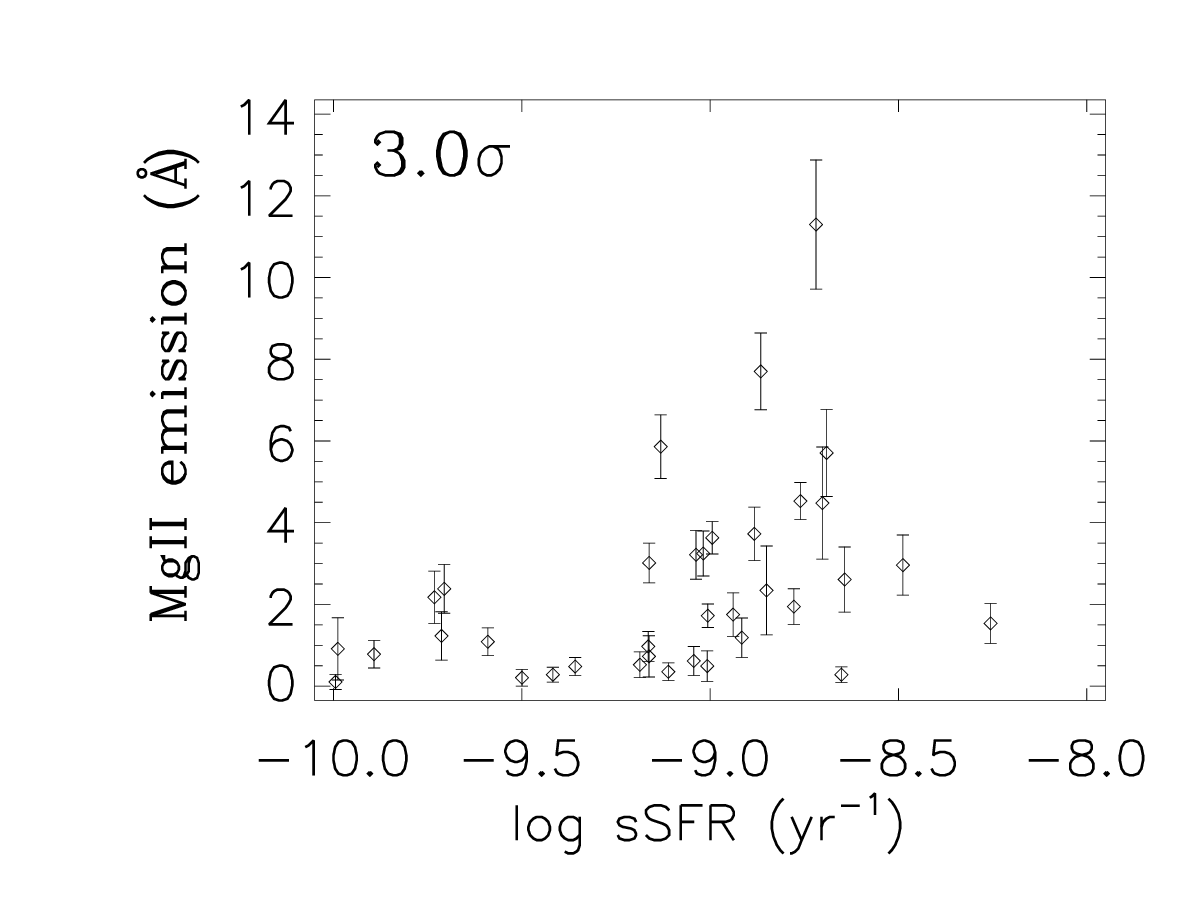} 
\includegraphics[width=3in]{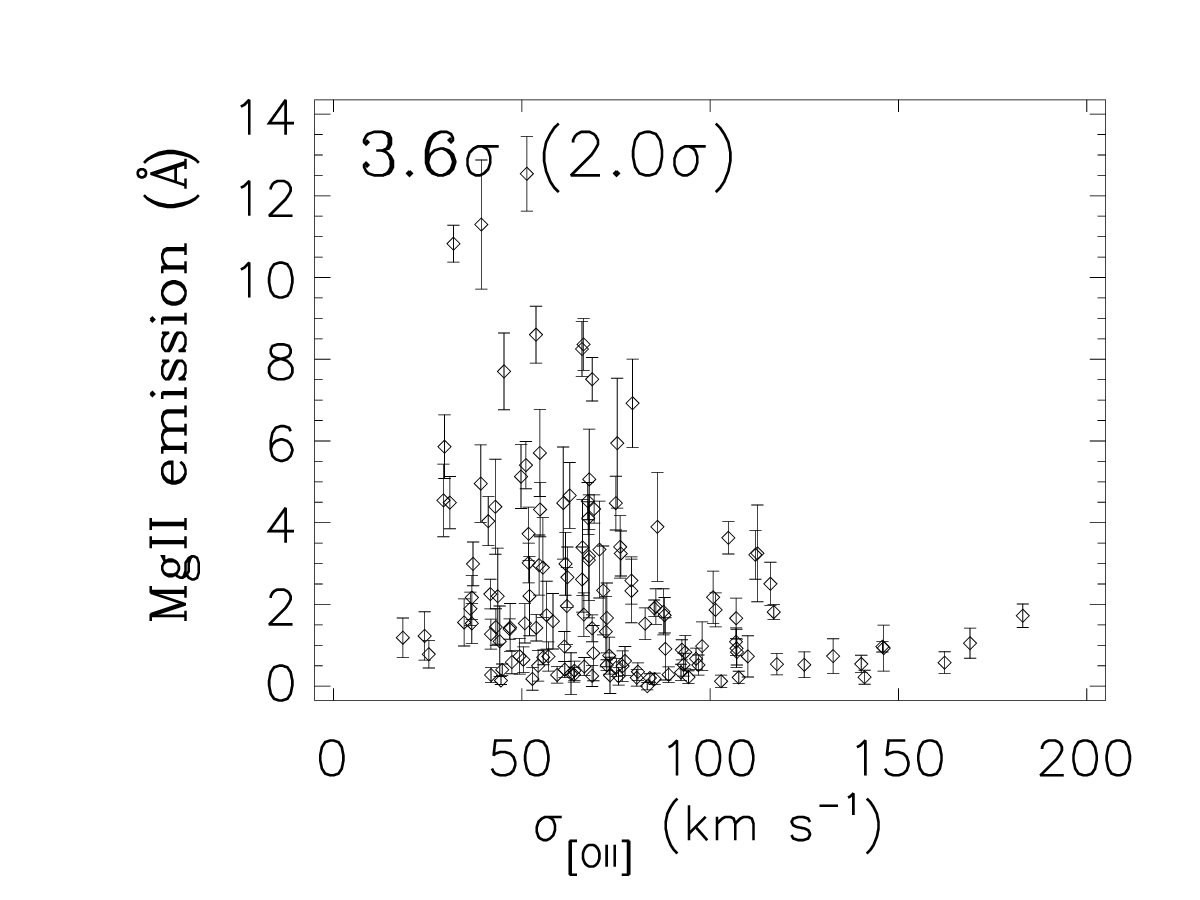} \\ 
\includegraphics[width=3in]{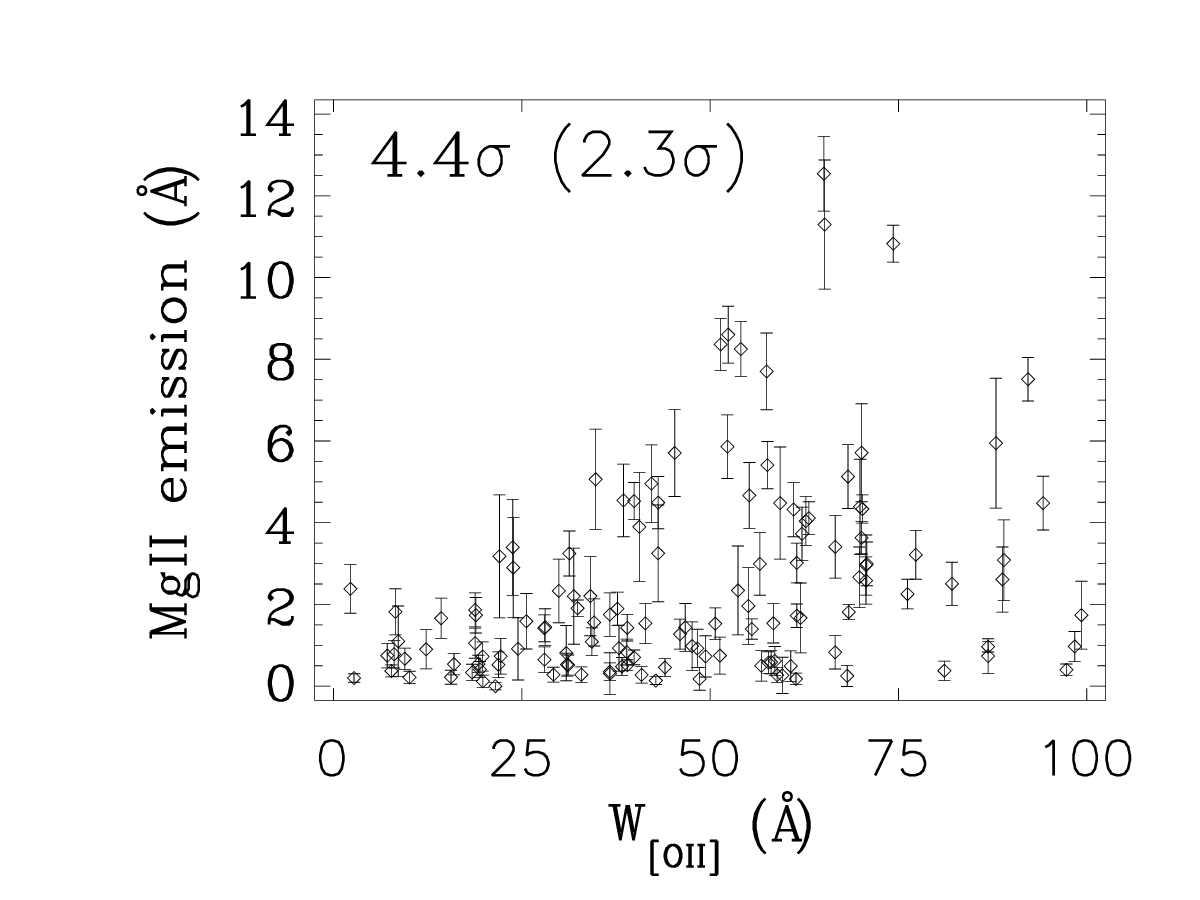} 
\includegraphics[width=3in]{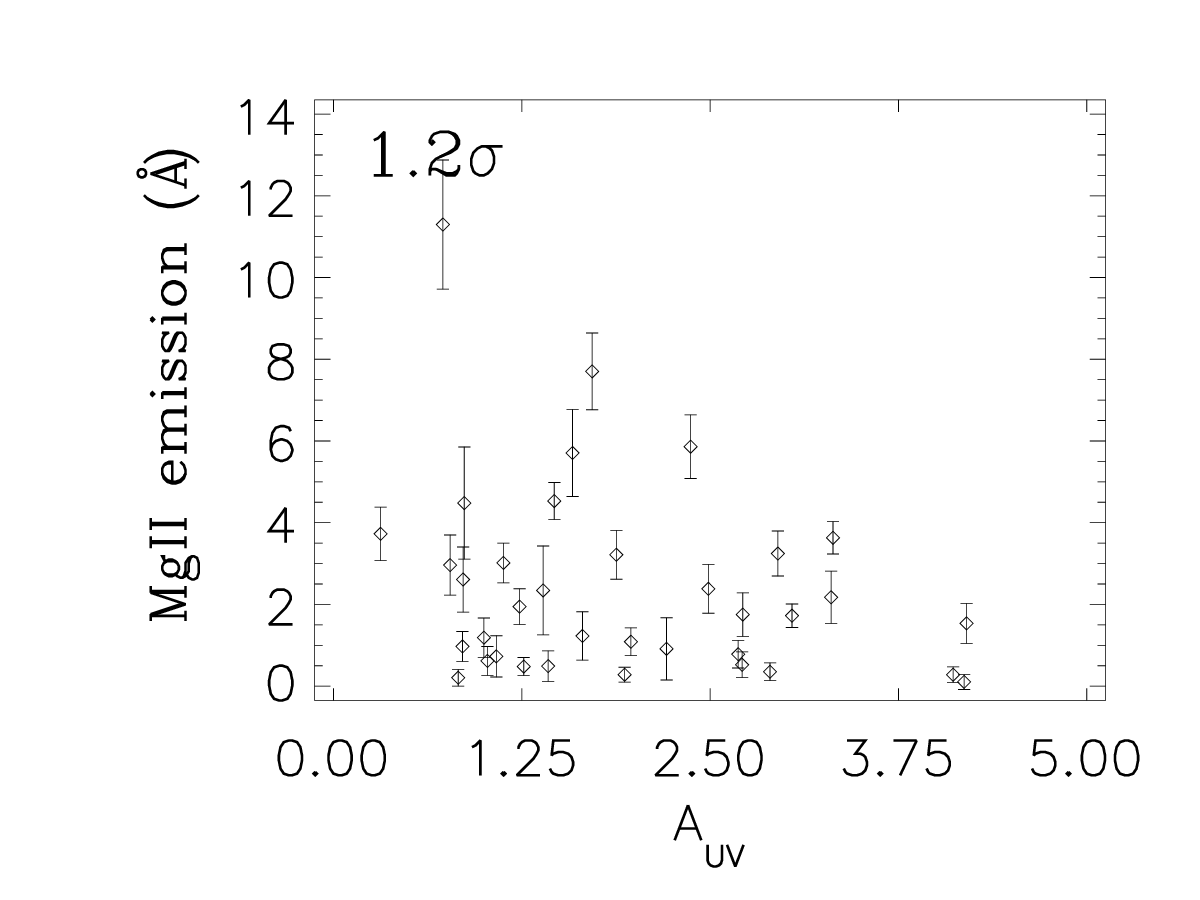} \\ 
\includegraphics[width=3in]{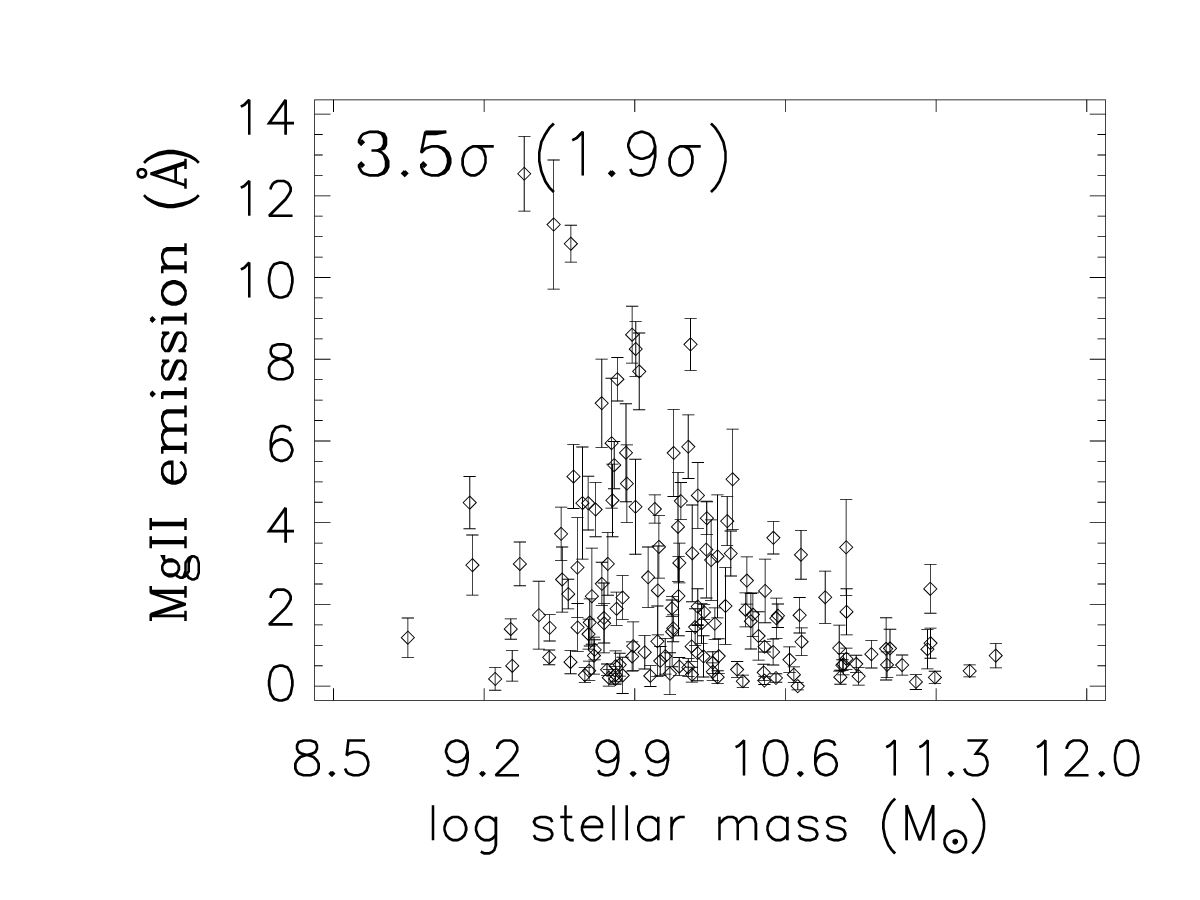} 
\end{array}$
\end{center}
\caption{Mg~II emission strength versus sSFR, $\sigma_{\rm [OII]}$, $W_{\rm [OII]}$, $A_{\rm UV}$, and $M_*$, respectively. Strong correlations ($\ge$ 3$\sigma$) are found between Mg~II emission strength and sSFR, $\sigma_{\rm [OII]}$, $W_{\rm [OII]}$, and $M_*$, such that objects with pronounced Mg~II emission tend to have larger sSFR, lower $\sigma_{\rm [OII]}$, larger $W_{\rm [OII]}$, and lower $M_*$. As fewer objects have sSFR and $A_{\rm UV}$ measurements, we also calculated correlation significances inclusive only of objects with these measurements. The results of these analyses are shown in parentheses; the strongest correlation, adjusted for sample size, is between Mg~II emission strength and sSFR. Four objects with high sSFRs and low $A_{\rm UV}$ values appear to drive the correlations between Mg~II emission and sSFR and $A_{\rm UV}$, respectively; these objects are otherwise unremarkable in both their spectra and images.}
\label{MgIIemission_stellarmass}
\end{figure*}

\subsection{Galaxy Properties Correlated with Mg~II Emission Strength}

The spectra in our sample exhibit a range of Mg~II profiles, with some objects showing only absorption and others presenting strong emission peaks in both the 2796 and 2803 \AA\ lines. Approximately 15\% of objects show robust Mg~II emission, defined as a combined 2796/2803 \AA\ emission significance of at least 6$\sigma$ above the scaled Fe~II absorption profile (Section \ref{sec: mgiiandproperties}). From the color-magnitude and color-mass diagrams in Figure \ref{CMD_MgII}, we conclude that Mg~II  emission is prevalent in bluer systems with lower $M_*$, consistent with the results of \citet{weiner2009}, \citet{erb2012}, and \citet{martin2012}. We do not find, however, that objects separate in luminosity space based on the presence or absence of Mg~II emission, as \citet{weiner2009} do. We acknowledge that our sample is substantially smaller than the \citet{weiner2009} data set and that we furthermore select objects showing Mg~II emission not above the continuum -- like \citet{weiner2009} do -- but rather above the scaled Fe~II absorption profile. 

We also examined the strength of Mg~II emission in a suite of composite spectra assembled according to 18 different galaxy properties. We find that Mg~II emission is stronger in objects characterized with lower $A_{\rm UV}$, lower $M_*$, higher sSFR, lower $\sigma_{\rm [OII]}$, and larger $W_{\rm [OII]}$. Mg~II emission is also stronger in galaxies with 1$\sigma$ outflows (as opposed to inflows), stronger Fe~II$^*$ emission, smaller $W_{\rm FeII}$, and smaller disk inclinations, although these four properties are not direct tracers of the stellar or H~II region properties of galaxies (Section \ref{sec: mgiiandproperties}). Considering the ensemble of five galaxy properties reflective of stellar environments, we assembled a variety of composite spectra  in order to determine if one property in particular was responsible for most of the variation in Mg~II emission. We constructed 40 spectra in total, five groups of eight composite spectra each. The groups consisted of composite spectra holding one property constant and modulating the remaining four properties (4 properties $\times$ a binary division of each = 8 composite spectra). In each of these 40 composite spectra, we systematically measured the strength of Mg~II emission and then calculated D$_{\rm MgII}$ for each pair of spectra (Section \ref{sec: mgiiandproperties}). We find that when $A_{\rm UV}$, $M_*$, $\sigma_{\rm [OII]}$, and $W_{\rm [OII]}$ are held constant, the largest D$_{\rm MgII}$ values are observed for the pairs of spectra divided by sSFR. In other words, sSFR appears to more strongly modulate Mg~II emission than $A_{\rm UV}$, $M_*$, $\sigma_{\rm [OII]}$, or $W_{\rm [OII]}$. Additional evidence that sSFR may drive the variation in Mg~II emission strength comes from the correlation significances between Mg~II emission and sSFR, $\sigma_{\rm [OII]}$, $W_{\rm [OII]}$, $A_{\rm UV}$, and $M_*$ shown in Figure \ref{MgIIemission_stellarmass}. When the sSFR, $\sigma_{\rm [OII]}$, $W_{\rm [OII]}$, $A_{\rm UV}$, and $M_*$ samples are normalized to a common size, we find the strongest correlation between Mg~II and sSFR (3.0$\sigma$). 

For completeness, we also investigated other galaxy properties besides sSFR that may be additionally modulating Mg~II emission strength. Based on the measurements of the composite spectra presented in Figure \ref{MgIIemitters_figure}, we find that the galaxy properties showing the strongest absolute variation in Mg~II emission strength are $M_*$ and Fe~II$^*$ emission strength. Each of these two properties are correlated with Mg~II emission at $>$ 6$\sigma$, while the remaining seven properties, including sSFR, are correlated with Mg~II emission at $\lesssim$ 5$\sigma$. Given that Fe~II$^*$ emission is likely modulated by $A_{\rm UV}$ (Section \ref{sec: modulatedbydust}), we investigate here how $M_*$ and $A_{\rm UV}$ are correlated with Mg~II emission strength. In Figure \ref{MgII_AUVM}, we show composite spectra assembled holding $M_*$ ($A_{\rm UV}$) constant and varying $A_{\rm UV}$ ($M_*$). If one galaxy property was responsible for controlling the bulk of Mg~II emission variation, we would expect to find only minimal changes in Mg~II emission strength when that property is held constant. Rather, we find that Mg~II emission strength changes when $M_*$ is held constant and also when $A_{\rm UV}$ is held constant. These results suggest that both $M_*$ and $A_{\rm UV}$ modulate Mg~II emission strength and that neither parameter dominates in controlling Mg~II emission. Even so, the composite spectra divided by $A_{\rm UV}$ show more consistent variation in their Mg~II profiles than the composite spectra divided by $M_*$, suggestive that $A_{\rm UV}$ modulates Mg~II more strongly than $M_*$. \citet{prochaska2011} develop models of galactic winds in which the presence of dust affects the incidence of Mg~II emission, where dustier systems show significantly less emission. These authors note that the Mg~II doublet is particularly susceptible to attenuation by dust since resonant photons scatter and experience longer path lengths -- and therefore more opportunities for encountering a dust grain -- than non-resonant photons, akin to the case of \lya \citep{kornei2010}. 

\begin{figure*}
\begin{center}$
\begin{array}{c}
\includegraphics[width=2.3in]{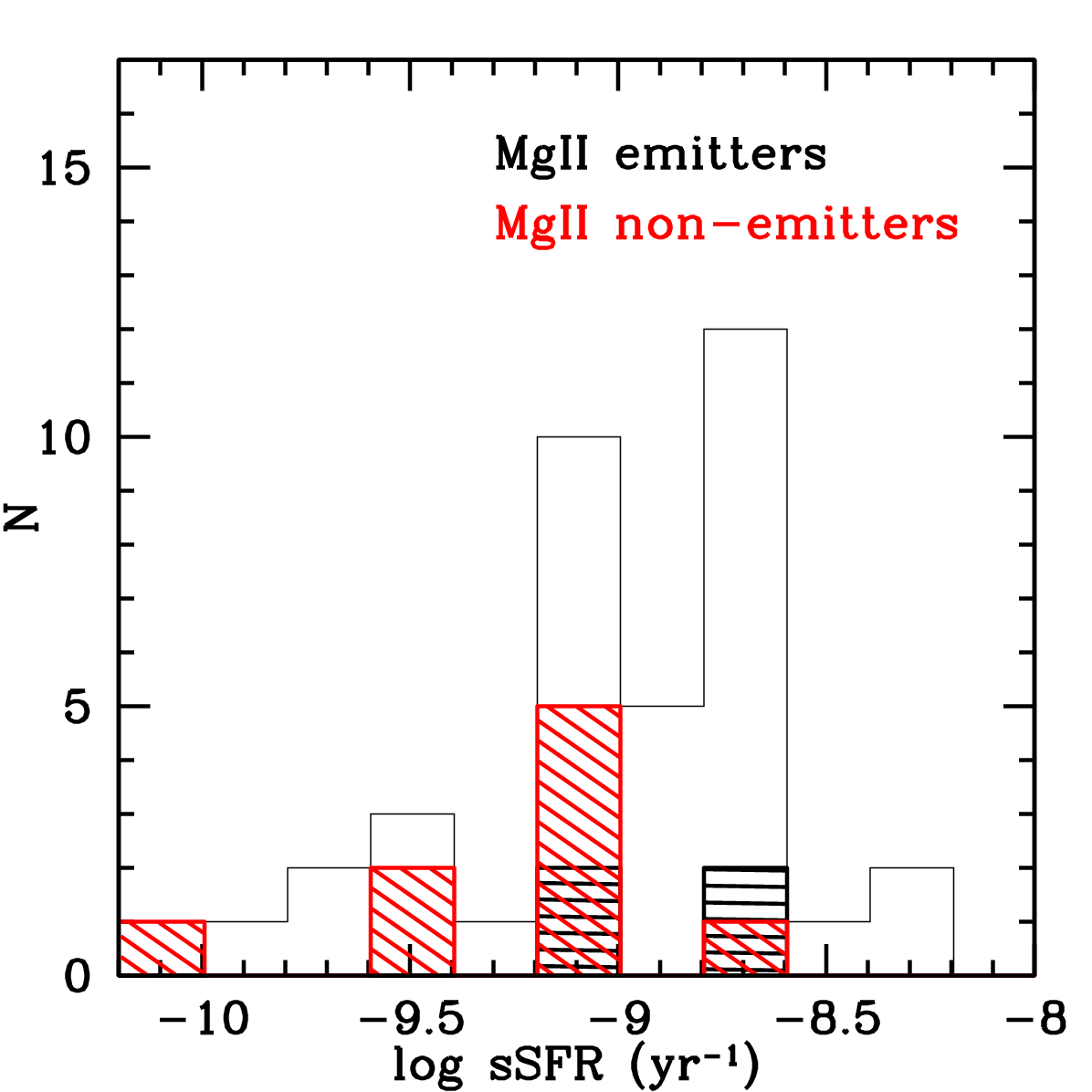} 
\includegraphics[width=2.3in]{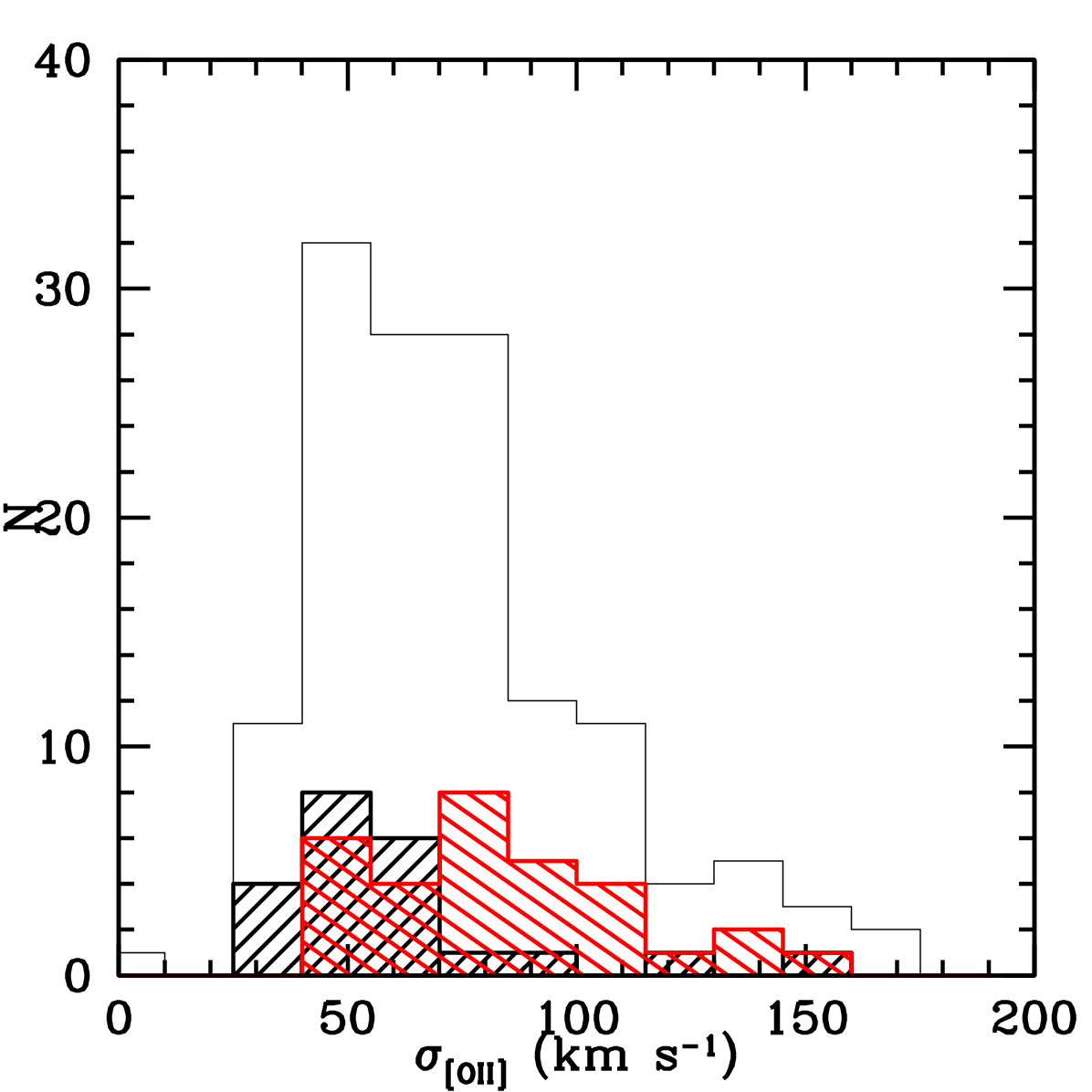} \\ 
\includegraphics[width=2.3in]{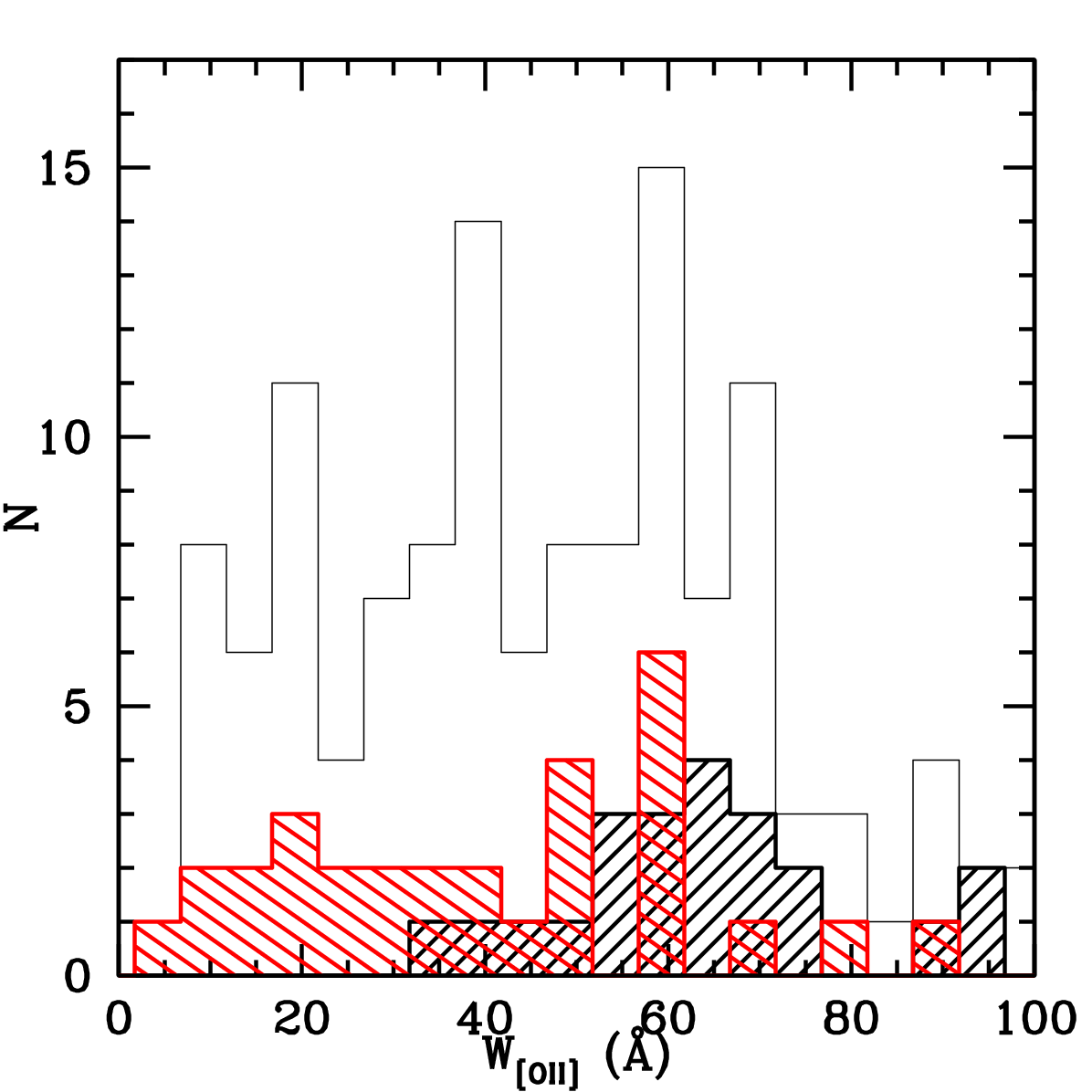} 
\includegraphics[width=2.3in]{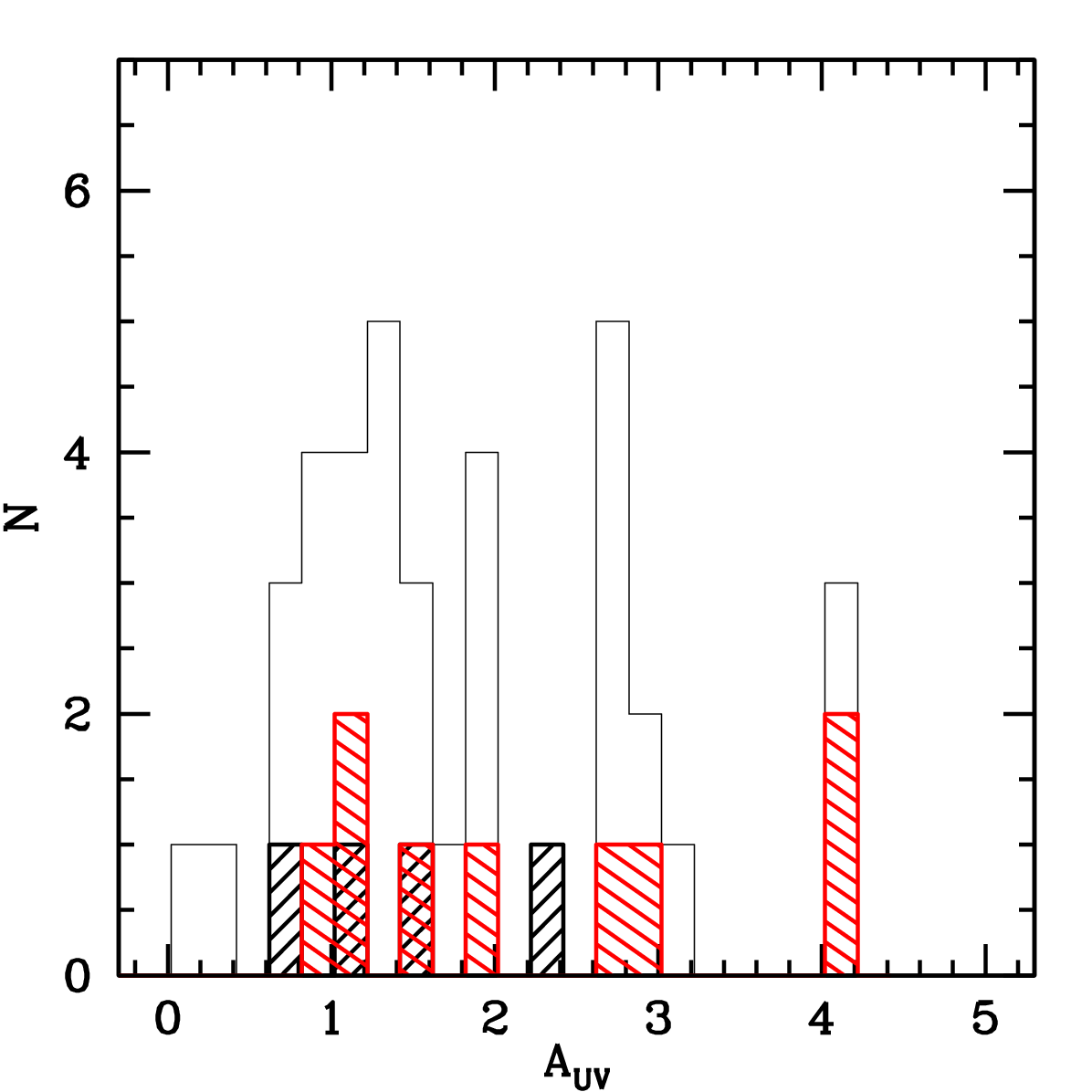} \\ 
\includegraphics[width=2.3in]{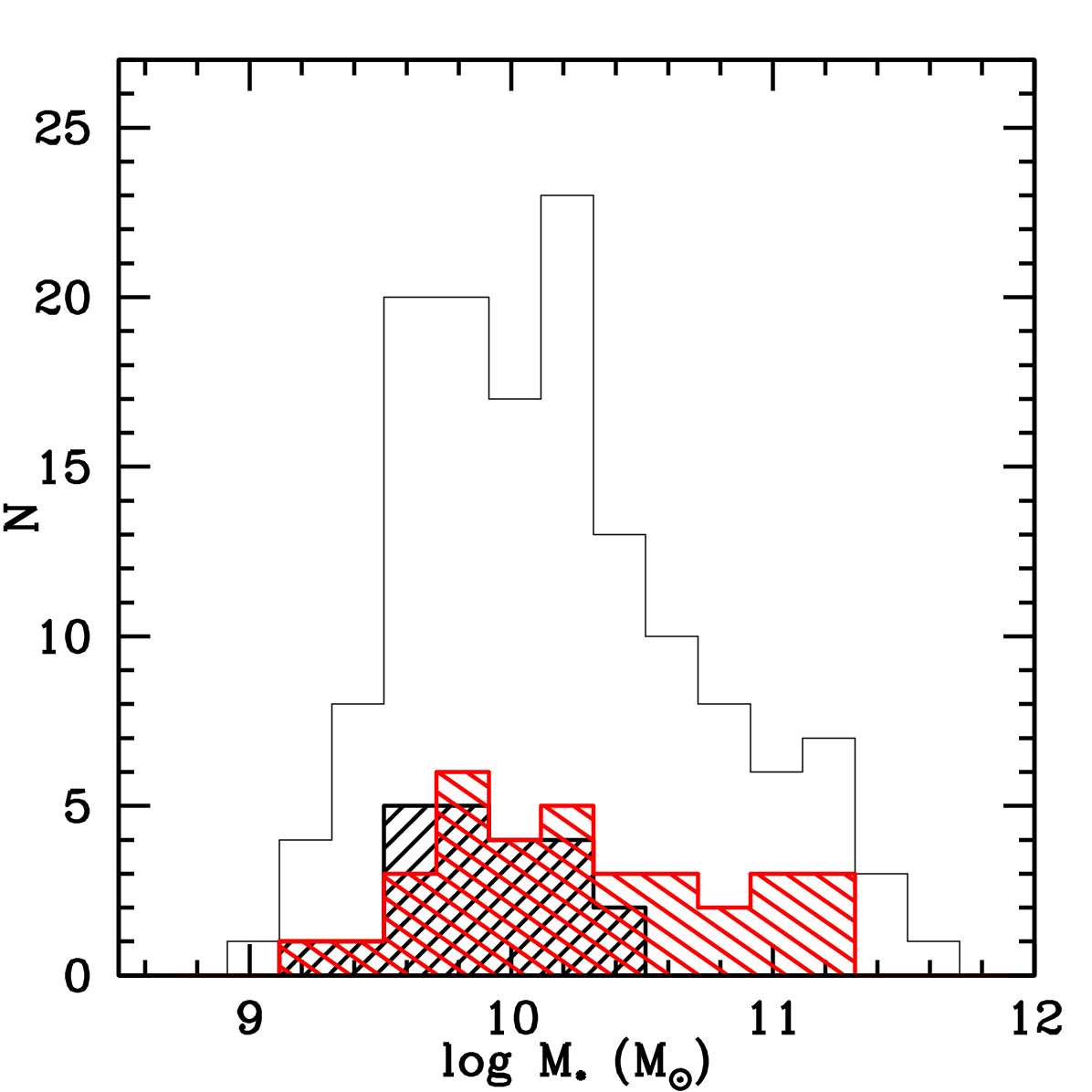} 
\end{array}$
\end{center}
\caption{Histograms of five key parameters modulating Mg~II emission strength: sSFR, $\sigma_{\rm [OII]}$, $W_{\rm [OII]}$, $A_{\rm UV}$, and $M_*$. In each panel, the open histogram shows the parent sample of objects with S/N $>$ 4.66. Mg~II emitters are shown in the black shaded histogram and Mg~II non-emitters are indicated with the red shaded histogram. Mg~II emitters are characterized by higher sSFR, lower $\sigma_{\rm [OII]}$, larger $W_{\rm [OII]}$, lower $A_{\rm UV}$, and lower $M_*$ than Mg~II non-emitters.}
\label{MgIIhistograms}
\end{figure*}

We conclude that a typical Mg~II emitter has some characteristic properties: it has a higher sSFR, a lower $A_{\rm UV}$, and a lower $M_*$ than the sample as a whole. These properties can be attributed to a population of highly star-forming, young galaxies that are still assembling the bulk of their stellar mass. These results suggest that galaxies exhibiting strong Mg~II emission may be undergoing a transformation from bursty (i.e., high-sSFR), minimally-attenuated, low-mass objects to a more mature population in which attenuation by dust precludes observations of  emission lines. Robust estimates of galaxy ages will be instrumental for testing the hypothesis that galaxies with strong Mg~II emission lines represent a stage of galaxy evolution that perhaps a large fraction of the galaxy population evolves through. Probing the geometry and morphology of extended Mg~II emission is additionally important for understanding the distribution of gas in and around galaxy halos. 

While objects with strong Fe~II$^*$ emission show some properties in common with Mg~II emitters -- i.e., larger $W_{\rm [OII]}$ and lower $A_{\rm UV}$ -- the dependencies of Fe~II$^*$ and Mg~II emission strengths on galaxy properties are not identical. In particular, Fe~II$^*$ emission is not as closely linked with sSFR and $M_*$ as Mg~II emission is. Since sSFR and $M_*$ trace star formation and the build-up of stellar mass, the tight relationship between these properties and Mg~II strength would be expected if Mg~II emission originated in star-forming regions \citep{erb2012}. On the other hand, if Fe~II$^*$ emission arises in galactic halos, as \citet{giavalisco2011} suggest, then it would be surprising if Fe~II$^*$ emission was strongly modulated by properties describing star-forming environments. Observations that Mg~II emission is spatially extended \citep{martin2013,erb2012} are not inconsistent with Mg~II emission originating in star-forming clumps, as Mg~II is a resonant line highly susceptible to scattering. 

\begin{figure*}
\begin{center}$
\begin{array}{c}
\includegraphics[trim=0in 0.05in 0in 0in,clip,width=6in]{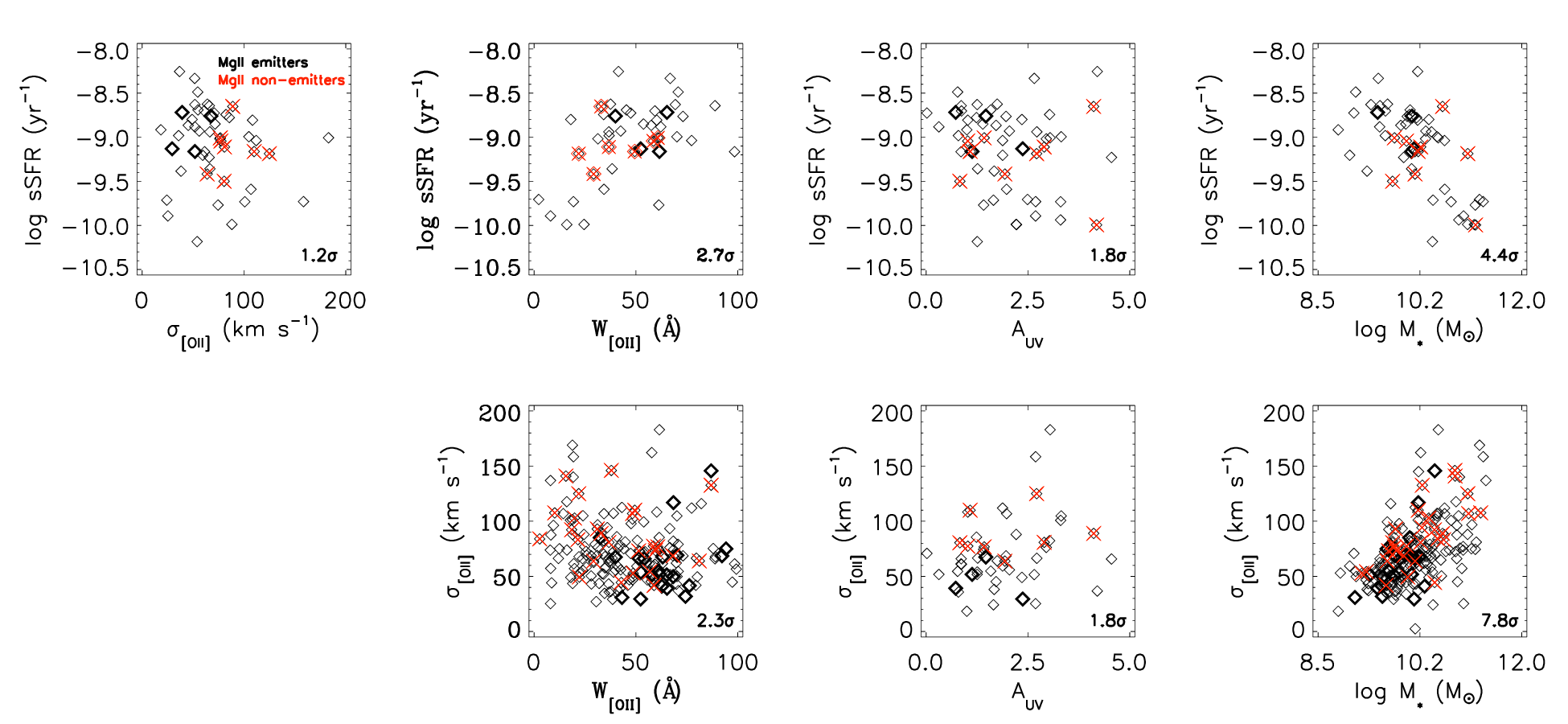} \\ 
\includegraphics[width=6in]{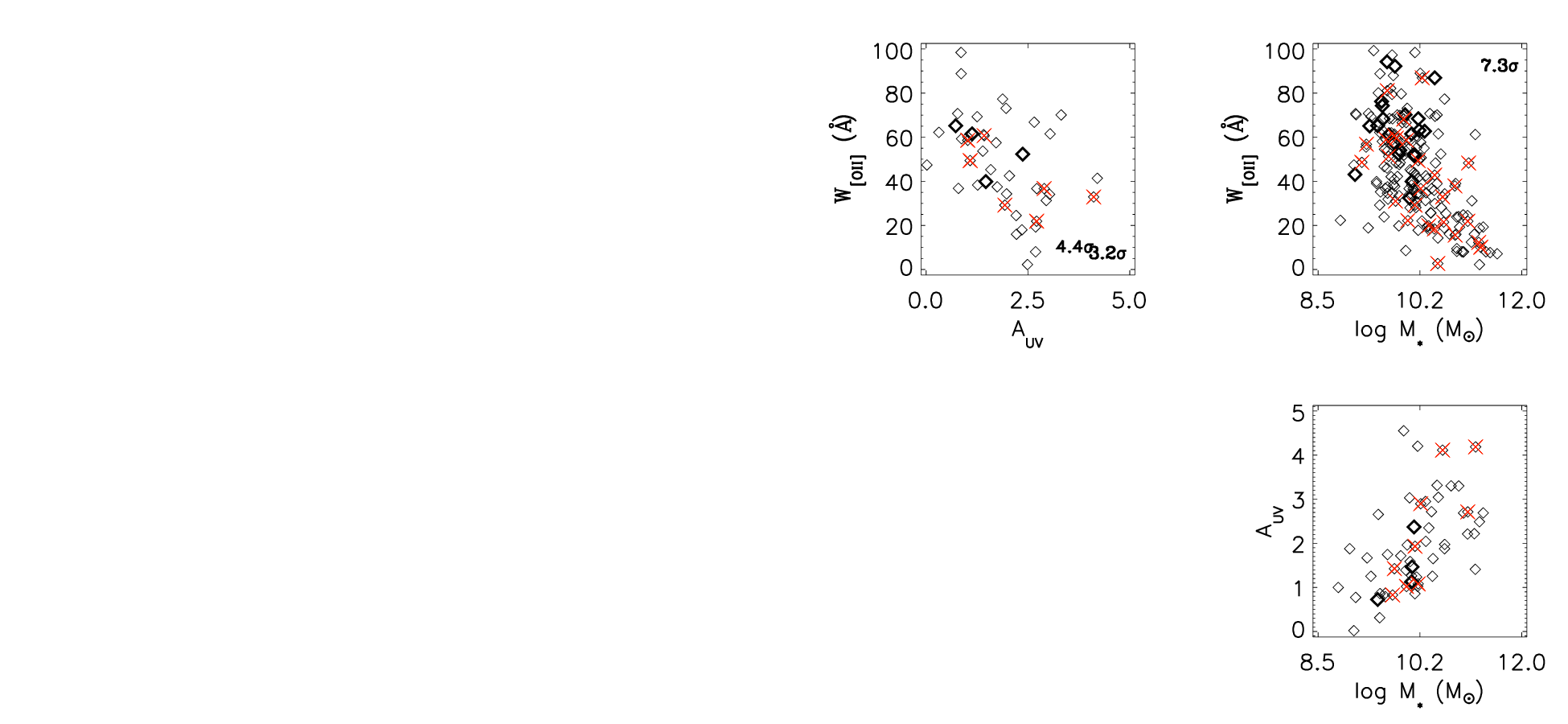} 
\end{array}$
\end{center}
\caption{Intercorrelations of sSFR, $\sigma_{\rm [OII]}$, $W_{\rm [OII]}$, $A_{\rm UV}$, and $M_*$. Mg~II emitters are indicated with thick black diamonds and Mg~II non-emitters are shown as red stars. We propose that both sSFR and $A_{\rm UV}$ strongly modulate Mg~II emission, such that objects with stronger Mg~II emission typically have lower sSFR and lower $A_{\rm UV}$ than objects with weak or absent Mg~II emission.}
\label{MgIIintercorrelations}
\end{figure*}

\subsection{The Absence of Fe~II$^*$ Emission in Local Samples} \label{sec: local}

\begin{figure*}
\centering
\includegraphics[width=7in]{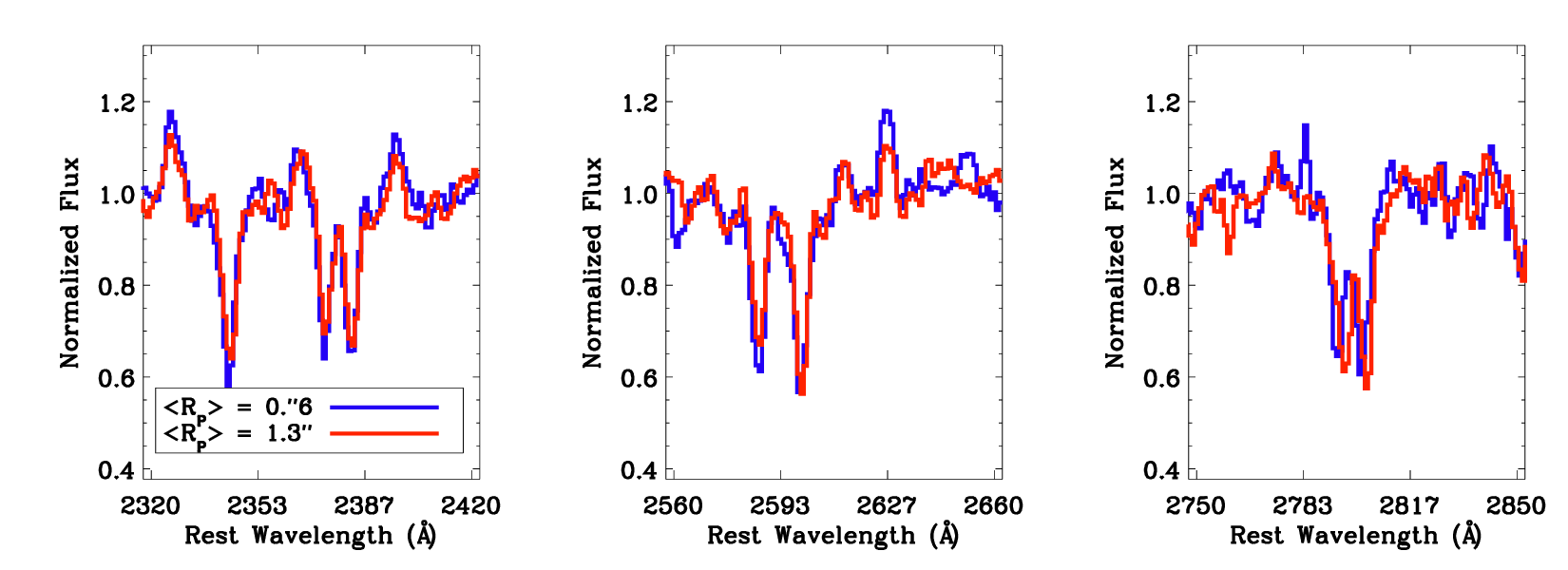} 
\caption{Composite spectra assembled on the basis of angular Petrosian radius. The stack of smaller objects ($\langle$R$_{\rm P}$$\rangle$ = 0.$''$6) is shown in blue while the composite spectrum comprised of larger objects ($\langle$R$_{\rm P}$$\rangle$ = 1.3$''$) is shown in red. Typical errors on both composite spectra are $\sim$0.04, in normalized flux units. Smaller objects show stronger 2396 and 2626 \AA\ Fe~II$^*$ emission than larger objects at the 1.8$\sigma$ level, an indication that our data are consistent with the theory of Fe~II$^*$ emission slit losses presented by \citet{giavalisco2011} and \citet{erb2012}. Smaller objects also exhibit more blueshifted Mg~II centroids, analogous to the results of \citet{law2012} for a sample of star-forming galaxies at \emph{z} = 1.5--3.6.}
\label{angular_size}
\end{figure*} 

While Fe~II$^*$ emission lines are prevalent in samples at \emph{z} $\geq$ 0.5, including star-forming and post-starburst galaxies, AGNs, and quasars \citep[][this work]{wang2008,coil2011,giavalisco2011,rubin2010b,rubin2010c,erb2012}, Fe~II$^*$ emission is conspicuously absent in local starbursts. Figure \ref{FeII_compare} contrasts the composite spectrum of \z1 star-forming galaxies from this work with a composite spectrum of \z0 star-forming galaxies from \citet{leitherer2010}. One immediately notices the lack of Fe~II$^*$ emission in the local sample\footnote{Mg~II emission above the continuum is also absent in the \citet{leitherer2010} composite spectrum, although obvious Mg~II emission is likewise not present in the composite spectrum of our own data.}. \citet{giavalisco2011} propose that spectra from nearby samples lack Fe~II$^*$ emission due to slit losses. Given the small physical-to-angular conversion valid for the local universe -- 170 pc/$''$ at the average redshift of the \citet{leitherer2010} sample -- spectroscopic observations at \z0 fail to encompass the halos of galaxies where \citet{giavalisco2011} suggest Fe~II$^*$ emission originates. Indeed, the \citet{leitherer2010} observations target only small H~II regions of size $\sim$100 pc. Spectroscopic observations of distant galaxies, in comparison, are inclusive of extended emission given typical slit widths of $\sim$ 1$''$ and physical-to-angular conversions of 8 kpc/$''$ at \emph{z} = 1. The hypothesis that fine-structure emission originates in extended galaxy halos is supported by the results of \citet{jones2012}. These authors used a sample of 81 Lyman break galaxies at \z4 to show that the equivalent width of a fine-structure Si~II$^*$ emission line was significantly less than the equivalent width of its paired resonant Si~II absorption line. One would expect that the equivalent widths of the fine-structure and resonant absorption lines to be equal if both features originated from similar spatial scales. The result that the emission has a smaller equivalent width is consistent with Si~II$^*$ emission being more spatially extended than the resonant absorption and therefore falling beyond the spectroscopic slit. Other authors have also invoked slit losses to explain the absence of Si~II$^*$ emission in local samples \citep[e.g.,][but see \citealt{france2010}]{schwartz2006} while high-redshift (\z3) observations show Si~II$^*$ \citep[e.g.,][]{shapley2003}. Since differences in $A_{\rm UV}$ are strongly linked to changes in Fe~II$^*$ emission strength, as discussed above, it is important to note that 
there is redshift evolution in $A_{\rm UV}$ at a fixed SFR, such that objects at higher redshifts show less dust attenuation, on average \citep{adelberger2000}. Therefore, the differences in Fe~II$^*$ emission strength between local and \z1 samples may be due both to slit losses and differences in dust attenuation. 

\begin{figure*}
\centering
\includegraphics[width=6in]{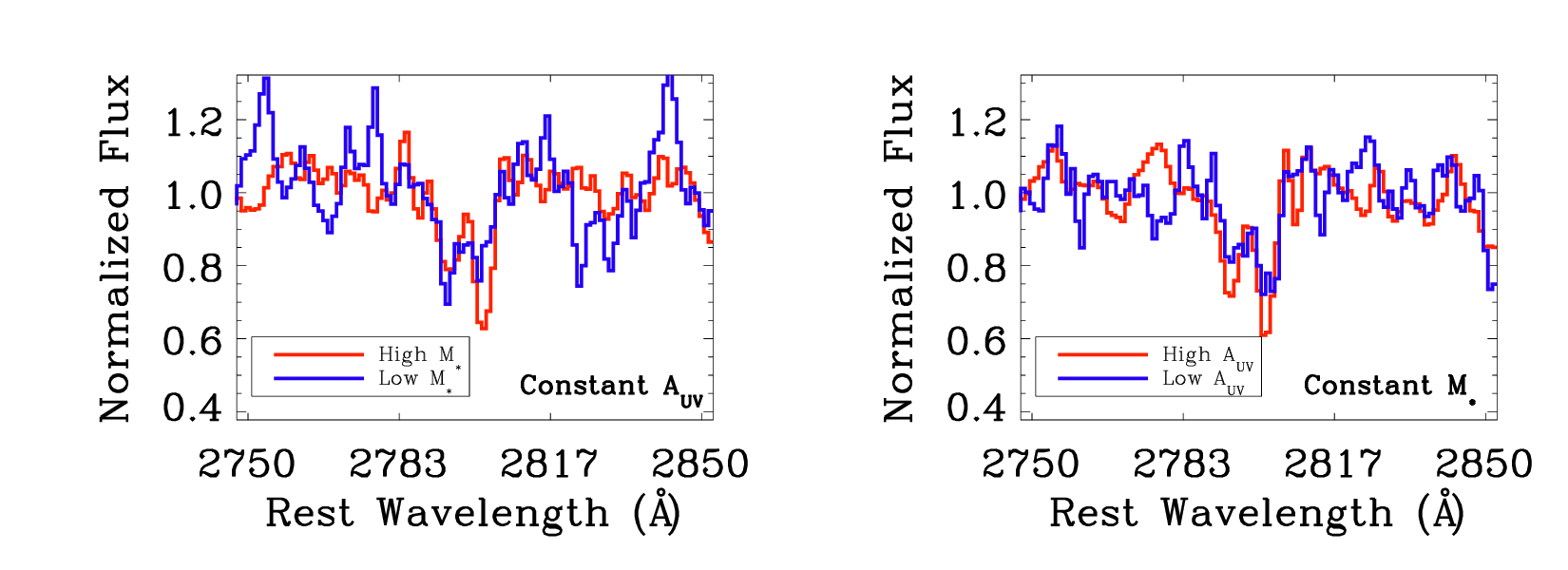} 
\caption{Mg~II profiles of composite spectra assembled holding $M_*$ ($A_{\rm UV}$) constant and varying $A_{\rm UV}$ ($M_*$). Mg~II variation is seen in both pairs of composite spectra, indicating that both $M_*$ and $A_{\rm UV}$ modulate Mg~II. However, the composite spectra divided by $A_{\rm UV}$ show stronger and more consistent variation in their Mg~II profiles (D$_{\rm MgII}$ = 0.71 \AA) than the composite spectra divided by $M_*$ (D$_{\rm MgII}$ = 0.00 \AA), suggestive that $A_{\rm UV}$ modulates Mg~II more strongly than $M_*$. It is important to acknowledge that these composite spectra were each assembled from fewer than 10 individual spectra and are accordingly of lower S/N than the majority of the composite spectra in this paper.}
\label{MgII_AUVM}
\end{figure*}

If the frequency of detecting Fe~II$^*$ emission depends on the spatial scale probed by spectroscopic observations, as \citet{giavalisco2011} suggest, one would expect to see variation in the strength of Fe~II$^*$ emission as a function of galaxy angular size. We used Petrosian radii -- measured in the \emph{V} band for objects at \emph{z} $<$ 1.10 and in the \emph{I} band for objects at \emph{z} $>$ 1.10 -- to divide the sample into two groups based on angular size. From composite spectra assembled from each group, we found that smaller objects ($\langle$R$_{\rm P}$$\rangle$ = 0.$''$6, where R$_{\rm P}$ is the Petrosian radius) showed $\sim$50\% stronger 2396 and 2626 \AA\ Fe~II$^*$ emission than larger ($\langle$R$_{\rm P}$$\rangle$ = 1.$''$3) objects, significant at the 1.8$\sigma$ level (Figure \ref{angular_size}). The average change in angular diameter distance between the two groups of objects is only ~12\%. While these results are consistent with Fe~II$^*$ emission arising from spatially-extended winds, we caution that solely dividing galaxies on the basis of angular size is a blunt tool for analysis given the diversity of galaxies populating each angular size bin. Angular size is correlated with redshift at the 2.7$\sigma$ level, and redshift is in turn correlated with SFR, $A_{\rm UV}$, and $W_{\rm [OII]}$ (Figure \ref{sfr_z}). In the absence of a larger sample, it is difficult to isolate objects that vary in angular size but do not vary significantly in other galaxy properties.  

The lack of Fe~II$^*$ emission in local samples has motivated the hypothesis that slit losses preclude observations of extended Fe~II$^*$ emission in nearby galaxies. \citet{prochaska2011} find that Fe~II$^*$ emission is spatially extended with non-zero surface brightness at large galactocentric radii; a 1$''$ slit covering a galaxy at \emph{z} $>$ 0.5 would include less than 50\% of the Fe~II$^*$ emission, assuming a spherically-symmetric wind. At lower redshifts, even less of the Fe~II$^*$ emission would fall into the spectroscopic slit. While \citet{prochaska2011} focus on Fe~II$^*$ emission arising in the presence of gas flows, Fe~II$^*$ lines are generated even in the absence of galactic winds. While not all objects in our sample show evidence for outflows, we used spatially-resolved imaging in \citet{kornei2012} and outflow fraction calculations in \citet{martin2012} to infer that the prevalence of outflows is likely modulated by galaxy inclination. All star-forming systems at \z1 may in actuality drive outflows, but only a fraction of objects exhibit outflow signatures depending on the orientation of the outflowing wind with respect to the observer.

\section{Summary and Conclusions} \label{sec: conclusions}

Fine-structure Fe~II$^*$ and resonant Mg~II emission lines, observable from the ground over a wide range of redshifts, are an important probe of gas flows. We have investigated the properties and prevalence of Fe~II$^*$ and Mg~II emission in a sample of 212 star-forming galaxies at \z1. We utilized LRIS rest-frame ultraviolet spectroscopy and a rich data set of \emph{GALEX}, \emph{HST}, and \emph{Spitzer} imaging. Our study focused on the kinematics of Fe~II$^*$ and Mg~II emission and how the strength of these lines vary as a function of star-forming, gas flow, interstellar gas absorption, stellar population, size, morphological, and redshift properties. We provide below a numbered list of our main conclusions:

1. Fe~II$^*$ emission is prevalent at \z1 in composite spectra assembled from a range of galaxy properties, although Fe~II$^*$ emission is not observed in local studies probing star-forming regions. This absence of Fe~II$^*$ emission in nearby samples may be due to slit losses. 

2. The centroids of the strongest Fe~II$^*$ emission lines are consistent with the systemic velocity of the galaxy and Fe~II$^*$ emission may consequently originate either in the disk of the galaxy or in a spatially-extended outflowing wind. 

3. Fe~II$^*$ emission is primarily modulated by $A_{\rm UV}$, where less dusty systems show stronger Fe~II$^*$ emission. Objects selected on the basis of strong Fe~II$^*$ emission also tend to show stronger Fe~II resonant absorption, although we caution that this effect may be primarily driven by redshift evolution as Fe~II$^*$ emitters are preferentially at higher redshifts than Fe~II$^*$ non-emitters in our sample. 

4. Mg~II emission is most pronounced in systems with higher sSFR, lower $A_{\rm UV}$, and lower $M_*$. We find the strongest correlation between Mg~II emission strength and sSFR.

We have demonstrated that galaxies with strong Fe~II$^*$ or Mg~II emission have typically higher sSFR, lower $A_{\rm UV}$, and lower $M_*$ than the sample as a whole. These objects may accordingly represent a bursty (i.e., high-sSFR), minimally-attenuated, low-mass stage of galaxy evolution. Future studies of emission in star-forming galaxies will benefit from targeted searches of galaxies with higher sSFR, lower $A_{\rm UV}$, and lower $M_*$ properties. As Fe~II$^*$ is thought to originate in extended galaxy halos and the $A_{\rm UV}$ values measured in this study reflect the attenuation toward H~II regions, upcoming work will be necessary for understanding the relationship between dust in star-forming regions and dust at larger galactocentric distances. 
 
Investigations of the spatial and morphological properties of galactic winds in emission -- information that is lacking from most current long-slit spectroscopic observations -- are critical for detailed modeling of outflows and testing of non-sperically symmetric models of galactic winds. Several studies thus far have offered tantalizing evidence that resonant and fine-structure emission may be spatially extended beyond the stellar continuum, although larger sample sizes are needed. New instrumentation such as KCWI and MUSE will yield data on the three dimensional structure of gas flows. Upcoming observations with these instruments will be important for understanding the enrichment of the circumgalactic medium and the connections between galaxies and their environments. 

\begin{figure}
\centering
\includegraphics[width=3.5in]{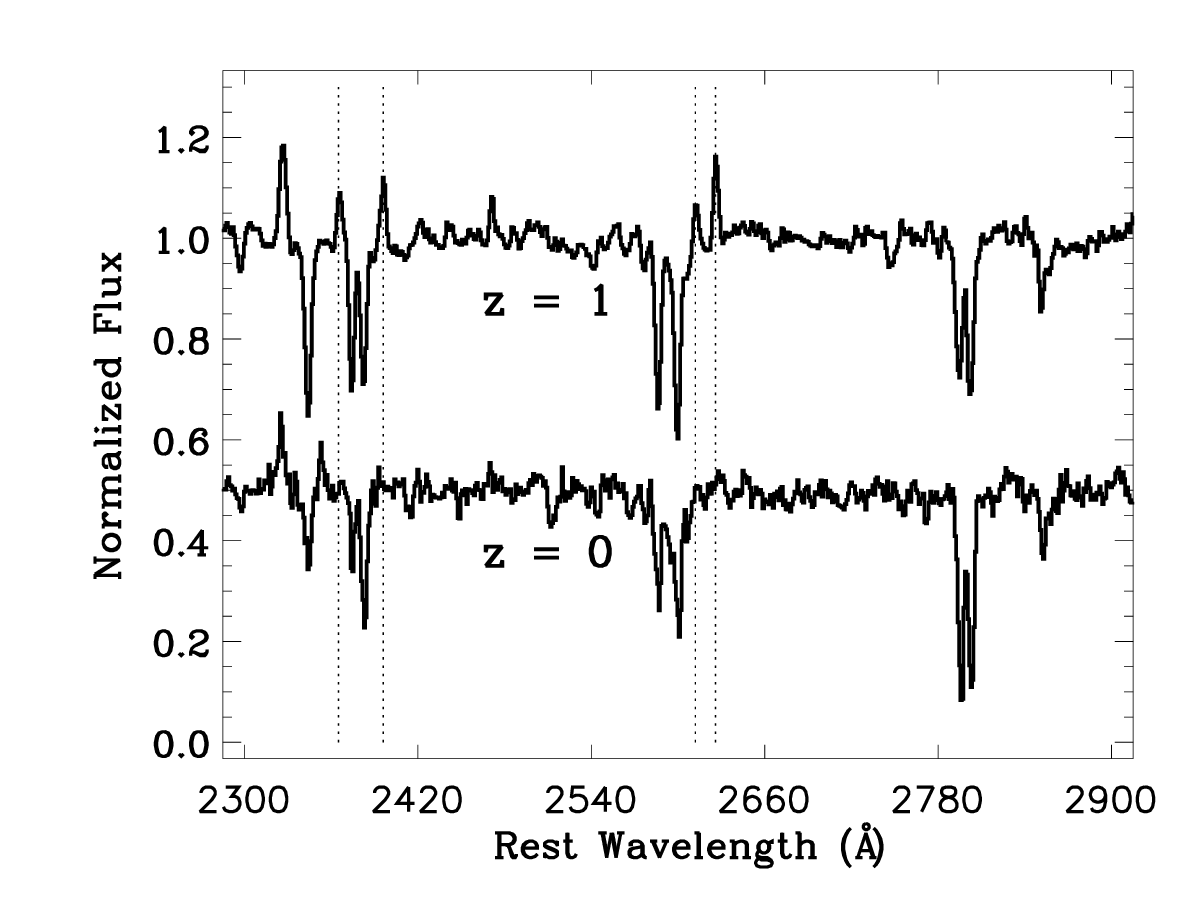} 
\caption{Comparison of composite spectra of star-forming galaxies at \z1 (this work) and \z0 \citep{leitherer2010}. Fe~II$^*$ emission (vertical dotted lines) is conspicuous in the \z1 sample while the \z0 sample (28 local star-forming galaxies observed in 46 unique pointings) does not show Fe~II$^*$ emission. \citet{giavalisco2011} and \citet{erb2012} hypothesize that slit losses may be responsible for the lack of Fe~II$^*$ emission in nearby samples. As the \citet{leitherer2010} pointings target very small spatial scales (starburst regions of size $\sim$100 pc), spatially-extended Fe~II$^*$ emission would be missed by these observations. Spectroscopic observations of galaxies at \z1, on the other hand, typically encompass the entire galaxy given the physical-to-angular conversion of $\sim$8 kpc/$''$.}
\label{FeII_compare}
\end{figure}

\begin{acknowledgements}
We thank Kevin Hainline for providing code used in our Monte Carlo simulations. K.A.K. is grateful for support from a UCLA Dissertation Year Fellowship. A.E.S. acknowledges support from the David and Lucile Packard Foundation. This study was supported in part by the NSF under contract AST--0909182 (C.L.M.). A portion of this work was completed at the Aspen Center for Physics (C.L.M.). The Alfred P. Sloan Foundation and an NSF CAREER award (AST-1055081) supported A.L.C. This study makes use of data from AEGIS, a multiwavelength sky survey conducted with the \emph{Chandra}, \emph{GALEX}, \emph{Hubble}, Keck, CFHT, MMT, Subaru, Palomar, \emph{Spitzer}, VLA, and other telescopes and supported in part by the NSF, NASA, and the STFC. We also recognize and acknowledge the very significant cultural role and reverence that the summit of Mauna Kea has always had within the indigenous Hawaiian community.  We are most fortunate to have the opportunity to conduct observations from this mountain.
\end{acknowledgements}

\nocite{wright2006}

\clearpage

\bibliography{../../../../Paper2/Outflow_refs}

\end{document}